\documentclass[a4paper,11pt]{article}
\pdfoutput=1 % if your are submitting a pdflatex (i.e. if you have
\usepackage{jheppub} % for details on the use of the package, please
\usepackage{hyperref}% add hypertext capabilities
\usepackage{booktabs}% more beautiful tables
\usepackage[T1]{fontenc} % if needed
\usepackage{siunitx} % numbers and units
\usepackage{xcolor,graphicx}% Include figure files
\usepackage{dcolumn}% Align table columns on decimal point
\usepackage{bm}% bold math
\usepackage[normalem]{ulem}
\usepackage{braket} % bra-kets for quantum states 
\usepackage{amsmath}
\usepackage{cancel}%to cross out parts of equations with the command \cancel{}
\usepackage{slashed}
\usepackage{multicol}
\usepackage[capitalise]{cleveref}
\usepackage{amssymb}
\DeclareFontFamily{U}{futm}{}
\DeclareFontShape{U}{futm}{m}{n}{
  <-> s * [.95] fourier-bb
  }{}
\DeclareSymbolFont{Ufutm}{U}{futm}{m}{n}
\DeclareMathAlphabet{\mathbbb}{U}{bbold}{m}{n}

\crefformat{subsubsection}{\S#2#1#3}

\counterwithin*{equation}{section}
\allowdisplaybreaks

\newcommand{\nnl}{\nonumber \\}

\newcommand{\cO}{\mathcal{O}}

\newcommand{\M}{{\mathcal M}}

\newcommand{\mpl}{M_{\mathrm Pl}}

\newcommand{\re}{{\mathrm{Re}} \,}

\newcommand{\hc}{\textrm{h.c.}}

\usepackage{etoolbox,siunitx}
\robustify\bfseries

\begin{document}

%=======================================================================
\title{Spinning waveforms of scalar radiation 
\\ in quadratic modified gravity}
%=======================================================================

\author{Adam Falkowski and}
\author{Panagiotis Marinellis}

\affiliation{Universit\'{e} Paris-Saclay, CNRS/IN2P3, IJCLab, 91405 Orsay, France}

\emailAdd{adam.falkowski@ijclab.in2p3.fr,
panagiotis.marinellis@ijclab.in2p3.fr 
          }

%=======================================================================
\abstract{
We study scalar-tensor gravitational theories using on-shell amplitude methods. 
We focus on theories with gravity coupled to a massless scalar via the Gauss-Bonnet and Chern-Simons terms.   
In this framework, we calculate the waveforms for classical scalar radiation emitted in scattering of macroscopic objects, including spin effects. 
To this end, we use the Kosower-Maybee-O'Connell formalism, with the 5-particle amplitude for scalar emission in matter scattering calculated at tree level using the unitarity-factorization bootstrap techniques. 
We also discuss in detail the dependence of that amplitude on the contact terms of the intermediate  4-particle scalar-graviton-matter amplitude. 
Finally, we discuss the conditions for resolvability of classical scalar radiation. 
}
%===================================================================
\maketitle

%=================================================
\section{Introduction} 
%================================================

%
The detection of gravitational waves by the collaborative efforts of LIGO, Virgo, and KAGRA \cite{LIGOScientific:2016aoc,LIGOScientific:2017vwq,KAGRA:2021vkt} represents a significant breakthrough in astrophysics. These collective endeavors have propelled gravitational wave astronomy into the forefront, urging us to investigate a broader range of gravitational phenomena with ever greater precision.

As the main source of gravitational waves are binary systems of compact astrophysical objects, there has been a lot of interest in the study of the problem of binary dynamics in the theory of general relativity (GR). Some of the most popular and successful approaches in that direction have been the post-Newtonian (PN) expansion, which consists of an expansion in weak gravitational fields and non-relativistic velocities \cite{Levi:2018nxp,Porto:2016pyg,Blanchet:2013haa} and the effective one-body formulation \cite{EOB1, EOB2, EOB3, EOB4, EOB5}. Recently, though, there has been a burst of approaches which focus on an expansion in the gravitational coupling $G$ only, which is referred to as post-Minkowskian (PM) expansion. These include worldline effective theory~\cite{WEFT1,WEFT2,WEFT3,WEFT4,WEFT5,WEFT6}, worldline quantum field theory \cite{WQFT1,WQFT2,WQFT3}, effective field theory methods~\cite{Solon1,Bern1,Bern2}, the exponential representation of the S~matrix~\cite{Exp1,Exp2}, heavy-mass effective theories \cite{Heavymass1,Heavymass2}, the Kosower-Maybee-O'Connell formalism (KMOC) \cite{Kosower:2018adc, Maybee:2019jus, Cristofoli:2021vyo}, the eikonal approach \cite{Eikonal1,Eikonal2,Eikonal3,Eikonal4} as well as methods generalising the eikonal with the use of KMOC \cite{EikonalKMOC}. 
As for now, results for spinless objects have been derived for the conservative dynamics up to 4PM and first self-force order in~\cite{WQFT3} (and in \cite{Bern:2024adl} for extremal black holes in $\mathcal{N}=8$ supergravity) and for gravitational radiation up to one loop from a variety of different approaches \cite{Jakobsen:2021smu, Riva:2021vnj, Brandhuber:2023hhy, Elkhidir:2023dco, Herderschee:2023fxh, Caron-Huot:2023vxl}. 
Spin effects have also been implemented in each of these approaches to study scattering data \cite{Ben-Shahar:2023djm, Liu:2021zxr, Bern:2020buy, Aoude:2022trd, Luna:2023uwd} as well as radiation \cite{Jakobsen:2021lvp, Riva:2022fru, DeAngelis:2023lvf, Aoude:2023dui, Brandhuber:2023hhl, Bohnenblust:2023qmy}. 
So far, much less efforts have been devoted to employ modern QFT methods for calculating effects {\em beyond} GR.\footnote{Certain scalar tensor theories were discussed in Refs.~\cite{Bhattacharyya:2024aeq, Bhattacharyya:2024kxj} using worldline quantum field theory methods.} 
Very recently, Ref.~\cite{Brandhuber:2024bnz} studied effects of cubic interactions in an effective field theory approach of gravity using the KMOC formalism. 
Furthermore, 
amplitude methods were used to calculate the conservative potential in scalar-tensor theories at second PM order~\cite{Davis:2023zqv}. 
.

In this article we apply the recent developments in QFT calculations of classical phenomena to theories beyond GR.  
We use the KMOC approach and we focus on scalar-tensor theories, studying their leading order effects.
In this class of theories, one introduces a new massless scalar degree of freedom  $\phi(x)$ coupled non-minimally to gravity.
 We consider the action of the form 
 
\begin{align}
S = \int d^4 x {\cal L}, 
\qquad 
{\cal L} = 
{\cal L}_{\rm EH}
+ {\cal L}_{\rm SGB/DCS}[\phi, g_{\mu \nu}] 
+ {\cal L}_{\rm matter}[\Phi,g_{\mu \nu}]
     ,\end{align}

where

\begin{align}
       &{\cal L} _{\rm EH} =  \frac{\mpl^2}{2} \sqrt{-g} R \\
    &{\cal L}_{\rm SGB/DCS} =  \sqrt{-g} \Big[
    \mpl \big (
    f_1(\phi) R_{\rm GB}^2 
    + f_2(\phi) R_{\rm CS}^2 )
    +\frac{1}{2} (\nabla^\mu \phi \nabla_\mu \phi)\Big] 
, \end{align}
and ${\cal L}_{\rm matter}$ describes the minimal coupling of matter $\Phi$ of arbitrary spin to gravity.
The Gauss-Bonnet invariant is given in terms of the Riemann tensor, Ricci tensor and Ricci scalar as by 
$R_{\rm GB}^2=R^{\mu \nu \rho \sigma}R_{\mu \nu \rho \sigma}-4R^{\mu \nu}R_{\mu \nu} + R^2$. 
The Chern-Simons invariant (Pontryagin density) 
is given by $R_{\rm CS}^2 = R^{\mu \nu \rho \sigma} \tilde{R}_{\mu \nu \rho \sigma}$ with $\tilde{R}^\mu_{ \ \nu \rho \sigma} = \frac{1}{2} \varepsilon_{\rho \sigma}^{\ \ \ \alpha \beta} R^{\mu }_{ \ \nu \alpha \beta}$.  
The Lagrangian above should be viewed as an effective theory with a cutoff $\Lambda$,
truncated to quadratic order in the Riemann tensor and to the leading order in derivative expansion of the scalar. 
We assume that the spacetime is asymptotically Minkowski, and  that the field $\phi$ vanishes fast enough at spatial infinity. 

In the following, we restrict our attention to function $f_i(\phi)$ that admit a Taylor expansion around small $\phi$ perturbations $f_i(\phi)=f_i(0) + f_i'(0) \phi + \mathcal{O}(\phi^2)$, 
and we define 
$f_1'(0)\equiv  \frac{\alpha}{\Lambda^2}$, 
$f_2'(0) \equiv \frac{\tilde{\alpha}}{\Lambda^2}$, such that the coupling constants $\alpha$ and $\tilde{\alpha}$ are dimensionless.
The terms proportional to $f_i(0)$ do not alter the equations of motion since they are topological. 
On the other hand, the terms proportional to $\alpha$ and $\tilde \alpha$ yield a non-minimal coupling between gravity and a scalar field, which arise in a class of modified gravity theories. 
which reduce to GR when taking the scalar field to vanish. 
In particular, if we fix $f_2(\phi) \to 0$, the choice of $f_1(\phi) = \frac{\alpha}{\Lambda} e^{\frac{\phi}{\Lambda}}$ defines the string-inspired Einstein-dilaton Gauss Bonnet (EdGB) theory~\cite{Kanti:1995vq, Maeda:2009uy}. Moreover, The simplest choice $f_1(\phi) = \frac{\alpha}{\Lambda^2} \phi $ is referred to as the shift-symmetric scalar Gauss-Bonnet theory~\cite{Sotiriou:2013qea, Sotiriou:2014pfa}. We will refer to all cases that induce that coupling with the Gauss-Bonnet invariant as scalar-Gauss-Bonnet theories (SGB). 
On the other hand, for $f_1(\phi) \to 0$, the choice $f_2(\phi) = {\tilde{\alpha} \over \Lambda^2} \phi$ defines the dynamical Chern-Simons gravity~\cite{Alexander:2009tp, Yunes:2009hc}~(DCS). 
These theories have been extensively studied in the literature in the context of gravitational wave emission, since they alter the dynamics of compact objects' binaries~\cite{Yagi:2011xp, Yagi:2012vf, Yagi:2013mbt, Witek:2018dmd, Julie:2019sab, Shiralilou:2021mfl, Julie:2024fwy, Loutrel:2018ydv}. 
In particular, the binary system may  lose energy faster than what is predicted in GR, as it can radiate away the scalar field quanta.  

In SGB and DCS gravity theories, it has been shown that black holes (BH) acquire scalar "hair", except for the case of non-spinning objects in DCS gravity~\cite{Kanti:1995vq, Yagi:2011xp, Sotiriou:2013qea, Sotiriou:2014pfa, Perkins:2021mhb, Alexander:2009tp, Yunes:2009hc}.
On the other hand, other astrophysical objects, e.g. neutron stars (NS) do not get scalarized in these theories. 
From the QFT perspective, scalar hair would induce a coupling of our matter fields (BHs) to the scalar.  
In the following we do not consider objects with scalar hair, leaving those to an upcoming publication,  and therefore matter is assumed not to couple directly to $\phi$.

In this paper we  focus on radiation emitted by arbitrary spinning objects with no scalar hair in the aforementioned scalar-tensor theories.
Using on-shell unitarity methods and spinor helicity variables, along with the appropriate classical limit of QFT amplitudes, we extract the classical piece of the amplitude describing  scattering of matter fields with emission of the scalar. Then, using the KMOC formalism we connect that amplitude to observable quantities, namely to the scalar waveform at any spin order, from which we can extract the radiated power due to  scalar emission. 
This can be interpreted as energy lost to exotic scalar radiation by astrophysical objects such as neutron stars or non-spinning black holes in the DCS theory (for black holes with scalar hair this source of radiation also exists, but is expected to be subleading). 

In order to derive the scalar emission amplitude we proceed recursively by "gluing" lower-point on-shell amplitudes, starting from 3-particle amplitudes at complex kinematics. 
Amplitudes for arbitrary masses and spins of this type have already been translated in spinor-helicity language in Ref.~\cite{Arkani-Hamed:2017jhn} and we are going to make use of this construction to address spin effects more easily. 
On our way,  we will face the four-point amplitude describing the scattering of a graviton off a massive spinning particle, producing a massless scalar particle in the final state 
$\M[1_{\Phi} 2_{\bar\Phi} 3^s_h 4_\phi ]$, where the $s$ superscript denotes helicity and the subscripts $h$ denotes the graviton. In the spinning case, the classical limit of
this amplitude cannot be uniquely fixed 
by unitary factorization, as it admits contact terms which survive the classical limit \emph{at any non-zero spin order}.  
The study of this amplitude bear some similarities to the one of the Compton amplitude in the case of QED and gravity. 
In that case, the helicity-flipping amplitudes constructed via unitarity methods and assuming minimal coupling, suffer from unphysical poles for spins $S>1$ and $S>2$ respectively. 
It is, thus, necessary to fix this problem by adding additional terms which cancel the unphysical pole~\cite{Chung:2018kqs,Falkowski:2020aso, Aoude:2022trd, Cangemi:2023bpe, Bjerrum-Bohr:2023iey, Azevedo:2024rrf, Bautista:2022wjf}. 
The ambiguity of this procedure corresponds to the freedom of adding contact terms to the amplitude.  
In our setting we will see that the amplitude we will construct will have no unphysical poles. 
Nevertheless, the techniques worked out in the context of Compton amplitudes can be adapted to derived the full set of classical contact terms for  $\M[1_{\Phi} 2_{\bar\Phi} 3^s_h 4_\phi ]$.
We will list the contact terms at the linear order in spin, and discuss how they affect the observable waveforms.
We will also 
describe how to systematically include all possible contact terms 
at higher orders in spin. 
We will remain, however, agnostic for the moment concerning the coefficients and the source of these contact terms, 
i.e. we do not match these contact terms to particular solutions in scalar tensor theories.

This article is organized as follows. 
In~\cref{sec:waveforms} we review the basics of the KMOC formalism and define the observable scalar waveforms that we compute. 
Before plunging into scalar-tensor theories, 
in~\cref{sec:dilaxion} we study a simpler toy model, with gravity replaced by QED, with the scalar coupled non-minimally to photons  via the interactions $\sim \phi F^{\mu \nu}F_{\mu \nu}$ and $\sim \phi F^{\mu \nu} \tilde{F}_{\mu \nu}$\footnote{Similarly to the case of the Riemann we define $\tilde{F}_{\mu \nu} = \frac{1}{2} \varepsilon_{\mu \nu \alpha \beta} F^{\alpha \beta}$}.
This allows us to introduce our amplitude methods and review the techniques of waveform integration in a somewhat simpler setting where the results are more compact. 
We mention that these interactions have also some physical relevance in the context of compact objects in Einstein-Maxwell-dilaton-axion theories~\cite{PhysRevD.43.3140,HOROWITZ1991197,Garcia:1995qz,Julie:2017rpw,Julie:2018lfp,Khalil:2018aaj}.
The main results are contained in~\cref{sec:gbcs}. 
We recursively construct the spinning 5-particle scalar emission amplitude in the shift-invariant SGB/DCS theory. 
We discuss the dependence of that amplitude on the contact terms of the 4-particle amplitude related to  it by unitary factorization. 
Finally we explicitly compute the scalar waveforms in hyperbolic scattering of matter particles up to linear order in spin, and discuss power emitted in scalar radiation. 
Some more technical material has been relegated to appendices.
In~\cref{app:INT} we describe the method we used to compute the integrals required to calculate the waveforms present in the main text. 
In~\cref{app:CONT} we discuss the construction of the classical contact terms at arbitrary order in spin. 
In~\cref{app:allorders} we give partial (in the sense of not including  contact terms) results concerning the waveforms at arbitrary orders in spin.

Before beginning, 
we briefly  summarize our conventions. 
We  work with the mostly minus metric $\eta_{\mu \nu} = (1,-1,-1,-1)$, 
and the natural units $\hbar = c = 1$.
The sign convention for the totally anti-symmetric Levi-Civita tensor $\epsilon^{\mu\nu\rho\alpha}$ is 
$\epsilon^{0123} = 1$. 
The Lorentz-invariant integration measure is 
$d\Phi_k \equiv {d^4 k \over (2\pi)^4 } \delta(k^2-m^2)\theta(k^0)$. 
For 3-vectors we use the bold notation: 
$\boldsymbol{x} \equiv \vec{x}$.  
We use the all-incoming convention for our on-shell amplitudes unless stated otherwise.
In the context of GR, the Christoffel connection is
$ \Gamma^\mu_{\ \nu \rho} =   
{1 \over 2}  g^{\mu \alpha} \left (
\partial_\rho g_{\alpha \nu} +  \partial_\nu g_{\alpha \rho} - \partial_\alpha g_{\nu \rho} \right ) 
$, 
the Riemann tensor is 
$R^\alpha_{\ \mu \nu \beta}  =   
 \partial_\nu \Gamma^\alpha_{\ \mu \beta} -  \partial_\beta \Gamma^\alpha_{\ \mu \nu}  
 +  \Gamma^\rho_{\ \mu \beta}  \Gamma^\alpha_{\ \rho \nu}
 - \Gamma^\rho_{\ \mu \nu}  \Gamma^\alpha_{\ \rho \beta}  
$, 
and the Ricci tensor is 
$R_{\mu \nu}  =  R^\alpha_{\ \mu \nu \alpha}$.

%==============================================
\section{Waveforms for scalar emission} 
\label{sec:waveforms}
%==============================================

In this section we review the connection between the waveform for classical scalar radiation and five-point amplitudes for scalar emission, following the KMOC formalism of Ref.~\cite{Cristofoli:2021vyo}. 
The asymptotic observable that we will focus on in this paper is the expectation value of the scalar field:  
 \begin{align}  
{\cal R}_\phi  \equiv  {}_{\rm out} \langle \psi | \phi(x) | \psi \rangle_{\rm out}
 .   \end{align}
Here, $\ket{\psi}_{\rm out}$ is an asymptotic out state describing the particles being scattered, and $\phi(x)$ is a real quantum field corresponding to the radiated scalar. 
Writing 
$\phi(x)  = \int d\Phi_k  \big [ a_ke^{-ikx}  +a_k^\dagger  e^{ ikx}  \big ]$,  
with the creation/annihilation operators satisfying
$[a_k, a_{k'}^\dagger] = 2 E_k (2 \pi)^3 \delta^3(\boldsymbol{k} - \boldsymbol{k'})$ , 
we have 
  \begin{align}  
{\cal R}_\phi  =  & 
\int d\Phi_k e^{-ikx}   {}_{\rm out} \langle \psi | a_k \ket{\psi}_{\rm out} + \hc 
=  \int d\Phi_k e^{-ikx}   \bra{\psi}  S^\dagger a_k  S \ket{\psi} +  \hc 
 ,    \end{align}
 where $\ket{\psi}$ is the asymptotic in state (with the "in" label suppressed), 
and $S$ is the unitary operator evolving the in states into out states, 
$\ket{\psi}_{\rm out} = S \ket{\psi}$. 
Restricting to a two-body scattering  state,  $\ket{\psi} = \ket{\Phi_1 \Phi_2}$,  and writing the $S$ matrix as $S = 1 + i T$,   
   \begin{align}  
{\cal R}_\phi  =  & 
  \int d\Phi_k e^{-ikx}   \bigg [
i \langle \Phi_1 \Phi_2 |  a_k  T |  \Phi_1 \Phi_2  \rangle 
+ \langle \Phi_1 \Phi_2 | T^\dagger a_k  T |  \Phi_1 \Phi_2  \rangle  \bigg ] +  \hc 
 .   \end{align}
Our analysis in this paper will be restricted to the leading order, where only the first term in the square bracket is relevant. 
Then the radiation observable can be written in the form 
\begin{align}  
\label{eq:SW_RphiOfJ}
{\cal R}_\phi^{\rm LO}   =  & 
 i \int d \Phi_k \big [ \tilde J(k) e^{-i k x}  - \tilde J(k)^* e^{i k x} \big ]  
. \end{align}
 with  the current defined as 
       \begin{align}  
\tilde J_\phi   =  & 
\langle \Phi_1 \Phi_2  k|   T |  \Phi_1 \Phi_2  \rangle 
 .   \end{align}
 Plugging in the wave packets for the $\Phi_i$ states, taking the leading term in the classical limit, and integrating over the wave packets assuming they are sharply peaked at the classical momentum $p_i$, one can express the current in the following form: 
  \begin{align}
  \label{eq:SW_JofM}
\tilde J_\phi   =  & 
  {1  \over 16 \pi^2 } 
  \int d \mu \M_5^{\rm cl}
 .   \end{align} 
Here, the integration measure is defined as 
\begin{align}  
\label{eq:SW_dmu}  
d\mu  = &  
\delta^4(w_1 + w_2 - k ) 
\Pi_{i=1,2} [e^{i b_i w_i}  d^4 w_i  \delta (p_i w_i) ]    
,    \end{align} 
where $b_i$ are the impact parameters of the $\Phi_i$ states. 
On the other hand, $\M_5$ in \cref{eq:SW_JofM} abbreviates the two-to-three amplitude describing scattering of $\Phi_1$ and $\Phi_2$ states with scalar emission:
  \begin{align} 
\M_5 \equiv 
\M(\Phi_1(p_1+w_1) \Phi_2(p_2+w_2) \to \Phi_1(p_1) \Phi_2(p_2) \phi(k) )
. \end{align} 
The ${}^{\rm cl}$ label in \cref{eq:SW_JofM} stands for the leading terms under the classical scaling: 
  \begin{align} 
  \label{eq:SW_ClassicalScaling}
p_i \to \hbar^0 p_i, \qquad 
w_i \to \hbar^1 w_i, \qquad  
k \to \hbar^1 k, \qquad 
a_i \to \hbar^{-1} a_i , \qquad 
b_i \to \hbar^{-1} b_i 
  . \end{align}
That is to say, we are interested in the limit where both the emitted scalar momentum $k$ and the exchanged momenta $w_i$ are soft. 
$a_i$ denote the classical spin vectors of the scattering particles normalized to their masses, 
$a_i^\mu \equiv S_i^\mu/m_i$. 
The spins are formally taken to infinity in the limit $\hbar \to 0$, with the product 
$S_i k$ remaining finite. 

Quite generally, Ref.~\cite{Cristofoli:2021vyo} 
proves that if an observable can be written in the form of \cref{eq:SW_RphiOfJ} with a current localized near $x =0$, then at large $\boldsymbol{x}$ that observable can be approximated as  
\begin{align}
{\cal R}_\phi = { W_\phi \over |\boldsymbol{x}| } +  \dots 
, \end{align}
where $W_\phi$ is called the {\em waveform}, which is a function of time and the unit vector $\boldsymbol{n} \equiv 
{\boldsymbol{x} \over |\boldsymbol{x}|} $ describing the direction of emitted radiation. 
The dots stand for terms vanishing faster at large $|\boldsymbol{x}|$, which do not contribute to radiation. 
Furthermore, the waveform can be related to the current as~\cite{Cristofoli:2021vyo}
\begin{align}
\label{eq:SW_WphiOfJ}
W_\phi = { 1 \over 4 \pi } \int_{-\infty}^\infty {d \omega \over 2 \pi}  \tilde J(\omega, \omega \boldsymbol{ \hat n}) e^{-i\omega t }  
\end{align}
where $t \equiv x^0 - |\boldsymbol{x}|$. 
It is convenient to introduce the {\em spectral waveform} $f_\phi$ as the Fourier transform of $W_\phi$:
$f_\phi \equiv \int_{-\infty}^\infty dt  W_\phi e^{i \omega t}$. 
The spectral waveform is a function of the radiated momentum $k^\mu = \omega n^\mu$, 
where $n^\mu \equiv (1, \boldsymbol{n})$ such that $n^2 = 0$. 
Comparing \cref{eq:SW_WphiOfJ} and \cref{eq:SW_JofM} one obtains a succinct expression for the spectral waveform for $\omega >0$: 
\begin{align}
\label{eq:SW_fphiOfM}
f_\phi = &  
{1  \over 64 \pi^3 }  \int d \mu \M_5^{\rm cl}|_{k = \omega n} 
. \end{align}
For $\omega < 0$, $f_\phi(\omega) = f_\phi(-\omega)^*$, as required by the reality of $\phi$ and $W_\phi$. 
Given the spectral waveform, one can calculate the waveform in the time domain via the inverse Fourier transform: 
\begin{align}
\label{eq:SW_WphiOWM}
W_\phi = &  \int_{-\infty}^\infty {d \omega \over 2\pi} f_\phi  e^{-i \omega t}
 = 
  \int_0^\infty {d \omega \over 2\pi} f_\phi  e^{-i \omega t}  + \hc  
\end{align} 

To calculate the radiated power we need another radiation observable: 
  \begin{align}  
{\cal R}_{\partial \phi}^\mu  = {}_{\rm out} \langle \psi | \partial^\mu \phi(x) | \psi \rangle_{\rm out} 
 .   \end{align} 
 From this definition, it is clear that the spectral waveform for $\omega>0$ is just \cref{eq:SW_fphiOfM} multiplied by $-i k^\mu$: 
   \begin{align}  
\label{eq:SW_spectralFormScalarDerivative}       
 f_{\partial\phi}^\mu  =  - i \omega n^\mu f_\phi  ,      \end{align} 
and consequently 
   \begin{align}  
W_{\partial \phi}^\mu = n^\mu \partial_t W_\phi 
  .      \end{align} 
The energy momentum tensor of the scalar field is 
$T^{\mu \nu} =  \partial^\mu \phi  \partial^\nu \phi  - {1\over 2}\eta^{\mu\nu} \partial^\alpha \phi  \partial_\alpha \phi $, 
in particular $T^{00} =  {1 \over 2} ((\partial_t \phi)^2 + (\boldsymbol{\nabla} \phi)^2  )$. 
Therefore, the angular distribution of the  power emitted in scalar radiation is given by  
  \begin{align}   
  \label{eq:SW_power}
{d P_\phi \over d \Omega} = {(W_{\partial\phi}^0)^2 + (W_{\partial\phi}^k)^2  \over 2}  =  
(\partial_t W_\phi)^2 
   .   \end{align}  
Total radiated power is 
$P_\phi = \int d \Omega (\partial_t W_\phi)^2$.

%===============================================
\section{Toy model: QED + dilaton/axion} 
\label{sec:dilaxion}
%================================================

Before delving into scalar radiation in gravitational theories, we first consider a toy model where gravity is replaced with electromagnetism. 
The model shares several features with the SGB/DCS scalar-tensor theory discussed in the next section, in particular calculation of scalar radiation waveforms follows a very similar template, however the results are somewhat more compact and transparent. 
In this sections we will discuss in some detail the technicalities involved in calculation of scalar emission amplitudes, so we can be more brief in the following section.

%--------------------------
\subsection{Setup} \label{SetupQED}

We consider a framework where a massless photon described by the vector field $A_\mu$ interacts with a massless spin-0 particle described by the real scalar field $\phi$. 
The system is described by the EFT Lagrangian 
\begin{align}  
\label{eq:DIL_Lagrangian}
 {\cal L} = & 
 - {1 \over 4}F_{\mu \nu} F_{\mu \nu}  
 + {1 \over 2} \partial_\mu \phi \partial_\mu \phi  
 - {\alpha \over 2 \Lambda } \phi F_{\mu\nu} F_{\mu\nu} 
 - {\tilde \alpha \over 2 \Lambda } \phi F_{\mu\nu} \tilde F_{\mu\nu}  
 , \end{align} 
 where $F_{\mu \nu} = \partial_\mu A_\nu - \partial_\nu A_\mu$, 
 and $ \tilde F_{\mu\nu} = {1 \over 2} \epsilon_{\mu\nu\alpha \beta } F_{\alpha\beta}$, 
 $\Lambda$ is the EFT cutoff, 
and $\alpha$ and $\tilde \alpha$ parametrize the strength of CP-even and CP-odd interactions.
In the  limit $\tilde \alpha \to 0$  the scalar is CP-even and will be referred to as the dilaton, for $\alpha \to 0$ the scalar becomes  CP-odd and will be referred to as the axion.

In the spinor language, the interactions between the scalar and the photon in \cref{eq:DIL_Lagrangian}  correspond to on-shell 3-point amplitudes of the form 
\begin{align}  
\label{eq:3.2}
\M \big [1_\gamma^- 2_\gamma^- 3_\phi \big ]  = & {\hat \alpha \over \Lambda} \langle 1 2 \rangle^2 
, \nnl 
\M \big [1_\gamma^+ 2_\gamma^+ 3_\phi \big ] = & {\hat \alpha^* \over \Lambda} [12]^2 
, \nnl  
\M \big [1_\gamma^- 2_\gamma^+ 3_\phi \big ]  = &0 , \nnl
\M \big [1_\gamma^+ 2_\gamma^- 3_\phi \big ]  = &0 
, \end{align} 
where $\hat \alpha = \alpha +  i \tilde \alpha$. 

The setup also contains charged massive matter particles $\Phi_i$, which will ultimately represent scattering classical chaged objects. 
We assume that the scalar is not directly coupled to matter. 
We also assume the photon coupling to matter is minimal,  that is all anomalous multipole moments vanish. 
The  minimal photon coupling to charged particles of arbitrary spin $S$ in the spinor language corresponds to the on-shell 3-point amplitudes of the form~\cite{Arkani-Hamed:2017jhn}
\begin{align}
\label{eq:SPD_M3pointQED}
\M \big [1_{\Phi_i} 2_{\bar \Phi_i} 3_\gamma^- \big ] = & 
- \sqrt 2 Q_i e {\langle 3 |p_1|\tilde{\zeta}]  \over [3 \tilde{\zeta}]} {[\bm 2 \bm1]^{2S} \over m^{2S}}
, \nnl 
\M  \big [ 1_{\Phi_i} 2_{\bar \Phi_i}  3_\gamma^+ \big ] = & 
- \sqrt 2 Q_i e {\langle \zeta|p_2|3]  \over \langle 3 \zeta \rangle}  { \langle \bm 2 \bm 1 \rangle^{2S} \over m^{2S} } 
, \end{align}
where $| \zeta \rangle$ and $| \tilde \zeta ]$ are arbitrary reference spinors (on-shell amplitudes do not depend on their choice). 
Massive spinors are denoted with the usual bold notation~\cite{Arkani-Hamed:2017jhn}, and we suppress their little group indices unless otherwise noted. 

Using the relation between the massive spinors in 3-particle on-shell kinematics: 
    \begin{align}
&| \bm 2 \rangle = 
- | \bm 1 \rangle  - \frac{|p_3|\bm 1]}{2m}   
, \nnl
&| \bm 2 ] = |\bm1] + \frac{|p_3|\bm 1\rangle}{2m} 
,    \end{align}
one can rewrite the minimal interactions as 
\begin{align} 
\M[1_{\Phi_i} 2_{\bar \Phi_i} 3_\gamma^-]  = &
- \sqrt 2 Q_i e {\langle 3 |p_1|\tilde{\zeta}]  \over [3 \tilde{\zeta}]}   
 \bigg  [ \mathbbb{1} +  { p_3 a_1   \over 2 S }      \bigg ]^{2S}
- \sqrt 2 Q_i e {\langle 3 |p_1|\tilde{\zeta}]  \over [3 \tilde{\zeta}]}  
 \bigg  [ \mathbbb{1}   -  { p_3 a_2  \over 2 S }      \bigg ]^{2S}
   ,   \nnl 
  \M[ 1_{\Phi_i} 2_{\bar \Phi_i} 3_\gamma^+]  = & 
- \sqrt 2 Q_i e {\langle \zeta|p_2|3]  \over \langle 3 \zeta \rangle}      
 \bigg  [  \mathbbb{1}   -   { p_3 a_1  \over 2 S }      \bigg ]^{2S}  
 =  
 - \sqrt 2 Q_i e {\langle \zeta|p_2|3]  \over \langle 3 \zeta \rangle}     
 \bigg  [  \mathbbb{1}  +  { p_3 a_2   \over 2 S }      \bigg ]^{2S}  
, \end{align}  
with the  spin vector defined as 
 \begin{align} 
   \label{eq:SPD_spinvector}
[a_n^\mu]^J_K = 
- S {\langle \bm n^J |\sigma^\mu| \bm n_K] 
+ [\bm n^J |\bar \sigma^\mu| \bm n_K \rangle    \over 2m^2} 
,  \end{align} 
for $n= 1,2$. 
Here, $J$ and $K$ refer to the $SU(2)$ little group indices of each particle\footnote{In the literature the more compact notation $a_n^\mu = 
- S {\langle \bm n |\sigma^\mu| \bm n]    \over m^2}$ is often used, where the symmetrization of the $SU(2)$ little group indices of the two matter particles is implicit.}. 
This definition of the spin vector is consistent with the identification $a^\mu = \frac{s^\mu}{m} = \frac{1}{m^2} \langle \mathbb{W}^\mu \rangle$, where $\mathbb{W}^{\mu}$ is the Pauli-Lubanski operator, obeying the commutation relations $[\mathbb{W}^\mu, \mathbb{W}^\nu] = 
- i  \varepsilon^{\mu \nu \rho \sigma} \mathbb{W}_\rho \mathbb{P}_\sigma$, with $\mathbb{P}^\mu$ representing the momentum operator, following the conventions in \cite{Maybee:2019jus}. With this definition, the spin vector is connected to the spin tensor via $S^{\mu \nu} = \frac{1}{m} \varepsilon^{\mu \nu \rho \sigma} p_\rho s_\sigma $ and it's subject to the Spin Supplementary Condition (SSC):

\begin{align}
    S^{\mu \nu} p_\nu = 0 .
\end{align}

The classical limit involves taking the limit of large $S$, cf.~\cref{eq:SW_ClassicalScaling}. 
In the limit $S \to \infty$, the 3-point amplitudes are recast in the exponentiated form \cite{Guevara:2018wpp,Guevara:2019fsj,Arkani-Hamed:2019ymq}: 
\begin{align} 
\label{eq:SPD_M3pointQEDexp}
\M^{\rm cl}[  1_{\Phi_i} 2_{\bar \Phi_i} 3_\gamma^-]  = &
- \sqrt 2 Q_i e {\langle 3 |p_1|\tilde{\zeta}]  \over [3 \tilde{\zeta}]} e^{p_3 a_1} 
= 
- \sqrt 2 Q_i e {\langle 3 |p_1|\tilde{\zeta}]  \over [3 \tilde{\zeta}]}    e^{-p_3 a_2}  
   ,   \nnl 
  \M^{\rm cl}[ 1_{\Phi_i} 2_{\bar \Phi_i} 3_\gamma^+]   = & 
- \sqrt 2 Q_i e {\langle \zeta|p_2|3]  \over \langle 3 \zeta \rangle}    e^{-p_3 a_1}   
=  
- \sqrt 2 Q_i e {\langle \zeta|p_2|3]  \over \langle 3 \zeta \rangle}  e^{p_3 a_2}  
.  \end{align}  
This form is particularly convenient for  calculations in the  classical limit. We will also assume from here onwards that $S$ is an integer for simplicity, to avoid tracing 
$(-)^{2S}$ factors.

We move to the 4-point amplitude describing matter annihilation into a photon and a scalar: 
${\cal M}_4^h \equiv {\cal M}[1_\Phi 2_{\bar{\Phi}} 3_\gamma^h 4_\phi ]$.
We split it as:

\begin{align}
    {\cal M}_4^h  = {\cal M}_U^h + {\cal M}_C^h \label{eq:3.8}
\end{align}

The first part satisfies the factorization property required by unitarity: 
${\rm Res}_{q^2\to 0} {\cal M}_U^h = 
- \sum_{h'} {\cal M} [1_\Phi 2_{\bar{\Phi}} (-q)_\gamma^{h'}] {\cal M} [q^{h'} 3^h 4_\phi]$, 
where the 3-particle on-shell amplitudes are the ones in \cref{eq:3.2}, \cref{eq:SPD_M3pointQED}. For this reason, we will refer to this piece as \emph{unitarity part} or \emph{unitarity amplitude}. 
The second part are the contact terms, which have the same little group properties as ${\cal M}_U^h$, but they do not have any kinematic poles, 
in particular 
${\rm Res}_{q^2\to 0} {\cal M}_C^h = 0$. 
We find:\footnote{%
Note the arbitrariness in defining ${\cal M}_U^h$ as it can be shifted by the contact terms.} 

    \begin{align}
&\M_U[1_\Phi 2_{\bar{\Phi}} 3^-_\gamma 4_\phi ] = - \frac{\sqrt{2} Q e \hat \alpha}{ \Lambda  } \frac{\langle 3| p_1 q | 3\rangle}{q^2} \frac{\langle \bm 2 \bm 1 \rangle^{2S}}{m^{2S}}  \nnl
&\M_U[ 1_\Phi 2_{\bar{\Phi}} 3^+_\gamma 4_\phi] = - \frac{\sqrt{2} Q e \hat \alpha^*}{ \Lambda  } \frac{[3|p_1 q | 3]}{q^2} \frac{[\bm 2 \bm 1 ]^{2S}}{m^{2S}} .
    \end{align}
Note that these amplitudes do not have a shift symmetry, in the sense they do not vanish as $p_4 \to 0$ (in that limit 
$q |3]/q^2 
= -p_4|3]/(2p_3p_4) 
= -|4\rangle/\langle 34 \rangle$).  
At four-points, we can again express the massive spinors in terms of one another. Explicitly, for $q^2<4m^2$~\cite{Aoude:2020onz, Cangemi:2023ysz}:
    \begin{align}
&|\bm 2 \rangle = - \frac{1}{\sqrt{1-\frac{q^2}{4m^2}}} \bigg [ | \bm 1 \rangle - \frac{|q|\bm 1]}{2m}   \bigg ] , \nnl
&| \bm 2 ] = \frac{1}{\sqrt{1-\frac{q^2}{4m^2}}}  \bigg [ |\bm 1] - \frac{|q|\bm 1\rangle}{2m}  \bigg ] . 
    \end{align}
where $q =  p_1 + p_2 = - p_3 - p_4$.  
Using this, the 4-point amplitude can be rewritten as:
  \begin{align}
&\M_U[1_\Phi 2_{\bar{\Phi}} 3^-_\gamma 4_\phi ] =  - \frac{\sqrt{2} Q e \hat \alpha}{ \Lambda  } \frac{\langle 3| p_1 q | 3\rangle}{q^2} \frac{1}{[1-\frac{q^2}{4m^2}]^S} \bigg [\bigg(1 -\frac{q^2}{4m^2} \bigg)\mathbbb{1} + \frac{q \cdot a_1}{2S}  \bigg]^{2S} \nnl
&\M_U[ 1_\Phi 2_{\bar{\Phi}} 3^+_\gamma 4_\phi] = - \frac{\sqrt{2} Q e \hat \alpha^*}{ \Lambda  } \frac{[3|p_1 q | 3]}{q^2} \frac{1}{[1-\frac{q^2}{4m^2}]^S} \bigg [\bigg(1 -\frac{q^2}{4m^2} \bigg)\mathbbb{1} - \frac{q \cdot a_1}{2S}  \bigg]^{2S}
    \end{align}
In the classical limit the amplitudes take the exponential form:
    \begin{align}
\label{eq:3.10}
&\M_U^{\rm cl}[ 1_\Phi 2_{\bar{\Phi}} 3^-_\gamma 4_\phi] = - \frac{\sqrt{2} Q e \hat \alpha}{ \Lambda  } \frac{\langle 3| p_1 q | 3\rangle}{q^2} e^{q \cdot a_1} \nnl
&\M_U^{\rm cl}[ 1_\Phi 2_{\bar{\Phi}} 3^+_\gamma 4_\phi] = -\frac{\sqrt{2} Q e \hat \alpha^*}{ \Lambda  } \frac{[3|p_1 q | 3]}{q^2} e^{- q \cdot a_1}  .
    \end{align}
Note that the classical scaling of the above is 
${\cal M}_U^h \to \hbar^0 {\cal M}_U^h$, 
which is the same as for the standard QED Compton amplitude $\M_U[1_\Phi 2_{\bar{\Phi}} 3^{s}_\gamma 4^{s'}_\gamma]$. 
In contrast to the helicity-flipping QED Compton amplitude with minimal couplings, the amplitude in \cref{eq:3.10} does not suffer from unphysical poles for any spin. 

Let us now discuss about the contact term part  ${\cal M}_C^h$. We are interested in the contact terms with the same classical scaling as ${\cal M}_U^h$, that is 
${\cal M}_C^h \to \hbar^0 {\cal M}_C^h$. We will give two points of view:1
1) by direct inspection (which is sufficient at low spin orders), 
and 2) a more systematic one based on the construction that first appeared in Ref.~\cite{Bautista:2022wjf} to identify all possible contact terms with a specific $\hbar$ scaling and was applied to the case of gravitational Compton scattering. 
The latter is discussed in \cref{app:CONT} in detail. 

We will build the contact terms order by order in powers of the spin vector: 
$\M_C = \sum_{n=0}^\infty \M_C^{(n)}$ 
where $\M_X^{(n)} \sim a^n$. 
At the zero-th order it is easy to see that no contact terms with required properties exist, 
$\M_C^{(0)} = 0$, 
because $|3\rangle \sim |3] \sim \hbar^{1/2}$, 
and its impossible to lower the classical scaling in the absence of kinematic poles or the spin vector. 
 However, once spin is included, that is indeed possible taking into account the classical scaling of the spin vector $a_i \sim \hbar^{-1} a_i$.
At linear order in spin we can readily check that we can construct only one independent  contact term, which we parametrize as:
    \begin{align}
&\M_C^{(1)}[ 1_\Phi 2_{\bar{\Phi}} 3^-_\gamma 4_\phi] = \hat C_1  \frac{\sqrt{2} Q e }{ \Lambda  } \langle 3 | p_1 a_1 |  3  \rangle  , \nnl
&\M_C^{(1)}[ 1_\Phi 2_{\bar{\Phi}} 3^+_\gamma 4_\phi] = - \hat C_1^*  \frac{\sqrt{2} Q e }{ \Lambda  } [3|p_1 a_1 | 3] , \label{eq:3.12}
    \end{align}
with $\hat C_1 = C_1 + i \tilde C_1$ a dimensionless Wilson coefficient. The relative sign between the two opposite helicity amplitudes is determined by crossing symmetry which implies $\M[1_\Phi 2_{\bar{\Phi}} 3^-_\gamma 4_\phi ] = \M[ (-2)_{\Phi} (-1)_{\bar{\Phi}}  (-3)^+_\gamma (-4)_\phi]^*$ . 
In %appendix
\cref{appQEDCONT} we prove that this is actually the only classical contact term generated at this order and we also describe the algorithm to build contact terms beyond the linear order. 
Notice that this contact terms does not respect the shift symmetry either. 
If one insist on the shift symmetry in the axion  limit, one needs to fix $\tilde C_1 =0$.

%----------------------------------------
\subsection{Scalar emission amplitudes} \label{sec:ScalarEmissionAmplitudesQED}

We now move on to construct  5-point amplitude, 
 at the leading order in classical  scaling, describing matter scattering 
with a scalar emission, which we denote as $\M_5 \equiv \M[ (p_1+w_1) _{\Phi_1}  (p_2+w_2) _{\Phi_2} (-p_1) _{\bar \Phi_1}  (-p_2) _{\bar \Phi_2} (-k)_\phi]$ from here on. 
This is an input needed to calculate the scalar waveforms via \cref{eq:SW_fphiOfM}. 
More precisely, tree level level radiation is determined by the residues of $\M_5$ at $w_i^2 \to 0$, which  can be bootstrapped from the 3- and 4-particle on-shell amplitudes discussed previously. 
Consequently,  the 5-particle contact terms  do not contribute to classical radiation at tree level\footnote{Formally, this follows from $\int d \mu \,  {\rm const}  \sim \delta(b)$, 
where $d\mu$ is the integration measure in \cref{eq:SW_dmu}.} and we will not track them. 
For the presentation purpose, it is convenient to split the 5-particle amplitude as 
$\M_{5}^{\rm cl} = 
\M_{5U}^{\rm cl} + \M_{5C}^{\rm cl}$. The two pieces are defined by their residues:

\begin{align}
 {\rm Res}_{w_2^2 \to 0}   \M_{5X}^{\rm cl} 
 = 
 - \sum_s \M_{X}^{\rm cl} \big [   (p_1+w_1) _{\Phi_1} (-p_1) _{\bar \Phi_1}  (w_2)_h^s   (-k)_\phi   \big ]
 \M^{\rm cl} \big [ (p_2+w_2) _{\Phi_2} (-p_2) _{\bar \Phi_2} (-w_2)_h^{-s}  \big ] 
,  \end{align} 
and $X = U,C$ refer to the unitarity and contact amplitudes  in \cref{eq:GBCS_M4pointPhiExp,eq:GBCS_M4pointPhiContact}.
The residues at $w_1^2 \to 0$ are simply obtained from the above by $1\leftrightarrow 2$. 

After some algebra, for the $\M_{5U}$ part we obtain 
\begin{align}
 \M_{5U} = -\frac{8 Q_1 Q_2 e^2 }{ \Lambda w_1^2 w_2^2} \bigg \{   &[(p_1 w_2)(p_2 w_1) - (p_1 p_2) (w_1 w_2)] [\alpha \cosh(  w_i a_i) + i \tilde{\alpha} \sinh(w_i a_i) ] \nnl
    & + \varepsilon_{\mu \nu \rho \sigma} p_1^\mu p_2^\nu  w_1^\rho w_2^\sigma [ \tilde \alpha \cosh(w_i a_i) - i \alpha \sinh( w_i a_i) ] \bigg \},
\end{align}
where $w_i a_i = w_1 a_1 + w_2 a_2$. 
We can further simplify the result by setting in the classical limit $p_i w_i \to 0$ 
(which is due to the on-shell condition $(p_i + w_i)^2 = m_i^2$ implying $p_i w_i \sim \hbar^2$).
Together with the relation $w_1 + w_2 = k$ this allows us to simplify
\begin{align}
     \M_{5U} = -\frac{8 Q_1 Q_2 e^2 }{ \Lambda w_1^2 w_2^2} \bigg \{   &[(p_1 k)(p_2 k) - (p_1 p_2) (w_1 w_2)] [\alpha \cosh(  w_i a_i) + i \tilde{\alpha} \sinh(w_i a_i) ] \nnl
    & + \varepsilon_{\mu \nu \rho \sigma} p_1^\mu p_2^\nu  k^\rho w_2^\sigma [ \tilde \alpha \cosh(w_i a_i) - i \alpha \sinh( w_i a_i) ] \bigg \} . \label{eq:3.24}
\end{align}
Note that this result is valid at all spin orders.

Finally, we discuss the effects of the 4-particle contact terms on the scalar emission amplitude. 
We do it order by order in spin: 
$\M_{5C} = \sum_{n=0}^\infty  \M_{5C}^{(n)}$. 
At the linear order in spin, 
the contact amplitude in \cref{eq:3.12} leads to the following contribution to the scalar emission amplitude:  
\begin{align}
\label{eq:DIL_M5C1}
    \M_{5C}^{(1)} =  \frac{4i  Q_1 Q_2 e^2}{ \Lambda w_2^2} \bigg \{ C_1 \varepsilon_{\mu \nu \rho \sigma} p_1^\mu p_2^\nu a_1^\rho w_2^\sigma  + 
     \tilde{C}_1 [(p_1 p_2) (w_2 a_1) - (p_1 k) (p_2 a_1)] \bigg \} + (1 \leftrightarrow 2) ,
\end{align}
where we relabeled $\hat C_1 = \frac{C_1+i \tilde{C_1}}{2}$. 
In principle, there is no conceptual difficulties to continue this procedure to higher order in spins, but for our QED toy model, we restrict to the linear order. 
The discussion of higher order contributions is performed in \cref{app:allorders} only in the context of the gravitational SGB/DCS framework, but the same logic described there can be applied for the QED toy model as well.

%---------------------------------------- 
\subsection{Waveforms} \label{sec:QED_waveforms}

We are now ready to derive the waveforms up to linear in spin order for the dilaton and the axion. To that end, we expand the part of the amplitude in \cref{eq:3.24} up to linear order in spin. Then we have:
\begin{align}
    \M_{5U}^{(0)} =   -\frac{8 Q_1 Q_2 e^2 }{ \Lambda w_1^2 w_2^2} \bigg \{   \alpha [(p_1 k)(p_2 k) - (p_1 p_2) (w_1 w_2)] 
     + \tilde \alpha \varepsilon_{\mu \nu \rho \sigma} p_1^\mu p_2^\nu  k^\rho w_2^\sigma  \bigg \} .
\end{align}
\begin{align}
    \M_{5U}^{(1)} = -\frac{8 i Q_1 Q_2 e^2 }{ \Lambda w_1^2 w_2^2} \bigg \{  \tilde{\alpha}   [(p_1 k)(p_2 k) - (p_1 p_2) (w_1 w_2)]  (w_i a_i) +\alpha \, \varepsilon_{\mu \nu \rho \sigma} p_1^\mu p_2^\nu  k^\rho w_2^\sigma ( w_i a_i)  \bigg \} .
\end{align}
The $\M_{5C}^{(1)}$ contribution is given at \cref{eq:DIL_M5C1}.
We will compute these three contributions separately for the dilaton and the axion case.

In doing so we will have to evaluate the waveforms using \cref{eq:SW_WphiOWM} and \cref{eq:SW_fphiOfM}, performing the intermediate integrals. We will use the method for computing these integrals, first presented in \cite{DeAngelis:2023lvf} and later in \cite{Brandhuber:2023hhl} to compute spinning gravitational waveforms. In \cref{app:INT} we review the main ideas of that method and we give an explicit formula for all integrals that may enter in the evaluation of any waveform at leading order and at any spin order. 
Below we simply quote the results after completing the integrals. 
The notation used for the presentation is also explained in \cref{app:INT}.

Up to  linear order in spin, the waveform for the dilaton reads 
\begin{align}  
\label{eq:DIL_Wphi}
 \hspace{-2cm}
 W_\phi =  & {Q_1  Q_2  e^2 \over 16 \pi^2 }  {1 \over b \Lambda}     \bigg \{ 
   { \alpha    \over  \sqrt{\gamma^2 - 1}  (\hat u_1n)  }
  { 1 \over \sqrt{ T_1^2 + 1} }   
   \bigg [ \gamma - 
 {  (\gamma^2-1)  (\hat u_2 n)   \big [  \gamma  (\hat u_2 n) - (\hat u_1 n)  +  (\tilde b n)  T_1  \big ]   \over
  [\gamma  (\hat u_2 n) - (\hat u_1 n)  +  (\tilde b n)  T_1 ]^2  +  (\tilde v n)^2 ( T_1^2 + 1)     }  
    \bigg ]  
  \nnl  \hspace{-2cm} -  & 
     {   \alpha  \over  (\hat u_1 n)  }       
{d \over  dz}   \bigg \{   {1    \over \sqrt{ z^2 + 1} } 
    \re \bigg [ {   (\tilde v n)   z   - i (\tilde b n)  \sqrt{ z^2   + 1 }  
 \over
  \gamma  (\hat u_2 n) - (\hat u_1 n)  +  (\tilde b n)  z + i (\tilde v n) \sqrt{ z^2 + 1}   } 
    \nnl \hspace{-2.5cm} & 
\bigg ( 
{  a_1^A  \over b}   \big [ n^A/(\hat u_1 n) -    \gamma    \hat u_2^A   - z \tilde b^A  -  i \sqrt{ z^2   + 1 }  \tilde v^A  \big ]
+ 
{ a_2^A   \over b}  \big [   - \hat u_1^{A}  + z \tilde b^{A}  +  i \sqrt{ z^2   + 1 }  \tilde v^{A}  \big ]
\bigg )   \bigg ] \bigg \}_{z =T_1}
  \nnl  \hspace{-2cm} -   & 
      C_1 { a_1^A  \over  b }  {\tilde v^A   \over (\hat u_1 n) (1+ T_1^2)^{3/2} }  
  \bigg \}  
  + (1 \leftrightarrow 2) 
,       \end{align} 
where 
    \begin{align} 
    T_i \equiv {t - b_i n \over   (\hat u_i n)  b }
    . \end{align}   

For the axion, the waveform in the time domain to linear order in spin reads 
 \begin{align}  
\label{eq:DIL_Wtildephi}
 \hspace{-2cm}
 W_{\tilde \phi}  =  & 
 {Q_1  Q_2  e^2  \over  16 \pi^2  } {1 \over b \Lambda}   \bigg \{   
    -   {\tilde \alpha  (\tilde v n)   \over   (\hat u_1 n) } 
 { 1  \over \sqrt{ T_1^2 + 1} }    
  {   T_1  [ \gamma  (\hat u_2 n) - (\hat u_1 n) ]      -  (\tilde b n) 
 \over
[  \gamma  (\hat u_2 n) - (\hat u_1 n)  +  (\tilde b n)  T_1 ]^2  +  (\tilde v n)^2 ( T_1^2 + 1)   }  
  \nnl  \hspace{-2cm} +  & 
 {  \tilde\alpha 
   \over  b  \sqrt{\gamma^2 - 1}  (\hat u_1 n)   }  
 {d \over d z}  \bigg \{ 
   { \gamma   \big [ - (a_1 n)/(\hat u_1 n) + \gamma   (\hat u_2 a_1)  + (\hat u_1 a_2)    +  z \tilde b^A ( a_1^A  - a_2^A)      \big ]     \over \sqrt{ z^2 + 1} }   
     \nnl    \hspace{-2.5cm} & 
+   {  (\gamma^2 - 1 )   (\hat u_2 n)   \over \sqrt{ z^2 + 1} }   \re \bigg [ {  
   (a_1 n)/(\hat u_1 n)    - \gamma  (\hat u_2 a_1) 
   - (\hat u_1 a_2)  - \big [ a_1^A - a_2^A \big ]   \big [  z \tilde b^{A}  +  i \sqrt{ z^2   + 1 }  \tilde v^{A}  \big ]  
 \over
  \gamma  (\hat u_2 n) - (\hat u_1 n)  +  (\tilde b n)  z + i (\tilde v n) \sqrt{ z^2 + 1}   } 
 \bigg ] 
  \bigg \}|_{z=T_1}
\nnl   \hspace{-2cm} + & 
 \tilde   C_1  { a_1^A   \over  b} 
 { T_1 \hat u_2^A  - \gamma \tilde b^A  
 \over   \sqrt{\gamma^2- 1}   (\hat u_1 n)   (1 + T_1)^{3/2} }  
    \bigg \} 
  + (1 \leftrightarrow 2) 
 .     \end{align} 

%---------------------------------
\subsection{Limits}

Below we discuss two limiting cases for the  waveforms in \cref{eq:DIL_Wphi,eq:DIL_Wtildephi}. 
We first consider the asymptotic behavior when $|t|$ is much larger than the impact parameter $b$.  
Expanding the waveforms for $t \to \pm \infty$ we obtain for the dilaton  
 \begin{align}   
  W_\phi = &   
 \pm  {\alpha Q_1 Q_2 e^2  \over 16 \pi^2   \Lambda  t } 
 \bigg \{ 
  {2 \gamma \over \sqrt{\gamma^2-1} }
 \big [ 1 + \cO(t^{-1}) \big ]
 \nnl & 
+     { a_1^\mu - a_2^\mu \over b} 
 \big [  (\hat u_1 n)^2 +  (\hat u_2 n)^2  - 2  \gamma  (\hat u_1 n)   (\hat u_2 n)  \big ]  
  { (\tilde v n) \tilde b^\mu  - (\tilde b n) \tilde v^\mu \over  (\tilde b n)^2  +  (\tilde v n)^2} 
 {b \over t}
\bigg \} 
   + \cO(t^{-3})
    .   \end{align} 
The large $t$ behavior probes the small frequency limit of the spectral waveform and we can read off 
    \begin{align} 
    f_\phi(\omega \to 0) \sim  
{Q_1 Q_2 e^2   \over \Lambda }  
\bigg (  1 + \# a \omega   \bigg ) 
+ {\cal O}(a^2) 
   .   \end{align}
 For the axion the asymptotic structure is slightly different  
    \begin{align}    
  W_{\tilde \phi} = &   
 \pm {Q_1  Q_2  e^2  \over  16 \pi^2  } {b \over  \Lambda t^2}  \bigg \{ 
  \tilde  \alpha  \big [  (\hat u_1 n)^2 +  (\hat u_2 n)^2  - 2  \gamma  (\hat u_1 n)   (\hat u_2 n)  \big ]  
 { (\tilde v n)  \over  (\tilde b n)^2  +  (\tilde v n)^2 }  
   \nnl   \hspace{-2.5cm} & +  
{  \tilde  \alpha \over b}  {\gamma \over  \sqrt{\gamma^2- 1}  } 
\big [ (a_1 n) + (a_2 n)  
- (\hat u_2 a_1)( \gamma  (\hat u_1 n) +  (\hat u_2 n) )  
- (\hat u_1 a_2)( (\hat u_1 n) +    \gamma   (\hat u_2 n) )  \big ]  
    \nnl   \hspace{-2.5cm} & 
   + {\tilde C \over b  \sqrt{\gamma^2- 1}  }  
     \big [   (\hat u_2 a_1) (\hat u_1 n) + (\hat u_1 a_2) (\hat u_2 n)    \big ] 
      \bigg \}  
    + \cO(t^{-3})
    .   \end{align} 
This corresponds to the  spectral function 
\begin{align} 
    f_\phi(\omega \to 0) \sim  
    {Q_1 Q_2 e^2   \over \Lambda } (\omega b)  
\bigg (  1 + \# {a \over b } \bigg )   + {\cal O}(a^2) 
 .   \end{align}  
We see that the spectral function vanishes at $\omega \to 0$. 
This is a  consequence of the soft theorems and the shift symmetry in the axion limit. 

\vspace{1cm}

We move to  another limit of the waveforms: $\gamma \to 1$. 
We define the velocity $\beta \equiv \sqrt{\gamma^2-1}$,  
and for simplicity we fix the particle two to be at rest: 
$u_1 = (\gamma,\beta \boldsymbol{\hat v})$,
$u_2 = (1,\boldsymbol{0})$, 
$b_1 = (0, b \boldsymbol{\hat b})$,
$n = (1, \boldsymbol{\hat n})$,
where $\hat x$ are unit vectors. 
Then, for $t \lesssim b$ at the zero-th order in spin we find 
\begin{align}  
\partial_t W_\phi \simeq & 
 - \alpha  {Q_1 Q_2 e^2  \over 16 \pi^2   \Lambda  b^2  } 
\bigg \{ 
{2 t \over b} + \boldsymbol{\hat{b}}\cdot\boldsymbol{\hat n}
\bigg \}  \beta^2 
   ,   \end{align} 
   \begin{align}  
\partial_t W_{\tilde \phi} = & 
- \tilde \alpha  {Q_1 Q_2 e^2  \over 16 \pi^2   \Lambda  b^2 }  
 (\boldsymbol{\hat{v}} \times \boldsymbol{\hat{b}})\cdot \boldsymbol{\hat{n}}
 \beta^3 
   .   \end{align} 
In the dilaton limit, $\partial_t W_\phi$ scales as the second power of velocity, and the emitted power is 
$P \sim {\beta^4 \over \Lambda^2 b^4}$.  
In the shift-symmetric axion limit the velocity suppression becomes more pronounced,  $P \sim {\beta^6 \over \Lambda^2 b^4}$,  and the emitted powers is approximately constant in time (for $|t| \lesssim {\rm few} \ b$ and in the non-relativistic limit).

%===============================================
\section{GR + SGB/DCS} 
\label{sec:gbcs}
%================================================

In this section we consider a scalar-tensor gravitational theory with a scalar coupled to gravity in a shift-symmetric way via the Gauss-Bonnet (GB) and Chern-Simons (CS) terms~\cite{Kanti:1995vq, Maeda:2009uy, Sotiriou:2013qea,Sotiriou:2014pfa,Alexander:2009tp, Yunes:2009hc, Serra:2022pzl}. 
The discussion follows the same template as for the toy model in \cref{sec:dilaxion}, and for this reason we will skip most of the derivations. The shift symmetry, however, will lead to an important qualitative difference, that we will comment on below.  

%----------------------------------------------
\subsection{Framework}
\label{sec:GBCS_framework}

We consider a system with massless spin-2 and spin-0 degrees of freedom,  described by the metric field $g_{\mu\nu}$ and a real scalar field $\phi$. 
Their kinetic terms and interactions are summarized by the effective Lagrangian:  
\begin{align}  
\label{eq:GBCS_Lagrangian}
{\cal L} = \sqrt{-g} \bigg \{ 
{\mpl^2 \over 2} R + {1 \over 2} (\partial_\mu \phi)^2
+  {\mpl \over \Lambda^2} \phi  \big [
\alpha R_{\rm GB}^2 
+  \tilde \alpha  R_{\rm CS}^2 \big ] 
\bigg \} 
+ \dots 
. \end{align}   
Here, 
 $\mpl \equiv (8\pi G)^{-1/2} 
 \simeq 2.44 \times 10^{18}$~GeV is the (reduced) Planck scale, 
 and $\Lambda$ is the cutoff scale of this EFT. 
The Gauss-Bonnet and Chern-Simons invariants are defined as 
$R_{\rm GB}^2 \equiv 
R^{\mu \nu  \rho \sigma}  R_{\mu \nu  \rho \sigma}
- 4R^{\mu \nu}  R_{\mu \nu} + R^2$,
$R_{\rm CS}^2  \equiv R^{\mu \nu  \rho \sigma}  \tilde R_{\mu \nu  \rho \sigma}$, 
and  
$\tilde R_{\mu \nu}{}^{ \alpha \beta } \equiv  
{1 \over 2 } \epsilon^{ \alpha \beta \rho \sigma}  R_{\mu \nu  \rho \sigma} $.
Because, in four dimensions, both of these invariants are total derivatives, the Lagrangian in \cref{eq:GBCS_Lagrangian} is 
invariant under the shift $\phi \to \phi + C$, where $C$ is a constant. 
The coupling strength between the scalar and gravity is parametrized by the dimensionless $\alpha$ and $\tilde \alpha$.  
Phenomenological constraints on these parameters are rather loose: 
${\sqrt{|\alpha|} \over \Lambda} \lesssim 0.31$~km~\cite{Julie:2024fwy}, 
while there is currently no constraints on $\tilde \alpha$ in the regime where the EFT is valid~\cite{Serra:2022pzl}. 
In other words, the scale $\Lambda$ can easily be macroscopic.
Moreover, the lack of causality violation within the EFT validity regime implies $|\hat \alpha| \lesssim 1$~\cite{Serra:2022pzl}, where $\hat \alpha \equiv \alpha + i \tilde \alpha$. 
The dots in \cref{eq:GBCS_Lagrangian} stand for other (self-)interaction terms, which are organized in an EFT expansion controlled by the scale $\Lambda$; these will not be relevant for the following discussion. 
For $\tilde \alpha =0$ ($\alpha =0$)  the scalar is CP-even(odd); for generic $\alpha$ and $\tilde \alpha$ parity is broken \footnote{Alternatively, one could also imagine a setup with two massless scalars coupling in the Lagrangian via ${\cal L} \ni \alpha \phi_1 R_{\rm{GB}}^2 + \tilde \alpha \phi_2 R_{{\rm CS}}^2$, in which case parity remains unbroken. The resulting waveforms can then be directly read off of from our results, however the emitted power will modified due to the lack of interference between the parity-even and -odd parts.}

In this framework, the scalar-graviton interactions lead to the 3-particle on-shell amplitudes:
\begin{align}  
\label{eq:GBCS_M3pointPhi}
\M \big [1_h^- 2_h^- 3_\phi \big ]  = & 
{2 \hat \alpha \over \Lambda^2 \mpl} \langle 1 2 \rangle^4 
, \nnl 
\M \big [1_h^+ 2_h^+ 3_\phi \big ] = &
 {2 \hat \alpha^* \over \Lambda^2 \mpl} [12]^4
, \nnl  
\M \big [1_h^- 2_h^+ 3_\phi \big ]  = &  0 .
 . \end{align}
The Einstein-Hilbert term in that Langrangian leads to on-shell 3-gravitons amplitudes, but these are not important for our discussion and are not displayed. 
 To calculate scalar emission amplitudes we will need the coupling of gravity to spin $S$ matter, 
 which we assume to be minimal. 
 The corresponding 3-particle on-shell amplitudes are:
\begin{align}
\label{eq:GBCS_M3pointGR}
\M \big [1_{\Phi} 2_{\bar \Phi} 3_h^- \big ] = & - 
 {\langle 3 |p_1|\tilde{\zeta}]^2  \over \mpl [3 \tilde{\zeta}]^2}   \frac{[\bm 2 \bm 1]^{2S}}{m^{2S}}
, \nnl 
\M  \big [ 1_{\Phi} 2_{\bar \Phi}  3_h^+ \big ] = & -{\langle \zeta|p_1|3]^2  \over \mpl \langle 3 \zeta \rangle^2} 
  \frac{\langle \bm 2 \bm 1 \rangle^{2S}}{ m^{2S}} ,
\end{align} 
and can be written as
\begin{align}
\label{eq:GBCS_M3pointGR-exp}
\M \big [1_{\Phi} 2_{\bar \Phi} 3_h^- \big ] = & - 
 {\langle 3 |p_1|\tilde{\zeta}]^2  \over \mpl [3 \tilde{\zeta}]^2}   \bigg  [ \mathbbb{1} +  { p_3 a_1   \over 2 S }      \bigg ]^{2S}
, \quad \M^{\rm cl} \big [1_{\Phi} 2_{\bar \Phi} 3_h^- \big ] =  - 
 {\langle 3 |p_1|\tilde{\zeta}]^2  \over \mpl [3 \tilde{\zeta}]^2}   e^{ p_3 a_1 }
, \nnl 
\M  \big [ 1_{\Phi} 2_{\bar \Phi}  3_h^+ \big ] = &
- {\langle \zeta|p_1|3]^2  \over \mpl \langle 3 \zeta \rangle^2}  \bigg  [ \mathbbb{1} -  { p_3 a_1   \over 2 S } \bigg ]  , \quad \M^{\rm cl}  \big [ 1_{\Phi} 2_{\bar \Phi}  3_h^+ \big ] 
= 
- {\langle \zeta|p_1|3]^2  \over \mpl \langle 3 \zeta \rangle^2}   e^{ -  p_3 a_1  }
, \end{align}
with the spin vector $a_1$ defined in \cref{eq:SPD_spinvector}. 
 We assume $\phi$ does not couple directly to matter. 

We want to calculate the scalar emission amplitude in scattering of spinning matter particles, 
$\M[1_{\Phi_1} 2_{\Phi_2} 3_{\bar \Phi_1} 4_{\bar \Phi_2} 5_\phi]$, within the framework  defined in \cref{sec:GBCS_framework}. 
The first step towards this goal is to derive the 4-particle amplitude 
$\M[1_{\Phi_i} 2_{\bar\Phi_i} 3^-_h 4_\phi ]$,
$i=1,2$, which can be viewed as a scalar and a graviton emission from a matter line. 
Up to contact terms, the amplitude must take the form 
\begin{align}
\label{eq:GBCS_4-point-h-phi}  
&\M_U[1_{\Phi_i} 2_{\bar\Phi_i} 3^-_h 4_\phi ] = { 2 \hat \alpha \langle 3| p_1 q | 3\rangle^2
\over 
s \mpl^2 \Lambda^2}  \frac{\langle \bm 2 \bm 1 \rangle^{2S}}{m^{2S}}  
\nnl
&\M_U[ 1_{\Phi_i} 2_{\bar\Phi_i} 3^+_h 4_\phi] = { 2 \hat \alpha^* [3|p_1 q | 3]^2 \over s \mpl^2 \Lambda^2}  \frac{[\bm 2 \bm 1 ]^{2S}}{m^{2S}} .
    \end{align}
These amplitudes satisfy 
${\rm Res}_{q^2 \to 0} 
\M_U \big [1_h^\pm    2_\phi  3_{\Phi_i} 4_{ \Phi_i^*} \big ]
= - \M \big [(-q)_h^\pm  1_h^\pm   2_\phi \big ]
\M \big [q_h^\mp  3_{\Phi_i} 4_{\Phi_i^*} \big ]$.  
The shift symmetry of the theory is reflected by the fact that the amplitudes vanish in the limit when the scalar momentum goes to zero. 
Rewriting  \cref{eq:GBCS_4-point-h-phi} in the exponentiated form and taking the classical limit, 
\begin{align} 
\label{eq:GBCS_M4pointPhiExp}
\M_U^{\rm cl} \big [1_{\Phi_i} 2_{\bar\Phi_i} 3^-_h 4_\phi \big ]  = & 
 { 2 \hat \alpha  \langle 3| p_1 q | 3\rangle^2 \over \mpl^2 \Lambda^2 s}  
  e^{ q a_1 }
,   \nnl 
\M_U^{\rm cl} \big [1_{\Phi_i} 2_{\bar\Phi_i} 3^+_h 4_\phi   \big ]  = &      
 {2 \hat \alpha^* ([3|p_1 q | 3]^2 \over  \mpl^2 \Lambda^2 s}  
   e^{- q a_1}
.\end{align}
Note that $\M_U^{\rm cl} \to \hbar^2 \M_U^{\rm cl}$ under the classical scaling in \cref{eq:SW_ClassicalScaling}.
It is important to notice that the scaling is different than for the analogous amplitude in the toy model in \cref{sec:dilaxion}, where we had  $\M_U \to \hbar^0 \M_U$.

We will organize our discussion of contact deformations of the amplitude the same way we did in \cref{SetupQED}. 
For that, we split 
$\M^{\rm cl}[ 1_{\Phi_i} 2_{\bar\Phi_i} 3^s_h 4_\phi ]$ as 
$\M^{\rm cl} = \M_U^{\rm cl} + \M_C^{\rm cl}$, where now the contact piece should have the same classical scaling as $\M_U^{\rm cl}$ and should vanish in the limit $p_4\to 0$. The contact terms are again organized in the spin expansion:
\begin{align}
    \label{eq:GBCS_M4pointPhiContact}
    \M_C^{\rm cl} = \sum_{n=0}^\infty \M_C^{(n)} ,
\end{align}
and it is again clear to see that $\M_C^{(0)}= 0$ .

At the linear order in spin, there is a single contact term with the same scaling as 
$\M_U^{\rm cl}$, 
that is 
$\M_C^{(1)} \to \hbar^2 \M_C^{(1)}$. 
We parametrize it as 
 \begin{align} 
\label{eq:GBCS_M4pointPhiContact1}
\M_C^{(1)} \big [1_{\Phi_i} 2_{\bar\Phi_i} 3^-_h 4_\phi \big ]  = & 
       -{  \hat C_1 \over \mpl^2 \Lambda^2} \langle 3| p_1 q | 3\rangle \langle 3| p_1 a_1 | 3\rangle 
 ,     \nnl 
 \M^{(1)}_C  \big [1_{\Phi_i} 2_{\bar\Phi_i} 3^+_h 4_\phi   \big ]  = &      
 {    \hat C_1^*  \over \mpl^2 \Lambda^2} [ 3| p_1 q | 3]  [ 3| p_1 a_1 | 3]
 ,  \end{align} 
 where $\hat C_1 \equiv C_1 + i \tilde C_1$ is a dimensionless Wilson coefficient.  
 The unique contact term at the linear order is automatically shift symmetric. 
 In fact it can be shown that this is indeed the only contact term which can be produced from the polynomial construction.
 We again refer the reader to \cref{app:CONT} for a the more systematic treatment.
We only remark that beyond the linear order it is possible to write down contact terms with a better classical scaling than $\M_U^{\rm cl}$. 
For example at the quadratic level we have 
 $\tilde {\M}_C^{(2)} \big [1_\Phi 2_{\bar{\Phi}} 3^-_h 4_\phi \big ]  =  
      {   \tilde C_2 \over \mpl^2}  \langle 3| p_1 a_1 | 3\rangle^2  $ scaling as 
  $\tilde{\M_C}^{(2)} \to  \hbar^0 \tilde{\M}_C^{(2)}$, much as the corresponding 4-particle amplitude in the toy model in \cref{sec:dilaxion}.   
  However this contact term does not vanish in the limit $p_4 \to 0$, and thus it does not appear in a shift-symmetric theory. 
  All shift-symmetric contact terms scale as  $\M_C^{\rm cl} \to \hbar^n \M_C^{\rm cl}$ with $n \geq 2$. 

%-------------------------------------------
\subsection{Scalar emission amplitudes} 
\label{sec:GBCS_amplitudes}

We move to the 5-particle amplitude for scalar emission in matter scattering, 
$\M_5 \equiv \M[ (p_1+w_1) _{\Phi_1}  (p_2+w_2) _{\Phi_2} (-p_1) _{\bar \Phi_1}  (-p_2) _{\bar \Phi_2} (-k)_\phi]$, 
which is needed to calculate the scalar waveforms via \cref{eq:SW_fphiOfM}. 

Once again we split
$\M_{5}^{\rm cl} = \M_{5U}^{\rm cl} + \M_{5C}^{\rm cl}$, where  
\begin{align}
 {\rm Res}_{w_2^2 \to 0}   \M_{5X}^{\rm cl} 
 = 
 - \sum_s \M_{X}^{\rm cl} \big [   (p_1+w_1) _{\Phi_1} (-p_1) _{\bar \Phi_1}  (w_2)_h^s   (-k)_\phi   \big ]
 \M^{\rm cl} \big [ (p_2+w_2) _{\Phi_2} (-p_2) _{\bar \Phi_2} (-w_2)_h^{-s}  \big ] \label{eq:4.10}
,  \end{align} 
$X = U,C$ refer to the amplitudes  in \cref{eq:GBCS_M4pointPhiExp,eq:GBCS_M4pointPhiContact}, and the residues at $w_1^2 \to 0$ are simply obtained from the above by $1\leftrightarrow 2$. 
For the first part we obtain 
 \begin{align}
\label{eq:GBCS_M5U}
 \M_{5U}^{\rm cl} = & 
    { 16 \over \mpl^3 \Lambda^2 w_1^2 w_2^2 }   
 \bigg \{ 
  \bigg  [ \alpha \cosh \big [ w_i a_i \big ]   + i \tilde  \alpha \sinh \big [ w_i a_i \big ]  \bigg ] 
   \nnl   \times & 
  \bigg [ 2(p_1 k )^2  (p_2 k )^2    + 
w_2^2 \bigg  ( 
2 (p_1 p_2)    ( p_1 k) (p_2k )  
 - m_1^2 (p_2 k)^2   
   -  (w_1 k)  \big [ (p_1 p_2)^2 -  { m_1^2 m_2^2   \over 2}  \big ]   \bigg  ) 
  \nnl   +&   
  w_1^2 \bigg  ( 
   2 (p_1 p_2)  ( p_1 k ) (p_2 k )   - m_2^2 (p_1 k)^2 
    -  (w_2 k)   \big [(p_1 p_2)^2 -  {m_1^2 m_2^2  \over 2}  \big ]    
  \bigg  ) 
 \bigg ]   
 \nnl  -   & 
  \bigg  [i \alpha \sinh \big [ w_i a_i  \big ]  
 - \tilde \alpha \cosh \big [ w_i a_i \big ]  \bigg ] 
  \bigg [ 2  ( p_1 k) (p_2 k)   + (p_1 p_2) (w_1^2 +  w_2^2)  \bigg ]  
  \epsilon_{\mu\nu \alpha \beta} p_1^\mu   p_2^\nu k^\alpha  w_2^\beta  
 \bigg \} 
, \end{align}
where $w_i a_i = w_1 a_1 + w_2 a_2$ and we have applied the same simplifications of the classical limit at leading order as in \cref{sec:ScalarEmissionAmplitudesQED}.

We move to  $\M_{5C}^{\rm cl}$, which we again expand in orders of spin,  
$\M_{5C}^{\rm cl} =  
\sum_{n=1}^\infty \M_{5C}^{(k)}$. 
For $n=1$,  plugging the expression in \cref{eq:GBCS_M4pointPhiContact1} into \cref{eq:4.10} we find 
 \begin{align} 
 \label{eq:GBCS_M5C1}
\M_{5C}^{(1)}  = &
    {   8 i  C_1  \over \mpl^3 \Lambda^2  w_2^2  }    \bigg \{  
        [ 2 (p_1 p_2)^2  -  m_1^2 m_2^2 ]      (a_1 w_2) (k w_2) 
 \nnl   &  
 - 2 (p_1 p_2)  (k p_1) [  (a_1 w_2)   (k p_2)   + (k w_2)    (a_1 p_2)   ] 
+  (k p_1)^2  [ 2    (a_1 p_2)(k p_2)  -    m_2^2  (a_1 w_1) ]  
   \bigg \} 
 \nnl    + &   
 {   8 i  \tilde C_1  \epsilon^{\mu\nu\alpha\beta} p_1^\mu p_2^\nu w_2^\beta \over \mpl^3 \Lambda^2  w_2^2  }   \bigg \{ 
  \big [   (a_1 p_2)(k p_1) - (p_1 p_2) (a_1 w_2) \big ]  k^\alpha  
+   \big [  (k p_1)  (k p_2) - (p_1 p_2) (k w_2) \big ]  a_1^\alpha      
   \bigg \}   
 \nnl    &    
   + (1 \leftrightarrow 2)          
. \end{align}
% 

%--------------------------------------
\subsection{Resolvability}
\label{sec:GBCS_resolvability}

We pause for a moment to discuss the physical consequences of the  classical scaling of the amplitudes calculated earlier in this section. 
Using the scaling rules in \cref{eq:SW_ClassicalScaling}, with the standing that $w_i$ scales in the same way as $k$, that is $w_i \to \hbar w_i$, we can easily see that the scalar emission amplitude in \cref{eq:GBCS_M5U,eq:GBCS_M5C1} scale as $\M_{5} \to \hbar^0 \M_{5}$.
This is unlike the analogous amplitude in the toy model of \cref{sec:dilaxion}, where we found $\M_{5} \to \hbar^{-2} \M_{5}$, which is also the behavior encountered in GR for graviton emission.
As a consequence, taking into account the scaling of the integration measure in \cref{eq:SW_fphiOfM}, we will find that the tree-level spectral waveform scales as  $f_\phi \to \hbar^2 f_\phi$ in the SGB/DCS framework. 
This however should not be interpreted as the waveform being quantum and vanishing in the classical limit. 
As stressed in Ref.~\cite{Bellazzini:2022wzv}, quantum-classical dichotomy can be misleading in theories introducing new scales. 
What matters is whether the effect is resolvable, that is whether it may dominate over quantum smearing. 
Therefore, in the following we discuss the conditions for resolvability of the SGB/DCS scalar radiation. 

For reference, let us briefly review the situation in GR, following the discussion of Ref.~\cite{Bellazzini:2022wzv}. 
We are interested in scattering processes with a hierarchy of scales $\ell_c \ll \ell_{Pl} \ll R_s \ll b$, with the Compton wavelength $\ell_c \sim \frac{1}{m}$, the Planck length $\ell_{Pl} \sim \frac{1}{\mpl}$, the Schwarzschild radius $R_{s} \sim \frac{m}{\mpl^2}$ and the impact parameter $b$.  
The classical deflection angle in GR scattering is  $\theta \sim {R_s \over b}$, 
while the classical $n$PM corrections scale as 
$\delta_n \theta \sim (R_s/b)^n$. 
The quantum smearing of the deflection angle $\Delta \theta$ can be estimated as 
${\Delta \theta \over \theta}{\Delta b \over b} \sim {1 \over \alpha_g}$, where 
$\alpha_g  \sim {m^2 \over \mpl^2} = m R_s$ 
is the strength of the gravitational interactions.
The $n$PM corrections are resolvable, 
${\delta_n \theta \over \Delta \theta}\gg 1$, 
for~\cite{Bellazzini:2022wzv}
\begin{align}
\alpha_g \bigg ( {R_s \over b}\bigg )^{n-1} \gg 1
. \end{align}  
For $n=1$ this is trivially satisfied for macroscopic objects, $m \gg \mpl$, 
while for $n>1$ and a given  $b/R_s$, this can be always satisfied for a sufficiently large mass $m$. 
Equivalently, for a fixed mass, there is a window of impact parameters where $n$PM effects are resolvable, 
$R_s \ll b \ll b_{\rm max}$, where 
$b_{\rm max} \sim R_s (m/\mpl)^{2 \over n-1}$. 
The leading quantum correction to an $n$PM effect would scale as 
$\Delta_n \theta \sim (R_s/b)^n {1 \over b m}$. 
In that case, by an analogous reasoning, the resolvability condition, ${\Delta_n \theta \over \Delta \theta}\gg 1$, reduces to $b \ll  R_s$; in other words quantum corrections are never resolvable.  

We move to discussing resolvability of the momentum kick due scalar radiation in the SGB/DCS theory.   
For concreteness, we assume $|\hat \alpha| \sim 1 $. 
For reference, the correction from the leading order graviton emission in GR is a 3PM effects, corresponding to $n=3$ in the discussion above. 
The additional $\hbar^2$ in classical scaling  of the scalar emission amplitude will translate to the extra 
$(\omega/\Lambda)^2$ 
factor in the spectral waveform compared to graviton emission, 
and thus the extra $(\omega/\Lambda)^4$ factor in the emitted power and momentum kick. 
Since $\omega \sim 1/b$, 
 \begin{align} 
 \delta_\phi  \theta \sim  
 \bigg ( {R_s \over b}\bigg )^3  {1 \over \Lambda^4 b^4 }  
. \end{align} 
The effect due to scalar emission is resolvable for 
 \begin{align} 
 \alpha_g \bigg ( {R_s \over b}\bigg )^2
 {1 \over \Lambda^4 b^4 } \gg 1 
 ,   \end{align} 
which can be rewritten as 
 \begin{align}  
\label{eq:GBCS_bmax}
R_s  \ll b \ll  b_{\rm max }
, \qquad 
b_{\rm max} = R_s  {\alpha_g^{1/6} \over (\Lambda R_s)^{2/3}}
= 
R_s {\mpl \over m^{1/3} \Lambda^{2/3}} 
.   \end{align} 
Clearly, if $\Lambda$ is a genuine quantum gravity scale, $\Lambda \sim \mpl$, this condition becomes 
$b_{\rm max}/R_s  \ll (\mpl/m)^{1/3} \ll 1$ and then scalar radiation is never observable.  
However, from the phenomenological point of view, 
 $\Lambda^{-1}$ can even be macroscopic, 
 $\Lambda \lesssim 0.1 {\rm km}^{-1} \sim 10^{-18}~$~GeV,
 leading potentially to a broad window of resolvability. 
 For example, for $\Lambda \simeq 2 \times 10^{-18}$~GeV and $m \simeq M_{\odot} \simeq 10^{57}$~GeV (corresponding to 
 $R_s \simeq 1.5 \times 10^{19}{\rm GeV^{-1}}$) we have $b_{\rm max} \sim 10^{11} R_s$.

In an analogous way, one can also derive similar constraints in the case of spinning bodies. 
The inclusion of spin introduces a new length scale in the problem which is $\frac{S}{m}$, where $S$ is the magnitude of the spin vector of the classical object. 

The corrections to the scattering angle are of the form 
$\delta_{n,k} \theta \sim 
 \left(\frac{S}{bm}\right)^k
 \big ( {R_s \over b} \big )^n $
where $k$ denotes the $k$-th order in a spin expansion of the amplitudes. 
Resolvability of such an effect amounts to $\alpha_g  \left(\frac{S}{bm}\right)^k \left(\frac{R_s}{b}\right)^{n-1} \gg 1$ which at the $n$PM order  and at linear order in spin, $k=1$, translates to 
$S\gg (\frac{b}{R_s})^{n} $.
This can be satisfied for sufficiently large $S$, provided $(b/R_s)^n$ does not exceed the extremality bound $S_{\rm max}$.\footnote{For Kerr black holes $S$ should be lower than the well known bounds that prevent the appearance of naked singularities \cite{Carroll:2004st} - in our conventions this implies 
$ S_{\rm max} = \frac{m^2}{8 \pi \mpl^2}$.
It is conceivable that solutions in modified GR theories may modify this bound.}. Similar arguments also hold for higher spin orders. In the example of the linear in spin waveforms in the modified theories we have described the correction will be $\delta^{(1)}_\phi  \theta \sim  
 \left( {R_s \over b}\right)^3 {1 \over \Lambda^4 b^4 } \left(\frac{S}{bm}\right)$, such that:

\begin{align} 
 \alpha_g \bigg ( {R_s \over b}\bigg )^2
 {1 \over \Lambda^4 b^4 } \left(\frac{S}{bm}\right) \gg 1 \implies S \gg  \frac{\Lambda^4 b^7}{R_s^3}
 ,   \end{align} 
 
For our benchmark point, 
$m \sim M_{\odot}$, 
$\Lambda \sim 0.1{\rm km}^{-1}$,  
this implies 
$S \gtrsim 10^6 (b/R_s)^7$, 
with $S_{\rm max} \sim 10^{76}$ for Kerr black holes in GR. 

%--------------------------------------
\subsection{Waveforms}
\label{sec:GBCS_waveforms}

\begin{figure}[ht]
    \centering
\includegraphics[width=0.49\linewidth]{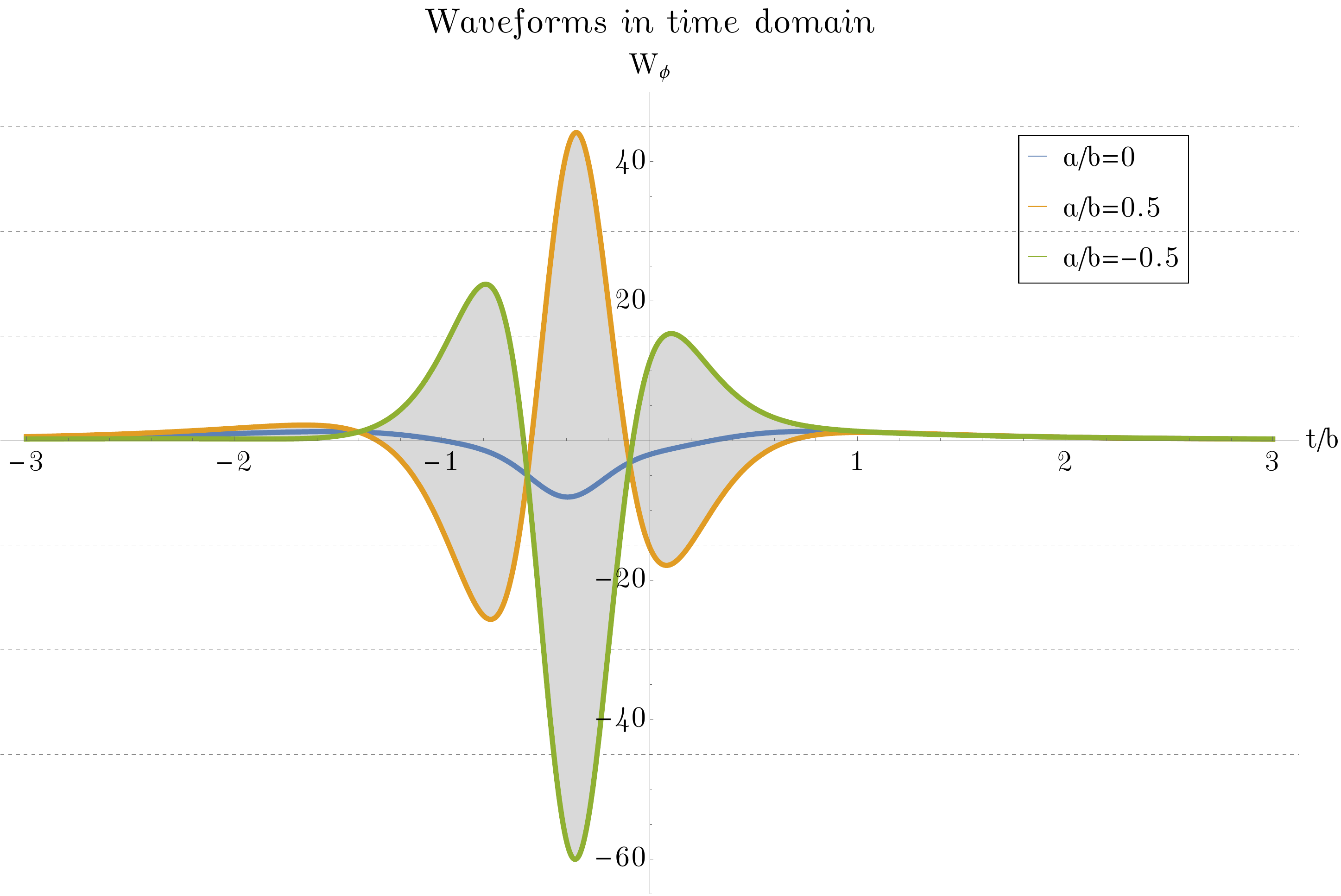} 
\includegraphics[width=0.49\linewidth]{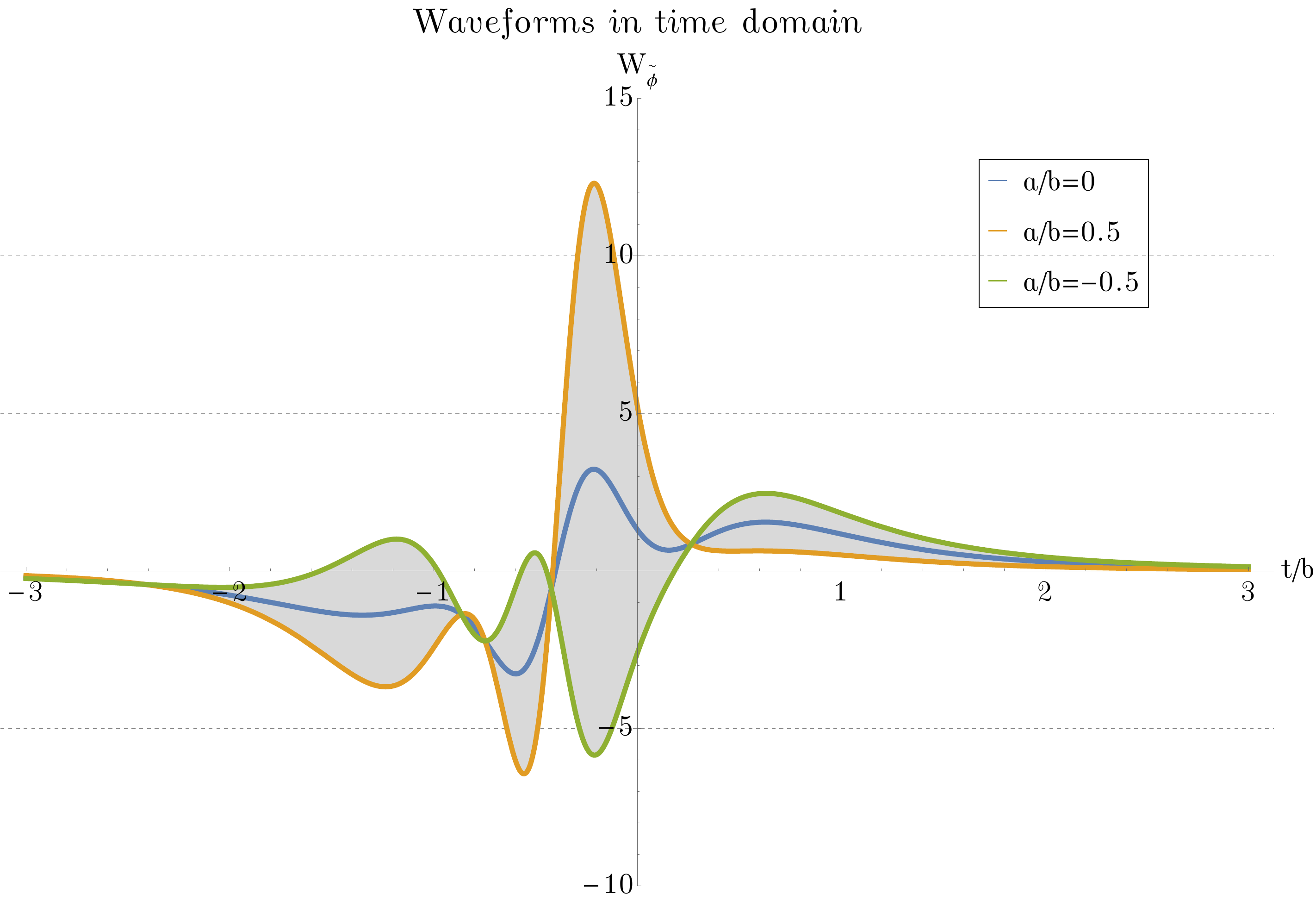}
    \caption{
    Scalar waveforms for the SGB (left) and DCS (right) cases up to linear order in spin plotted for different values of the spin magnitude $a \in [-b/2,b/2]$.}
    \label{fig:GBCSspins}
\end{figure}

Having discussed the issue of resolvability, we are ready to present the main results of this work: the tree-level waveforms for scalar radiation emitted in scattering of two  massive objects (with no scalar hair) in the SGB/DCS theory.
The spectral waveform are straightforwardly obtained by plugging the scalar emission amplitudes in \cref{eq:GBCS_M5U,eq:GBCS_M5C1} into \cref{eq:SW_fphiOfM},
and then integrating over the measure $d\mu$ using the master integrals in \cref{eq:INT_overw2sq-zbrep,eq:INT_overw1sqw2sq-zbrep}. 
The waveforms in the time domain are then obtained by a Fourier transformation of the spectral waveform.  
Much as for gravitational radiation in GR~\cite{Jakobsen:2021lvp, Riva:2022fru, DeAngelis:2023lvf, Aoude:2023dui, Brandhuber:2023hhl}, and analogously to the toy model in \cref{sec:dilaxion}, the spectral waveforms are only known in the integral form, however the waveforms in the time domain can be expressed by elementary functions.  
In the GB limit, $\tilde \alpha = \tilde C_k = 0$, to linear order in spin one obtains
 For the GB scalar, 
\begin{align} 
\label{eq:GBCS_Wphi}
\hspace{-2cm}    
  W_\phi = &   
  { m_1 m_2   \over 8 \pi^2 b^3  \mpl^3 \Lambda^2   (\hat u_1 n)^2 \sqrt{\gamma^2- 1}   }   
 \bigg ( 
 \alpha  
 \bigg \{ 
  - {d^2 \over  d z^2} \bigg [  {1 \over \sqrt{z^2   + 1 } }
  {   2 (\gamma^2- 1 )^2  (\hat u_2 n)^2  \big [   \gamma  (\hat u_2 n) - (\hat u_1 n)  +  (\tilde b n)  z \big ] 
  \over
 \big [   \gamma  (\hat u_2 n) - (\hat u_1 n)  +  (\tilde b n)  z \big ]^2 +  (\tilde v n)^2 (z^2 + 1)   } 
 \bigg ]
  \nnl   \hspace{-2cm} & 
   +  \big [  (\hat u_1 n)  +  \gamma  \big (2 \gamma^2 -  3   \big )  (\hat u_2 n)  \big ] 
   {2 z^2 - 1 \over (z^2 + 1)^{5/2} } 
  +  \big ( 2 \gamma^2 - 1 \big )  (\tilde b n) {3  z \over (z^2 + 1)^{5/2} } 
 \bigg \} 
\nnl   \hspace{-2cm}    + & 
{ 2 \alpha  \over  b}     {d^3  \over  dz^3} 
\re \bigg \{ {  1 \over \sqrt{ z^2 + 1} }  
{  (\hat u_2 n)   -  \gamma  (\hat u_1 n)   + \gamma   (\tilde b n)  z +  i \gamma   (\tilde v n) \sqrt{ z^2 + 1} 
 \over 
 \gamma  (\hat u_2 n) - (\hat u_1 n)   +  (\tilde b n)  z +  i (\tilde v n) \sqrt{ z^2 + 1}  } 
 \bigg [    z (\tilde v n)   -  i \sqrt{ z^2  + 1 }  (\tilde b n) \bigg ]  
\nnl  \hspace{-2cm}   &  
\times \bigg [ 
 {(a_1 n) \over  (\hat u_1 n) }  - \gamma    (a_1  \hat u_2) -  (a_2 \hat u_1) 
     - (a_1^A - a_2^A )  ( z \tilde b^{A}  +  i \sqrt{ z^2   + 1} \tilde v^{A})      \bigg ]    
 \bigg \}  
 \nnl   \hspace{-2cm}    & 
 -      \sqrt{\gamma^2-1} {  C_1    \over b } 
  {d^3 \over dz^3} \bigg \{ 
   { 1 \over   \sqrt{z^2   + 1 }   }  
  \re \bigg [ 
   \bigg (  
     z  \big [   (\tilde v n )a_1^A + (a_1 \tilde v) n^A  \big ] 
   - i \sqrt{ z^2   + 1}  \big [   ( \tilde b n) a_1^A   + (a_1 \tilde b) n^A  \big ]       \bigg )  
      \nnl \hspace{-2.5cm}   & 
   \times  \bigg ( 
    \hat u_2^A     -  \gamma \hat u_1^A   + \gamma   \big [   z \tilde b^A  +  i \sqrt{ z^2   + 1} \tilde v^A  \big ] 
\bigg ) \bigg ]  \bigg \} 
\bigg ) \bigg |_{z = T_1}  
  + (1 \leftrightarrow 2)  + \cO(a^2) 
     .      \end{align}   
     
In the CS limit, $\alpha = C_k = 0$,  we find 
    \begin{align}
\label{eq:GBCS_Wtildephi}     
    \hspace{-2cm}  
   W_{\tilde \phi}   =   &    
      {   m_1 m_2   \over 8 \pi^2  \mpl^3 \Lambda^2  (\hat u_1 n)^2  b^3 }
      \bigg (          
 2 \tilde \alpha  (\tilde v n )           
{d^2 \over dz^2 }  \bigg \{ { 1  \over \sqrt{ z^2 + 1} }      
\bigg [    \gamma     z 
-  (\gamma^2 - 1) (\hat u_2 n)   
{    z    \big [   \gamma  (\hat u_2 n) - (\hat u_1 n)   \big ] -      (\tilde b n ) 
 \over
 \big [  \gamma  (\hat u_2 n) - (\hat u_1 n)  +  (\tilde b n)  z \big ]^2  +  (\tilde v n)^2  (z^2 + 1)    } 
  \bigg ] 
        \bigg \}
        \nnl  \hspace{-2cm}  + &
 { \tilde \alpha    \over b  \sqrt{\gamma^2 - 1} }    \re 
   {d^3 \over d z^3 } \bigg \{ {  1   \over \sqrt{ z^2 + 1} }  
  \bigg (  {(a_1 n) \over  (\hat u_1 n) } - \gamma (a_1 \hat u_2)   - (a_2 \hat u_1) 
    + ( a_2^A - a_1^A)  \big [   z \tilde b^A + i \sqrt{ z^2   + 1} \tilde v^A  \big ]     
     \bigg )      
    \nnl \hspace{-2cm}  & 
\bigg (  
 { 2 (\gamma^2-1)^2 (\hat u_2 n)^2    
 \over
  \gamma  (\hat u_2 n) - (\hat u_1 n)  +  (\tilde b n)  z + i (\tilde v n) \sqrt{ z^2 + 1}   } 
-  (\hat u_1 n )  
 - \gamma \big (2 \gamma^2 - 3 \big )   (\hat u_2 n) 
 +  \big (2 \gamma^2 -  1  \big )    \big [ z (\tilde b n )  + i \sqrt{ z^2   + 1} (\tilde v n)  \big ] 
  \bigg  )   
 \bigg \}   
   \nnl \hspace{-2cm}  &
      - {1  \over \sqrt{\gamma^2- 1}   }   
    { \tilde C_1   a_1^A    \over  b  }       
   {d^3 \over d z^3}     \bigg \{  
 { 1 \over   \sqrt{z^2   + 1 }   }   \bigg [    (2 \gamma^2  -  1 ) \big [ z^2 (\tilde b n)   \tilde b^A     -   (z^2+1)    (\tilde v n)   \tilde v^A \big ]  
   -  (\gamma^2-1) n^A  
     \nnl \hspace{-2cm}  &
      +   \gamma  (\gamma^2-2)  (\hat u_1 n)      \hat u_2^A   
-  ( \gamma^2  - 2) (\hat u_2 n)  \hat u_2^A 
+   z \gamma (\tilde b n)  \hat u_2^A 
  - z \gamma^2   (\hat u_1 n)   \tilde b^A     
   +    z \gamma   ( \hat u_2 n) \tilde b^A   
   \bigg ]   \bigg \} 
 \bigg ) \bigg |_{z= T_1} 
 + (1 \leftrightarrow 2)   + \cO(a^2)
       .       \end{align} 

 In the general case, when the SGB/DCS coupling $\hat \alpha$ and the Wilson coefficients $\hat C_k$ are complex, 
the waveform is simply given by the sum of $W_\phi$ and $W_{\tilde \phi}$. 

In this section we restrict our discussion up to  linear order in spin. 
All-orders-in-spin contributions to the waveform originating from the $\M_{5U}$ amplitude are discussed in \cref{app:allorders}. 

\begin{figure}[h!]
    \centering
    \includegraphics[width=0.48\linewidth]{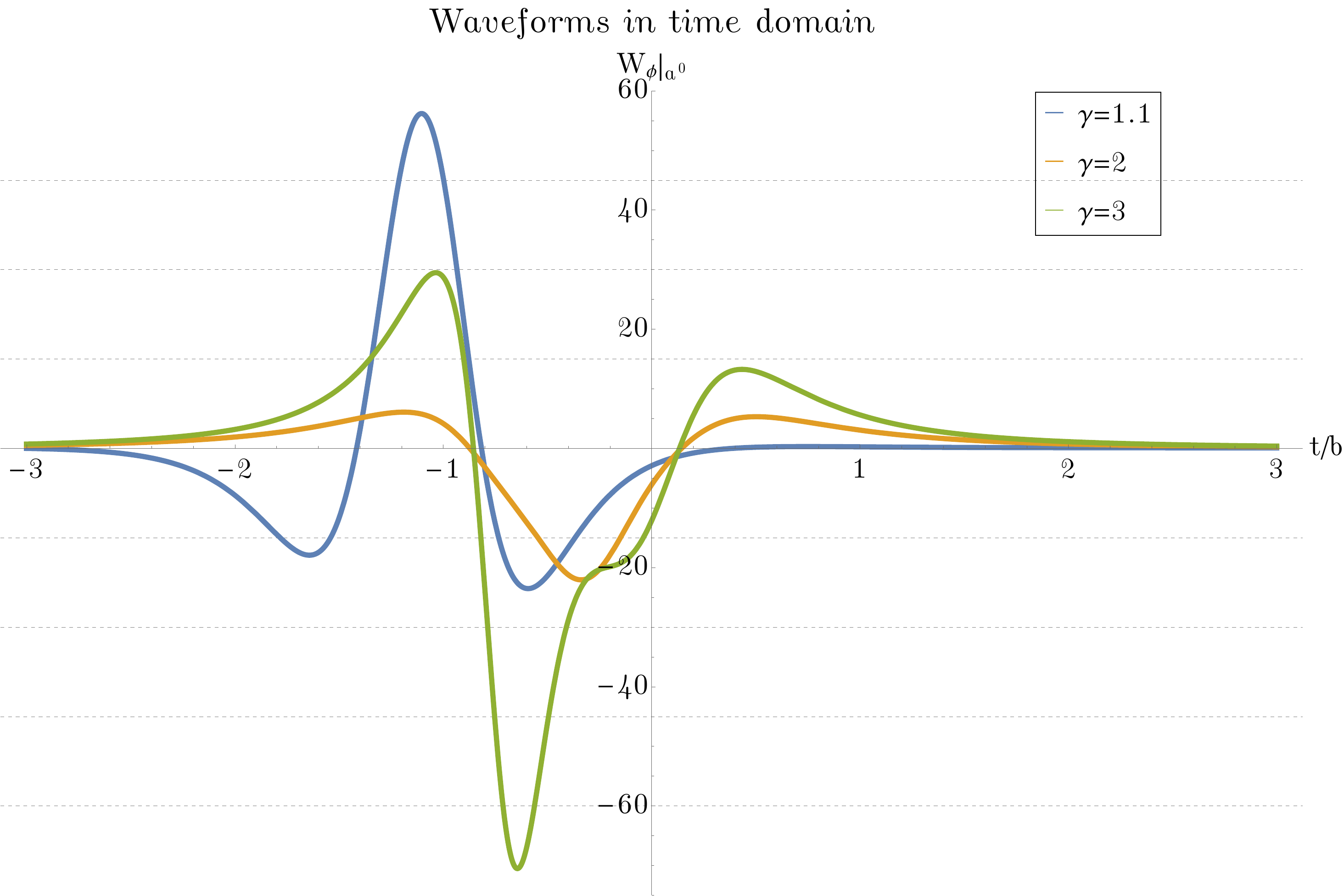}
        \includegraphics[width=0.48\linewidth]{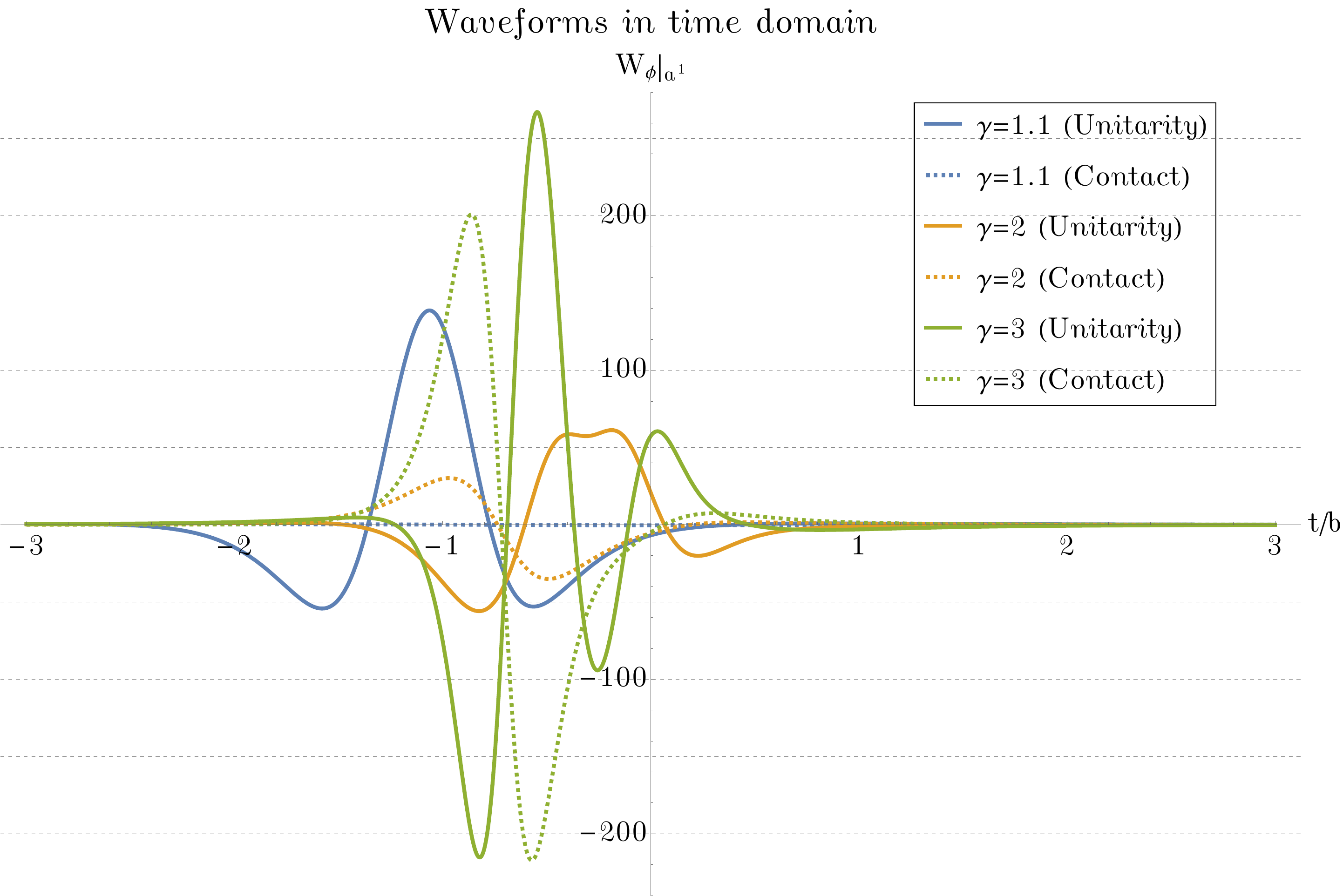}
    \caption{Left: Scalar piece of the SGB waveform for different values of $\gamma$. Right: Linear in spin part of the SGB waveform plotted for different values of $\gamma$. The unitarity and contact pieces are denoted with solid and dashed lines respectively. }
    \label{fig:GBgammafactor}
\end{figure}

\begin{figure}[h!]
    \centering
    \includegraphics[width=0.48\linewidth]{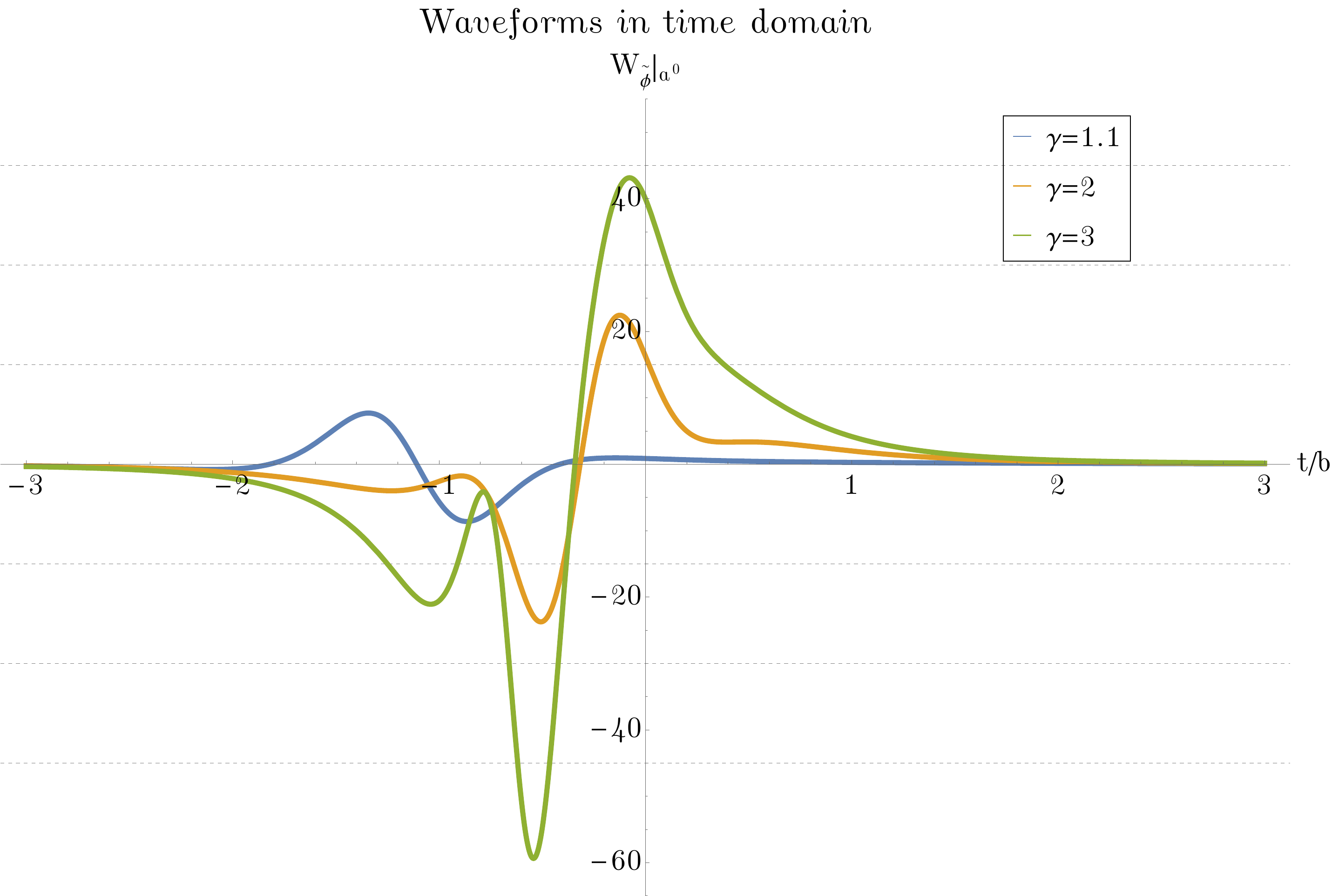}
        \includegraphics[width=0.48\linewidth]{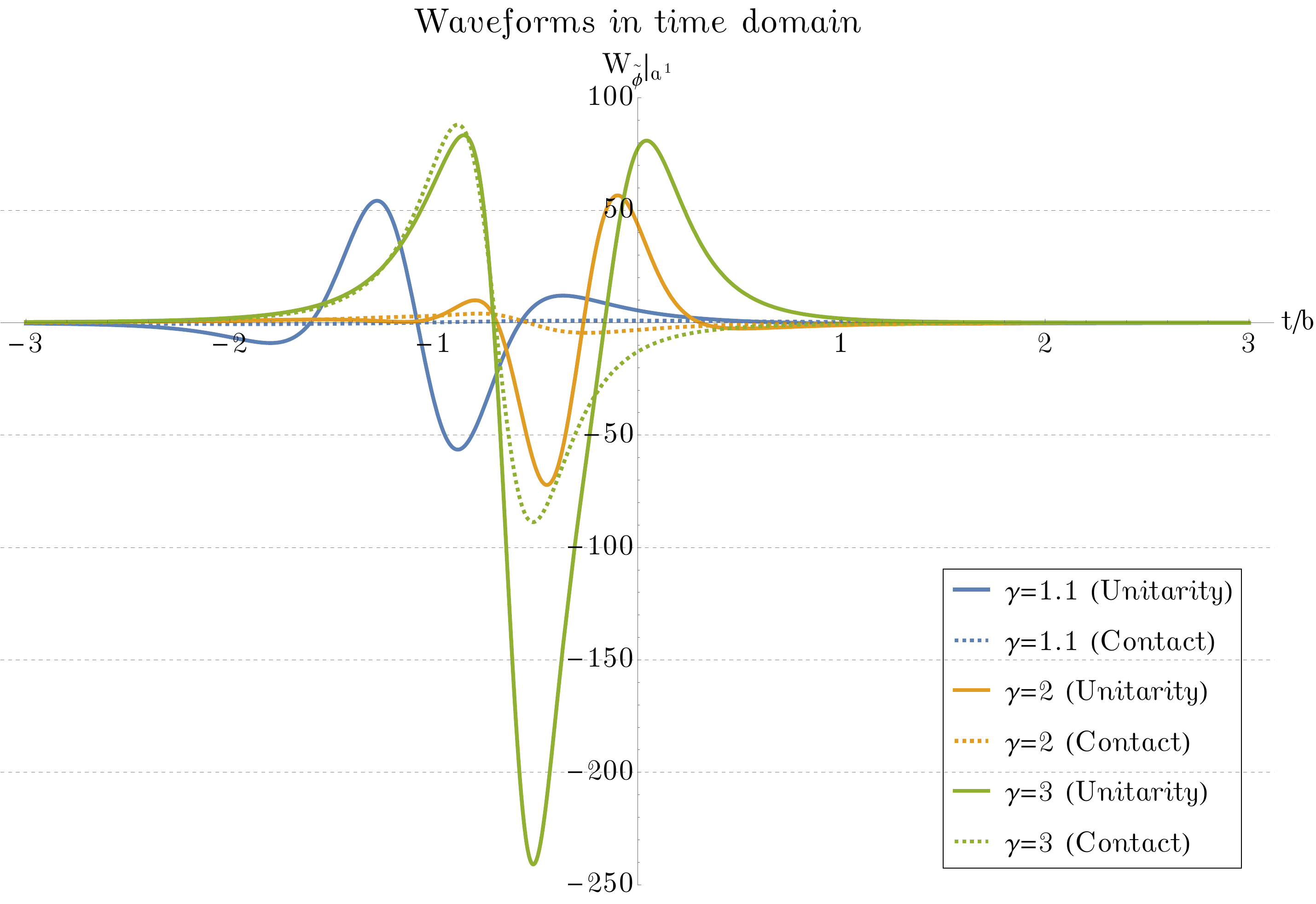}
    \caption{Left: Scalar piece of the DCS waveform for different values of $\gamma$. Right: Linear in spin part of the DCS waveform plotted for different values of $\gamma$. The unitarity and contact pieces are denoted with solid and dashed lines respectively.}
    \label{fig:CSgammafactor}
\end{figure}

We also plot the full waveforms shown in \cref{eq:GBCS_Wphi} and \cref{eq:GBCS_Wtildephi} as functions of the retarded time $t$ to highlight their behaviour. For that purpose, we choose to do the evaluation in a specific frame where 
$u_1^\mu = (1,0,0,0)$, 
$u_2^\mu = (\gamma, \sqrt{\gamma^2-1},0,0)$, 
$b_1^\mu= b(0,0,0,1)$, 
$b_2^\mu = (0,0,0,0)$, 
$a_1 = a(0,0,\cos \psi, \sin \psi)$, 
$a_2^\mu = (0,0,0,0)$, 
and we parametrize $n^\mu = (1, \sin \theta \cos \phi, \sin \theta \sin \phi, \cos \theta)$. 
We also restrict our attention to the the specific choice $\psi= \theta = \phi = \frac{\pi}{4}$. 
For convenience we set $\alpha = \tilde \alpha =  C_1 = \tilde C_1 =1 $, and waveforms are displayed in units of  ${ m_1 m_2   \over 8 \pi^2 b^3  \mpl^3 \Lambda^2  }=1 $.
 
In \cref{fig:GBCSspins} we can read off the full waveform for the two cases, with $\gamma=\frac{3}{2}$ and for different values of the spin magnitude. Note that the choice $\frac{|a|}{b}= 0.5$ corresponds to the limiting case, $|a|=\frac{S_{max}}{m}$ and $b=R_s$. 
In \cref{fig:GBgammafactor,fig:CSgammafactor} we also plot the waveforms at the zero-th and linear order in spin for different values of the $\gamma$ factor. In the linear in spin waveforms we have also taken out a factor of $\frac{|a|}{b}$ in front of the waveform. 

%------------------------------------
\subsection{Discussion}

To highlight the physics content of the  waveforms in \cref{eq:GBCS_Wphi,eq:GBCS_Wtildephi}, below we discuss two limits.
We first consider the asymptotic behavior when $|t|$ is much larger than the impact parameter $b$.  
Expanding the waveforms for $t \to \pm \infty$ we obtain for the GB scalar  
 \begin{align}  
 \label{eq:GBCS_WphiAss}
  W_\phi = &   
 \pm  { \alpha m_1 m_2   \over 4 \pi^2   \mpl^3 \Lambda^2  t^3 }   
  \bigg \{   
  { (\hat u_1 n)^2 + (\hat u_2 n)^2
  + 2 \gamma (2 \gamma^2 -3 )  (\hat u_1 n) (\hat u_2 n) \over     \sqrt{\gamma^2- 1}  }
  + {3  (a_1^\mu - a_2^\mu) \over  t }   
  { (\tilde b n)\tilde v^\mu - (\tilde v n)\tilde b^\mu  \over 
  (\tilde b n)^2 + (\tilde v n)^2} 
  \nnl   \times  & 
  \bigg [ 
 \big [ 
  \gamma (\tilde b n)^2 + (\tilde v n)^2)   -2 (\gamma^2-1) (\hat u_1 n) (\hat u_2 n)  \big ] \big [ (\hat u_1 n)^2 +  (\hat u_2 n)^2 \big ] 
  + 4 \gamma  (\gamma^2-1) (\hat u_1 n)^2   (\hat u_2 n)^2 
  \bigg ] 
  \bigg \}  
    \nnl    & + \cO(t^{-5})
    .   \end{align} 
 This is a good approximation for $|t| \gg b/\beta$.
The large $t$ behavior probes the small frequency limit of the spectral waveform and we can read off 
    \begin{align} 
    f_\phi(\omega \to 0) \sim  
{m_1 m_2   \over \mpl^3 }
{\omega^2 \over \Lambda^2}  \bigg (  1 + \# a \omega  + {\cal O}(a^2 \omega^2) \bigg ) 
   .   \end{align}
There is an $\omega^2/\Lambda^2$ suppression compared to the analogous spectral functions in the toy model in \cref{sec:dilaxion} or in GR. 
The vanishing of $f_\phi$ for $\omega =0$ is a consequence of the soft theorems and the vanishing of the scalar emission amplitudes when the scalar momentum goes to zero, which is due to the shift symmetry. This is also the reason why there is no memory effect at leading order for the scalar emission. In particular, mimicking the steps of the gravitational case \cite{Herderschee:2023fxh, Brandhuber:2023hhl},  we can define the scalar memory as $\Delta \phi = W_\phi (t = \infty) - W_\phi ( t = - \infty)$, which can be further expressed in terms of the soft limit of the four-point amplitude:

\begin{align}
  \Delta^{(0)} \phi \equiv & \int_{-\infty}^\infty dt \frac{d}{dt} W_\phi 
  = 
  -i \int_{0}^\infty d \omega \delta(\omega) \omega f_\phi (\omega) + \hc 
  = 
  - \frac{i}{2} \lim_{\omega \to 0^+} [ \omega f_\phi (\omega) ] + \hc 
  \nnl&= 
  \frac{-i}{128 \pi^3} \int d \mu S^{\rm cl}_\phi (\omega, q) \M_{4, \rm SGB/DCS}^{\rm cl} (q) +\hc ,
\end{align}
where we made use of \cref{eq:SW_WphiOWM} and \cref{eq:SW_fphiOfM} with the latter being valid at leading order. 
In the last step, the $S_\phi^{\rm cl}$ is the classical piece of the soft factor for the emission of a scalar and $\M_{4, \rm SGB/DCS}^{\rm cl}$ stands for the classical part of the amplitude describing the scattering of 2 massive particles through the SGB/DCS interactions. 
However, at tree-level there are no corrections to the scattering of two massive particles through these interactions - in fact, the first corrections come at 2 loops. 
Thus, without any knowledge of the soft factor, we can easily conclude that at leading order there is no memory effect:
 \begin{align}
 \Delta^{(0)} \phi = 0 . 
\end{align}

For the CS scalar the asymptotic structure is slightly different  
    \begin{align}   
     \label{eq:GBCS_WtildephiAss}
  W_{\tilde \phi} = &   
   \pm {3 m_1 m_2 b   \over 8 \pi^2   \mpl^3 \Lambda^2 t^4}  
    \bigg \{ 
   2 \tilde \alpha  (\tilde v n) 
    \bigg [  \gamma   ( (\hat u_1 n)^2  + (\hat u_2 n)^2)
    \nnl & 
    - { 2 (\gamma^2 -1) (\hat u_1 n) (\hat u_2 n)  \over    (\tilde b n)^2 +  (\tilde v n)^2 }
    \bigg ( (\hat u_1 n)^2 +   (\hat u_2 n)^2-  2 \gamma  (\hat u_1 n) (\hat u_2 n)  \bigg  ) \bigg ] 
    +  {\tilde \alpha \over b \sqrt{\gamma^2-1} } 
      \nnl  \times  & 
 \bigg [       
 \big (   (a_1 \tilde b) (\tilde b n) +  (a_1 \tilde v) (\tilde v n) \big )  
\big ( (2 \gamma^2-1) (\hat u_2 n)^2 - (2 \gamma^2+1) (\hat u_1 n)^2
- 2 \gamma (2 \gamma^2-3)   (\hat u_1 n)  (\hat u_2 n)  \big ) 
 \nnl  & 
 +  \big ( (a_2 \tilde b) (\tilde b n) +  (a_2 \tilde v) (\tilde v n) \big )  
\big ( (2 \gamma^2-1) (\hat u_1 n)^2 - (2 \gamma^2+1) (\hat u_2 n)^2
- 2 \gamma (2 \gamma^2-3)   (\hat u_1 n)  (\hat u_2 n)  \big ) 
 \nnl  & 
 - 2 \big (  (a_1 \hat u_2) (\hat u_2 n) +   (a_2 \hat u_1) (\hat u_1 n) \big ) 
\big (  (\hat u_1 n)^2  + (\hat u_2 n)^2 + 2 \gamma (2 \gamma^2-3)   (\hat u_1 n)  (\hat u_2 n) \big ) 
    \bigg \} 
   \nnl + & 
 { \tilde C_1 \over b \sqrt{\gamma^2-1} }    
 \bigg [ 
2  (a_1 \hat u_2)  (\hat u_1 n)^2 ( (\hat u_2 n) - \gamma  (\hat u_1 n) )
+ 2  (a_2 \hat u_1)  (\hat u_2 n)^2 ( (\hat u_1 n) - \gamma  (\hat u_2 n) )
 \nnl  &  
- ( (a_1 \tilde b) (\tilde b n) (a_1 \tilde v) (\tilde v n) )  (\hat u_1 n)^2 
- ( (a_2 \tilde b) (\tilde b n) + (a_2 \tilde v) (\tilde v n) ) (\hat u_2 n)^2 
 \bigg ]  \bigg \} 
    + \cO(t^{-5})
    .   \end{align} 
This corresponds to the  spectral function 
\begin{align} 
    f_\phi(\omega \to 0) \sim  
{m_1 m_2   \over \mpl^3 }
{\omega^2 \over \Lambda^2}  \bigg (  \omega b + \# a \omega  + {\cal O}(a^2 \omega^2) \bigg ) 
   .   \end{align}  
We see that the spectral function vanishes faster, as $\omega^3$,  at small frequencies, and that the zero-th order and linear orders exhibit the same asymptotic behaviors. The linear pieces may dominate if $S > b m$. With the same reasoning as above we can see that also in this case there is no memory effect at leading order, $\Delta^{(0)} \tilde \phi =0$.

\vspace{1cm}

We move to  another limit of the waveforms: $\gamma \to 1$, or equivalently 
$\beta \equiv \sqrt{\gamma^2-1} \to 0$. 
This is the relevant limit for discussing the physical situations when the matter scattering process represents typical astrophysical objects, such as neutron stars, black holes, etc. 
For simplicity, let us fix 
$u_1 = (\gamma,\beta \boldsymbol{\hat v})$,
$u_2 = (1,\boldsymbol{0})$, 
$b_1 = (0, b \boldsymbol{\hat b})$,
$a_1  =  L.(0, {S_1\over m_1} \boldsymbol{\hat{a}_1})$, 
$a_2  =  (0, {S_2\over m_2} \boldsymbol{\hat{a}_2})$
$n = (1, \boldsymbol{\hat n})$,
with 
$\boldsymbol{\hat v}^2 = \boldsymbol{\hat b}^2= \boldsymbol{\hat a}_i^2 =  \boldsymbol{\hat n}^2 = 1$, 
and $L$ the Lorentz boost from the rest frame of the particle $1$. 
Then, for $|t| \lesssim b/\beta$  we find 
\begin{align}  
\label{eq:GBCS_WphiNR}
\partial_t W_\phi \simeq & 
-  { 3 m_1 m_2   \over 4 \pi^2 b^4  \mpl^3 \Lambda^2 }
\bigg \{ 
3 \alpha 
(\boldsymbol{\hat{v}}\cdot\boldsymbol{\hat n})
(\boldsymbol{\hat{b}}\cdot\boldsymbol{\hat n})\beta^3 
  - 2 \alpha  {1 \over b} 
(\boldsymbol{\hat{v}} \times \boldsymbol{\hat{b}})  \cdot 
   \bigg (
4 (\boldsymbol{a_-} \cdot \boldsymbol{\hat{b}}) \boldsymbol{\hat n}
+ (\boldsymbol{\hat{n}} \cdot \boldsymbol{\hat{b}}) \boldsymbol{a}_-
\bigg   ) \beta^2 
\bigg \}  
   ,   \end{align} 
   
  \begin{align}  
   \label{eq:GBCS_WtildephiNR}
\partial_t W_{\tilde \phi} \simeq  & 
-3  { m_1 m_2   \over 8 \pi^2 b^4 \mpl^3 \Lambda^2 }
\bigg \{ 
4 \tilde \alpha 
(\boldsymbol{\hat{v}} \times \boldsymbol{\hat{b}}
)\cdot\boldsymbol{n} \beta^3 
\nnl & 
 + \tilde C_2 {S \over b m_2} \bigg [
3(\boldsymbol{\hat{v}}\cdot\boldsymbol{\hat{n}})
(\boldsymbol{\hat{v}}\cdot\boldsymbol{a}_+) 
-4 (\boldsymbol{\hat{b}}\cdot\boldsymbol{\hat{n}})
(\boldsymbol{\hat{b}}\cdot\boldsymbol{a}_+)
+ (\boldsymbol{\hat{v}} \times \boldsymbol{\hat{b}}
)\cdot\boldsymbol{n}  \, 
(\boldsymbol{\hat{v}} \times \boldsymbol{\hat{b}}
)\cdot\boldsymbol{a}_+
\bigg ] \beta^2 
\bigg \} 
   ,   \end{align} 
  where 
   $\boldsymbol{a}_\pm = 
   {S_1 \over m_1} \boldsymbol{\hat a}_1
\pm   {S_2 \over m_2} \boldsymbol{\hat a}_2 $.
At each order in spin we displayed the leading term in velocity expansion. 
For non-spinning objects $\partial_t W_\phi$ scales as the third power of velocity, and the emitted power is 
$P \sim {\beta^6 \over b^8}$. 
Applying this to closed circular orbits and using Kepler's law 
$\beta^2 \sim 1/b$, one finds the power emitted in scalar radiation of  objects with no scalar hair  is suppressed as  $P \sim  \beta^{22}$. 
This is a larger suppression compared to $P \sim \beta^8$ for black holes with scalar hair~\cite{Yagi:2011xp}. 
On the other hand, for spinning objects $\partial_t W_\phi$ scales as the second power of velocity at the linear order in spin, but it contains another $1/b$ factor. 
By the same logic, for closed orbits it leads to 
$P \sim \beta^{24}$. 

\begin{figure}[tbh]
    \centering
    \includegraphics[width=0.48\linewidth]{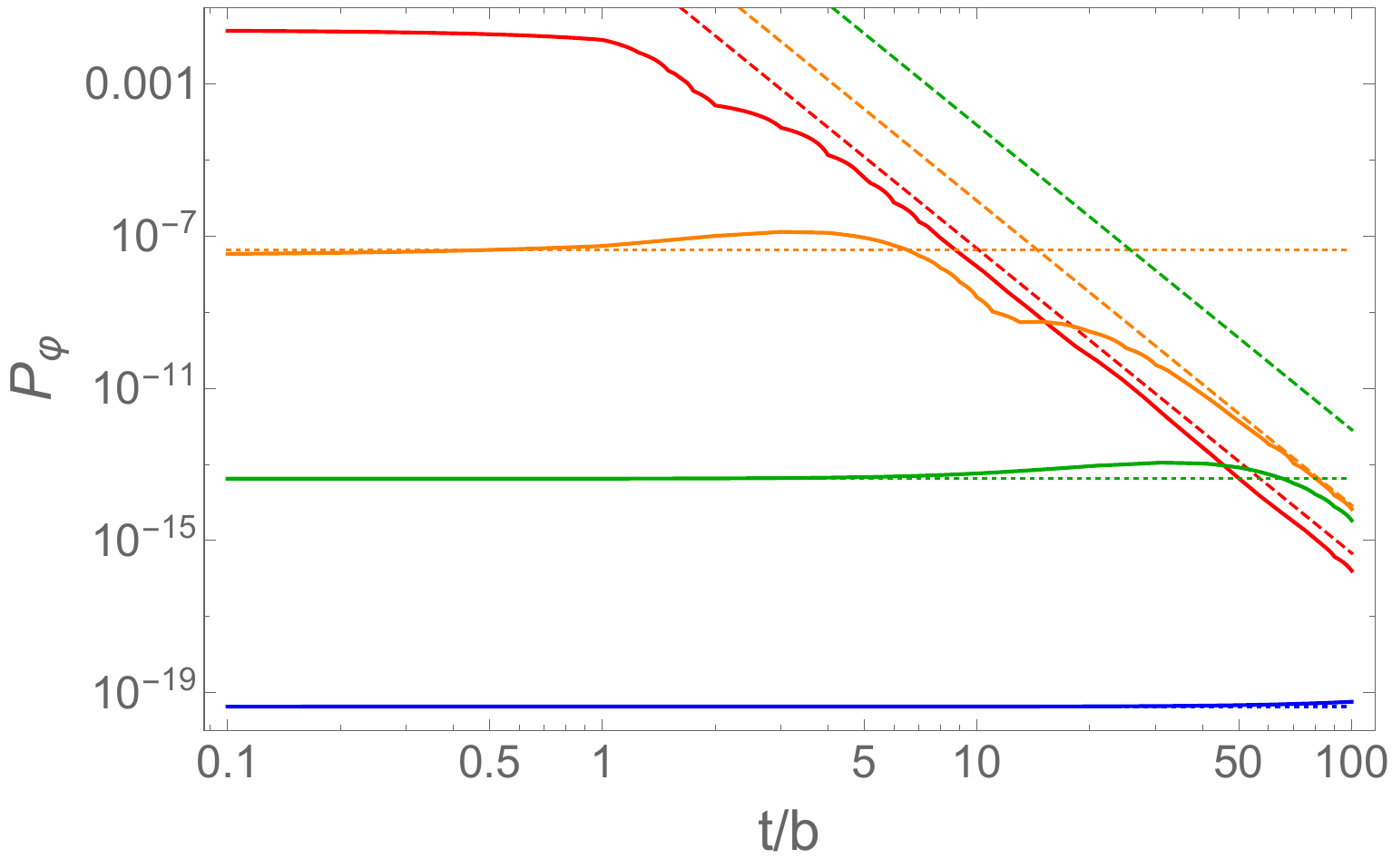}
        \includegraphics[width=0.48\linewidth]{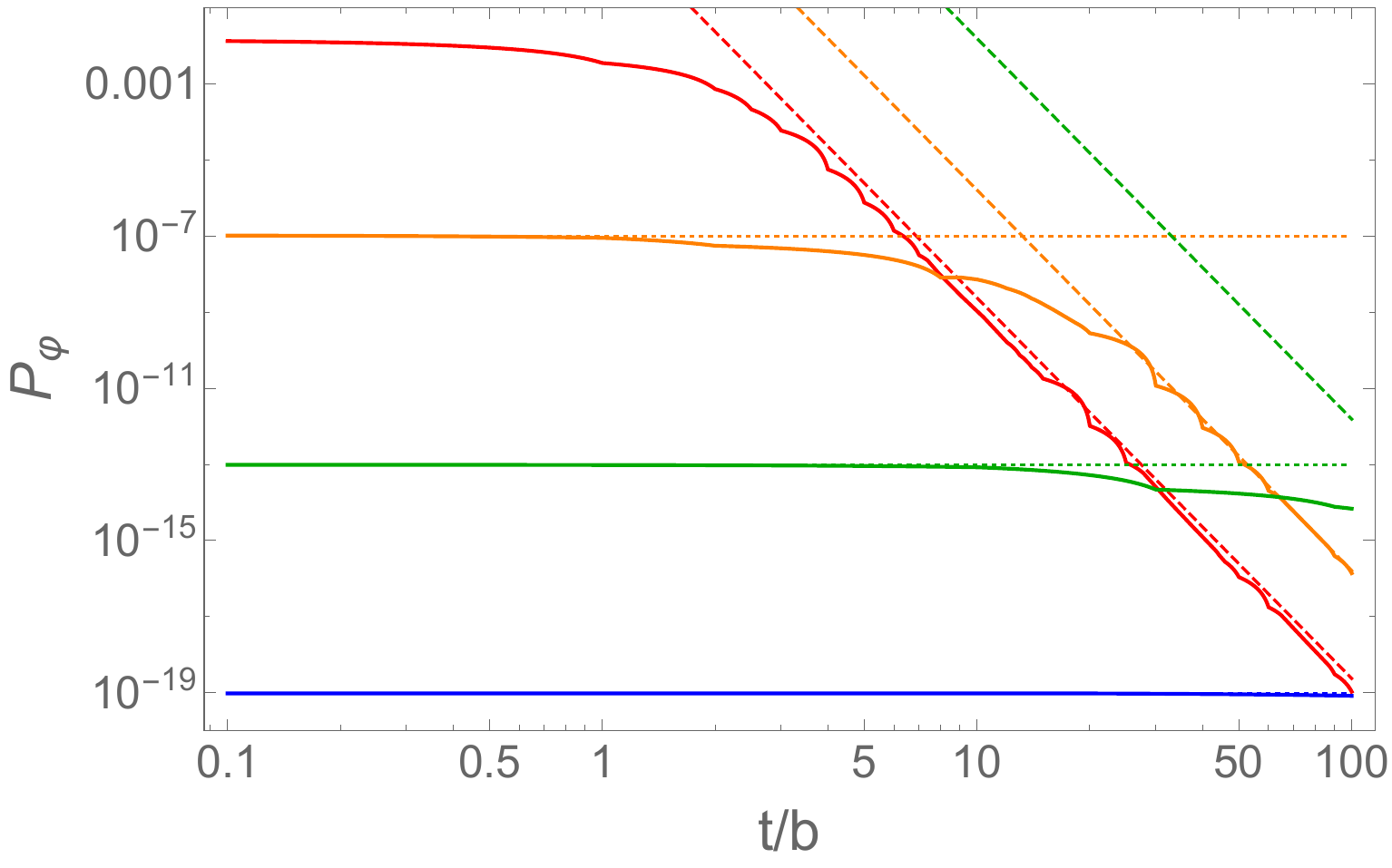}
    \caption{Left: total power (in units of 
    ${\alpha^2 m_1^2 m_2^2 \over \Lambda^4 \mpl^6 b^8}$) emitted in scalar radiation in the SGB limit at the zero-th order in spin for $\beta = 0.99$ (red), $\beta=0.1$ (orange), $\beta=0.01$ (green), $\beta=0.001$ (blue). 
The solid lines are obtained by integrating the complete waveform in \cref{eq:GBCS_Wphi}, 
the dotted line uses the small-velocity approximation in \cref{eq:GBCS_WphiNR}, and the dashed line uses the large $t$ approximation in \cref{eq:GBCS_WphiAss}. 
Right: same in the DCS limit with $\alpha \to \tilde \alpha$.  }
    \label{fig:powera0}
\end{figure} 

The total power  emitted in scalar radiation at the leading order in spin is shown in \cref{fig:powera0} for different values of $\beta$ for the SGB and DCS cases.
The solid curves are obtained by numerically integrating the tree-level waveforms  \cref{eq:GBCS_Wphi,eq:GBCS_Wtildephi} over the angular variables.  
For $\beta$ not too large, the observable interpolates between the constant value at small $t$ captured by the approximation in \cref{eq:GBCS_WphiNR,eq:GBCS_WtildephiNR}, and the power law decay at large $t$ described by \cref{eq:GBCS_WphiAss,eq:GBCS_WtildephiAss}.  

%===============================================
\section{Conclusions} 
\label{sec:conclusions}
%================================================

In this paper we studied scalar-tensor gravitational theories using on-shell amplitude methods. 
We focused on gravity minimally interacting with matter and self-interacting via the usual Einstein-Hilbert Lagrangian, while  coupled to a massless scalar via the GB and CS terms, 
cf.~\cref{eq:GBCS_Lagrangian}.   
Our setup possesses a shift symmetry acting on the scalar; in particular we assumed for simplicity the scalar does not have self-interactions and is not coupled directly to matter. 
In such a framework, we calculated the waveforms for classical scalar radiation in hyperbolic scattering of matter particles, using the KMOC formalism. 
Given the waveforms, the power emitted in scalar radiation can be readily calculated using \cref{eq:SW_power}. 
This can be interpreted as a component of energy emitted in encounters of macroscopic objects with no scalar hair (e.g. neutron stars) in the SGB/DCS scalar-tensor theory. 

Our main results are contained in \cref{sec:GBCS_waveforms} where we display the waveforms in the time domain at tree level and up to linear order in spin of the scattering particles, while a partial all-spin-orders formula for the unitarity piece is provided in \cref{app:allorders}. 
In \cref{sec:GBCS_amplitudes} we also write down the ingredients needed to derive that results, namely the classical limit of the 5-particle amplitude describing a singe scalar emission in matter scattering. The amplitude in our case transforms as 
${\cal M}_5^{\rm cl} \to \hbar^0 {\cal M}_5^{\rm cl}$  the classical scaling in \cref{eq:SW_ClassicalScaling}, 
unlike for the analogous graviton emission amplitude in GR, where one finds 
${\cal M}_5^{\rm cl} \to \hbar^{-2} {\cal M}_5^{\rm cl}$. 
This raises the question of resolvability of scalar radiation, which we address in~\cref{sec:GBCS_resolvability}.  
The conclusion is that, as long as the scale $\Lambda$ suppressing the SGB/DCS scalar-graviton interaction is small enough, 
there exists a window of impact parameters given
in \cref{eq:GBCS_bmax} where scalar radiation is resolvable.
In particular, 
for $\Lambda/|\hat \alpha|^{1/2} \sim 0.1~$km, 
which are the values probed by contemporary observations of gravitational waves in black hole inspirals, the window of resolvability is very broad for astrophysical mass objects.

The scalar emission amplitude depends in particular on the contact terms of the 4-particle matter-matter-graviton-scalar amplitude. 
We write the linear in spin contact term in the main text, and give more systematic discussion beyond the linear order in \cref{app:CONT}.
It would be interesting to find a matching between these contact terms and specific solutions in SGB/DCS theories, in analogy to what was done for GR in Ref.~\cite{Bautista:2022wjf}, which would involve studies of black hole perturbation theory in the context of modified gravity theories. It is also conceivable that a matching can be done at the level of the observables with a worldline EFT formalism e.g. along the lines of \cite{Kuntz:2019zef, Brax:2021qqo}. 

Much as in the GR case, the waveforms in the time domain are given by elementary functions, and in some interesting limiting cases the expressions are very compact.
In particular, the non-relativistic limit of the waveforms, where the relative velocity $\beta$ between the two bodies is small, is displayed in \cref{eq:GBCS_WphiNR,eq:GBCS_WtildephiNR}. 
In the limit $\beta \ll 1$,  the waveforms are approximately constant during the closest approach, that is for $|t| \lesssim {\rm few} \times b$. 
This is due to the shift symmetry, and it is  also a feature in the axion limit of the toy model in \cref{sec:dilaxion}.  
Furthermore, at the zero-th order in spin, 
the emitted power is suppressed by $\beta^{22}$, regardless of whether the SGB or DCS couplings dominates. 
The velocity suppression is however reduced at higher orders in spin. 

As mentioned earlier, it would be interesting to match our waveforms to specific solutions in the SGB/DCS theory, so as to provide an interpretation of the contact term dependence of the waveforms. 
Furthermore, it would be interesting to extend our results to include the scalar coupling directly to matter. 
This would allow us to connect to scattering of black holes in the SGB/DCS theory, as black holes are known to acquire a scalar charge. 
Finally, the on-shell amplitude calculations could be extended beyond tree level. 
These avenues will be explored in a future work.

%@@@@@@@@@@@@@@@@@@@@@@@@@@@@@@@@@@@@@@@@@@@@@@
\section*{Acknowledgements}

We thank Fabian Bautista for helpful discussions about the contact terms' construction. We would also like to thank Christos Charmousis, Stefano De Angelis, Nicolas Lecoeur for useful discussions. 
AF has received funding from the Agence Na- tionale de la Recherche (ANR)  grant ANR-19-CE31-0012 (project MORA) and from the European Union’s Horizon 2020 research and innovation programme  the Marie Skłodowska-Curie grant agreement No 860881-HIDDeN. 

\appendix

%================================================
\newpage
\section{Integrals} 
\label{app:INT}
%================================================

We now present the method for computing all the master integrals relevant for calculating scalar radiation waveforms in our framework. 
We follow the method first presented in \cite{DeAngelis:2023lvf} and we refer the reader to that reference for more details. 

We define the integral measure: 
\begin{align}
d \mu \equiv &  \delta^4(w_1 + w_2 - k)     \Pi_{i=1,2} [e^{i b_i w_i}  d^4 w_i \delta (u_i w_i)  ] 
, \end{align}
where it is implied that $k^2 = 0$. 
Note that this measure differs by the one defined in \cite{DeAngelis:2023lvf} by the exponential factor $e^{i b_i w_i}$. 
We then define the general single pole integral:
\begin{align}
I_i^{\mu_1 \dots \mu_n}  \equiv   \int d \mu   {w_i^{\mu_1} \dots  w_i^{\mu_n}   \over w_i^2 } , \label{SINGLE:POLE:INT}
. \end{align}  
and the general double pole pole:
 \begin{align}
J_i^{\mu_1 \dots \mu_n} \equiv &  \int   d \mu   { w_i^{\mu_1} \dots  w_i^{\mu_n}   \over w_1^2 w_2^2 }  \label{DOUBLE:POLE:INT}
 .  \end{align}  

These integrals are the ones needed for the computation of the spectral waveform in \cref{eq:SW_fphiOfM}, which is necessary to get the time domain waveform in \cref{eq:SW_WphiOWM}. We treat the amplitudes $\M_5^{rm cl}$ appearing in the computation as functions of the $w_i^\mu$ variables and we take out of the integral all other kinematic dependence, leaving us only with integrals of the form \cref{SINGLE:POLE:INT} and \cref{DOUBLE:POLE:INT}. 
We note that we can ignore the poleless - or contact - parts of the 5-point amplitude, as it was shown in \cite{DeAngelis:2023lvf} that only the residues in the factorization channels contribute with their prescription.

We are going to focus on the $I_2^{\mu_1 \dots \mu_n}$, $J_2^{\mu_1 \dots \mu_n}$ and comment at the end how to get integrals with $w_1$ insertions as well. The way to calculate the integrals is by first integrating over $w_1$ using the momentum conserving delta function and then applying the following parametrization which was first presented in \cite{Cristofoli:2021vyo}:
\begin{align}
    w_2^{\mu} = z_1 u_1^{\mu} + z_2 u_2^{\mu} + z_\upsilon \tilde{\upsilon}^\mu + z_b \tilde{b}^\mu, 
\end{align}
where
\begin{align}
     \centering \upsilon^\mu \equiv \varepsilon^{\mu \alpha \beta \rho} u_{1 \alpha} u_{2 \beta} \tilde{b}_\rho \ , \qquad \qquad \tilde{\upsilon}^\mu = \frac{\upsilon^\mu}{\sqrt{-\upsilon^2}} ,
\end{align}

\begin{align}
    \centering b \equiv \sqrt{-(b_1-b_2)^2} \ , \qquad \qquad \tilde{b}^\mu = \frac{b_1^\mu - b_2^\mu}{b} .
\end{align}

The Jacobian for this change of variables is $|\varepsilon_{\mu \nu \rho \sigma} \tilde{\upsilon}^\mu u_1^\nu u_2^\rho \tilde{b}^\sigma|=\sqrt{\gamma^2-1}$. 
A generic momentum $l^\mu$ can be written in terms of this parametrization:
\begin{align}
l^\mu = - [(\hat{u}_1 l) - \gamma(\hat{u}_2 l)] \hat{u}_1^\mu - [(\hat{u}_2 l) - \gamma (\hat{u}_1 l)] \hat{u}_2^\mu - (\tilde{b} l) \tilde{b}^\mu - (\tilde{\upsilon} l) \tilde{\upsilon}^\mu , 
\end{align}
with
\begin{align}
    \hat u_i \equiv {u_i \over \sqrt{\gamma^2-1} }
. \end{align}
Using this decomposition for $l =k$, the fact that $k^2=0$ imposes a constraint which can be expressed as:
\begin{align}
    (\tilde{b} k)^2 + (\tilde{\upsilon} k)^2 = 2 \gamma (\hat{u}_1 k)(\hat{u}_2 k) - (\hat{u}_1 k)^2 - (\hat{u}_2 k)^2 .
\end{align}

With the above, the measure becomes
\begin{align}
    d \mu = d^4z e^{i b_1 k} \sqrt{\gamma ^2-1} \delta(z_2+z_1 \gamma) \delta(z_1 +z_2 \gamma - u_1 \cdot k) e^{i (b_1 k + b z_b)} 
. \end{align}

We then proceed by performing the trivial $z_1$, $z_2$ integrations using the remaining delta functions setting their values to $z_1 = - \frac{(\hat{u}_1 k)}{\sqrt{\gamma^2-1}}$ and $z_2 = \frac{\gamma (\hat{u}_1 k)}{\sqrt{\gamma^2-1}}$. 
The integration over the $z_\upsilon$ variable can be performed by using the prescription used in \cite{DeAngelis:2023lvf}: 

\begin{align}
    \int_{- \infty}^{\infty} d z_\upsilon f(z_\upsilon, z_b) = \frac{1}{2} \bigg (  \int_{C^{(+)}} d z_\upsilon   f(z_\upsilon, z_b) - \int_{C^{(-)}} d z_\upsilon f(z_\upsilon, z_b) \bigg ) 
    ,\end{align}
where $C^{(+)}$ ( $C^{(-)}$) denotes the deformation of the integration contour in the upper (lower) complex half plane and $f(z_\upsilon, z_b)$ denotes the remaining function  the integral after the $z_1, z_2$ integrations.
This integration technique allows one to compute the integral by only picking up the residues on $z_\upsilon$ as the two contributions from the arcs at infinity of the two integrals cancel each other. 
These residues are actually associated to the factorization channels of the amplitude \cite{DeAngelis:2023lvf}. 
After completing that step, there is only one integration remaining over the variable $z_b$. 
We perform appropriate shifts and rescale $z_b \to (\hat{u}_i k) z $ with $i=\{1,2\}$.  
We then leave the integrals in an unintegrated form as a one dimensional integral over the variable $z$. 

Following these lines, for the single pole integrals one finds
\begin{align}
 \label{eq:INT_overw2sq-zbrep} 
\hspace{-2cm}
I_2^{\mu_1 \dots \mu_n}  = & - {\pi  e^{i b_1 k}  (\hat u_1 k)^n  \over   \sqrt{\gamma^2- 1}} 
 \int_{-\infty}^\infty { d z e^{ i  (\hat u_1 k) b z}    \over   \sqrt{z^2   + 1 }   } 
  \re \bigg \{ 
  \nnl & 
  \big [ \gamma    \hat u_2^{\mu_1} - \hat u_1^{\mu_1}  + z \tilde b^{\mu_1}  + i \sqrt{ z^2   + 1} \tilde v^{\mu_1}  \big ] 
  \big [ \mu_1 \to \mu_2 \big ] \dots  \big [ \mu_1 \to \mu_n \big ]    \bigg \} 
. \end{align}  

In this case, the $z$ integral can be evaluated into Bessel functions $K_n$, but for the purpose of calculating waveforms in the time domain it is convienient to leave to in an unintegrated form. 
If the $k$-th insertion is $w_1^{\mu_k}$ instead, then the $k$-th square bracket should be replaced by 
$ \big [ k^{\mu_k}/  (\hat u_1 k) + \hat u_1^{\mu_k}  - \gamma    \hat u_2^{\mu_k}   - z \tilde b^{\mu_k}  - i \sqrt{ z^2   + 1} \tilde v^{\mu_k}  \big ] $.

Moving to the double pole family, the result is

 \begin{align} 
   \label{eq:INT_overw1sqw2sq-zbrep}
J_2^{\mu_1 \dots \mu_n}     =  & 
  {\pi \over 2 \sqrt{\gamma^2 - 1}}  \int_{-\infty}^\infty   { d z   \over \sqrt{ z^2 + 1} }    \bigg \{ 
  \nnl & 
(\hat u_1 k)^{n-1}   e^{i b_1 k +   i (\hat u_1 k)  b z }   
\re \bigg [ {  \big [ \gamma    \hat u_2^{\mu_1}  - \hat u_1^{\mu_1}  + z \tilde b^{\mu_1}  +  i \sqrt{ z^2   + 1 }  \tilde v^{\mu_1}  \big ]  
\dots  \big [ \mu_1 \to \mu_n \big ] 
 \over
  \gamma  (\hat u_2 k) - (\hat u_1 k)  +  (\tilde b k)  z + i (\tilde v k) \sqrt{ z^2 + 1}   } 
 \bigg ] 
\nnl + & 
(\hat u_2 k)^{n-1}  e^{  i b_2 k + i (\hat u_2 k)  b z } 
\re \bigg [ 
{  \big [  {k^{\mu_1}\over \hat u_2 k}
 +     \hat u_2^{\mu_1}    - \gamma  \hat u_1^{\mu_1}      + z \tilde b^{\mu_1}  - i \sqrt{ z^2  + 1 } \tilde v^{\mu_1} \big ] 
\dots  \big [ \mu_1 \to \mu_n \big ]  
\over  
\gamma  (\hat u_1 k) - (\hat u_2 k)   - (\tilde b k)  z  +  i (\tilde v k) \sqrt{ z^2 + 1}  }
\bigg ] 
  \bigg \} 
 .  \end{align}
In the case of $w_1^{\mu_k}$ insertions, one should replace the $k$-th square bracket in the first line  by 
 $ \big [ {k^{\mu_k} \over (\hat u_1 k)} + \hat u_1^{\mu_k} -    \gamma    \hat u_2^{\mu_k}    - z \tilde b^{\mu_k}  -  i \sqrt{ z^2   + 1 }  \tilde v^{\mu_k}  \big ]$, 
 and in the second line by  
 $ \big [  \gamma  \hat u_1^{\mu_k}  -     \hat u_2^{\mu_k}    -    z \tilde b^{\mu_k}  +  i \sqrt{ z^2  + 1 } \tilde v^{\mu_k} \big ] $.

In order to compute the waveform in the time domain, as given in \cref{eq:SW_WphiOWM}, we have to complete the last two integrations over $z$ and $\omega$. 
We proceed by first integrating over the time domain, after extracting the correct powers of $\omega$ from the amplitude. 
In our case, at the $k$-th order in spin expansion we get $\sim \int_0^{\infty} d \omega \omega^{2+k} e^{- i \omega (t- (b_i n) - b (\hat{u_i} n) z)}$, which scales with an extra factor of $\omega^2$ compared to the leading order gravitational waveform. 
  
We can then exploit the general result:
\begin{align}
	  \int_0^{\infty} d\omega e^{-i \omega \tau} \omega^n= -  i^{n+1} \text{PV}^{(n)} \Big(\frac{1}{\tau} \Big ) + i^n \pi \delta^{(n)} (\tau) ,
\end{align}
to perform the frequency integration. 
Since the amplitude at leading order is a real function,  only the parts containing delta functions survive after adding the hermitian conjugate in \cref{eq:SW_WphiOWM}. 
This will be apparent in the main results of this work (see \cref{sec:QED_waveforms,sec:GBCS_waveforms}) and from the final form of the waveforms from the unitarity terms at all spin orders in \cref{app:allorders}. 
Then, the final integration of $z$ becomes trivial and the results amount  to  computing derivatives over $z$ and setting $z=T_i$ (after extracting appropriate factors out of the delta functions), where:
\begin{align}
    T_i = \frac{t - b_i \cdot n  }{(\hat{u}_i \cdot  n) b} ,
\end{align}
as introduced in the early works of Kovacs and Thorne~\cite{Kovacs:1977uw, Kovacs:1978eu}.

%================================================
\newpage
\section{Construction of classical contact terms} 
\label{app:CONT}
%================================================

\subsection{General contact terms for $2$ massive $\to$ $2$ massless scattering}

\label{app:GeneralCONT}

In this appendix we review the construction of arbitrary spinning 4-particle amplitudes in the case where two of the particles are massive (with equal masses) and two are massless in the classical limit. We will base our discussion on section 3.3 of \cite{Bautista:2022wjf} where the authors discuss how to build the most general contact terms in the case of the gravitational Compton amplitudes. We will be slightly more general and discuss general contact terms from the 4-point scattering of two massive-two massless particles.
For the massless particles a choice has to be made for the absolute value of their helicity, but we will keep the discussion as helicity independent as possible. We will assume factorization of the scalar and spinning part of the amplitude, that is also consistent with assuming minimal coupling - which leads to exponentiation of the amplitude part constructed from unitarity as seen already in all the examples studied in the literature. In order to start the discussion concerning the construction of our amplitude we repeat the necessary requirements we will impose to our contact deformations.
These are:

\begin{itemize}
\item	Proper factorization properties at the physical channels. This means, that any contact term that we add should respect the fact that its residue at any physical pole must vanish.

\item The amplitude should have no spurious poles, i.e. no poles apart from the ones appearing from the factorization property, dictated by unitarity.

\item Little group transformation properties of the amplitude should be satisfied. These are set from the helicities of the massless particles.

\item Correct dimensions, consistent with the amplitudes for 4-particle scattering $[\M_4]=0$.

\item Correct $\hbar$ scaling. In fact, the way we will construct the contact terms such that they do not change the $\hbar$ scaling of the amplitude. 
\end{itemize}

We consider scattering of four particles with momenta $p_1^\mu, p_2^\mu, p_3^\mu, p_4^\mu$, where $1$ and $2$ are massive particles and $3$ and $4$ are massless. We take all the momenta to be incoming. We first start by discussing all the possible four-vectors that enter in the problem and their $\hbar$ scaling. These are:

\begin{itemize}
\item $p_1^\mu \sim \hbar^0$,

\item $p_3^\mu \sim \hbar$ ,

\item $p_4^\mu \sim \hbar$. From this and $p_3^\mu$ we can build the linear dependent momentum exchange momenta: $q^\mu=-p_1^\mu - p_2^\mu = p_3 + p_4^\mu$ , which scales as $\hbar$ as well (that way we express $p_2^\mu$ always in terms of the rest of the momenta, from momentum conservation),

\item $\epsilon^\mu_{3, -} = \frac{\langle 3 | \sigma^\mu | \tilde{\zeta}]}{\sqrt{2}[3 \tilde{\zeta}]}=\epsilon^{* \mu}_{3,+} =  \tilde{\epsilon}^{ \mu}_{3,+} $, where obviously $\epsilon^{\mu}_{3,+} = \frac{\langle \zeta | \sigma^\mu | 3]}{\sqrt{2}\langle 3 \zeta \rangle }$ (notice the notation we use here for the complex conjugation: this notation will be used from here and onwards in this section). Here, $| \tilde{\zeta}]$ and $| \zeta \rangle$ are reference spinors. We can make a specific choice of reference spinors by specifying the gauge and the answer should not change. A convenient choice we will use is $| \zeta \rangle = |4 \rangle$ and $| \tilde{\zeta} ] = | 4 ]$. Polarization vectors can be checked that they do not scale with $\hbar$,

\item $\epsilon^\mu_{4, -} = \frac{\langle 4 | \sigma^\mu | \tilde{\xi}]}{\sqrt{2}[4 \tilde{\xi}]}=\tilde{\epsilon}^{ \mu}_{4,+}$, where obviously $\epsilon^{\mu}_{4,+} =\frac{\langle \xi | \sigma^\mu | 4]}{\sqrt{2}\langle 4 \xi \rangle }$ and now we choose $| \xi \rangle = | 3 \rangle$ and $| \tilde{\xi} \rangle= |3]$. Notice that now $\epsilon^\mu_{3, -} \sim \tilde{\epsilon}^{\mu}_{4 , -}$ and $\epsilon^\mu_{3, +} \sim \tilde{\epsilon}^{\mu}_{4 , +}$,

\item $a^\mu$, where $a^\mu = \frac{s^\mu}{m}$, where $s^\mu$ is the Pauli-Lubanski vector for particle 1. Recall that $a^\mu \sim \frac{1}{\hbar}$.

\end{itemize}

Notice that in this gauge we have $q \cdot \epsilon_i = p_3 \cdot \epsilon_i = p_{4} \cdot \epsilon_i = 0$ for any particle $i$ and any helicity. This basically tells us that the spin independent - or scalar - piece of the amplitude has to necessarily be proportional to $(p_1 \cdot \epsilon_i)^n$ or $(p_1 \cdot \tilde{\epsilon}_i)^n$ when written in terms of four-vectors (or covariant tensors more generally) in order to respect little group covariance. The power $n$ has to be dictated by the little group transformations. 

Note also that for higher spins that one, we can always construct polarization tensors by sewing together polartization vectors: $\epsilon^{\mu_1 \mu_2 ...\mu_n} = \epsilon^{\mu_1} \epsilon^{\mu_2} ... \epsilon^{\mu_n}$.

Now, we also write down the Mandelstam invariants for this problem:

\begin{align}
t &= (p_1 + p_3)^2 = m^2 + 2(p_1 p_3) , \nnl
%= m^2 + 2 \hbar (p_1 p_3) 
%\end{align}
%\begin{align}
u &= (p_2 + p_3)^2 = m^2 + 2 (p_2 p_3) = m^2 + 2(p_1 p_4)= m^2 - 2 %\hbar
(p_1 p_3) + 2 %\hbar^2 
(q p_3) ,  \nnl
%\end{align}
%\begin{align}
s &= (p_3+ p_4)^2 = 2 (p_3 p_4)
%= 2 \hbar^2 (p_3 p_4) .
 \label{eq:B.1}
\end{align}
%where in the last equalities we made explicit the dependence of the Mandelstams on $\hbar$ factors.

Pay attention to the fact that in the classical limit, $\hbar \to 0$, $u-m^2 \approx - (t - m^2) \sim \hbar$ .

We are also going to define the optical parameter $\xi$, as a combination of the Mandelstam variables that scales as $\hbar^0$:

\begin{align}
\xi^{-1} \equiv   \frac{m^2 s}{(t-m^2)(u-m^2)} \sim \hbar^0 
\end{align}

This definition of $\xi$ has an overall minus sign compared to the one in \cite{Bautista:2022wjf}. Our point here is to discuss spin effects and in order to do that we want to track down what is the minimum amount of contractions of the spin vector with other four-vectors in the problem that is needed to describe all the spin effects at arbitrary order.

To proceed for that task we have in mind the independent four-vectors (or tensors) that we discussed earlier and we are now going to introduce the vector $w^\mu= \frac{(t-m^2)}{2(p_1 \epsilon_{3,-})} \epsilon_{3,-}^\mu = \frac{(p_1 p_3)}{(p_1 \epsilon_{3,-})} \epsilon_{3,-}^\mu \to  \frac{(u_1 p_3)}{(u_1 \epsilon_{3,-})} \epsilon_{3,-}^\mu$, where the last equality is true in the classical limit (imposed from peaked wavepackets to the classical momentum value). The reason for the introduction of this vector is that 1) if we want to factorise the little group dependence, we do not want to work directly with $\epsilon_{3,-}^\mu$ because it scales with the little group of the particles (while the denominator inside the definition of $w^\mu$ takes care of that fact) and 2) because we want to work with vectors that scale as $\sim \hbar$ so that we get something $\sim \hbar^0$ when we "marry" them with $a^\mu$. 

One might think that we should, similarly, include $\tilde{w}^\mu = \frac{(t-m^2)}{2(p_1 \epsilon_{3,+})} \epsilon_{3,+}^\mu$
in the following discussion. 
However, one can show that:
\begin{align}
(w a)+ (\tilde{w} a) =- \frac{\xi}{1 - \xi} \big ( p_3 \cdot a - p_4 \cdot a \big ) , \label{eq:B.3}
\end{align}
showing that $(\tilde{w} a)$ can be written in terms of $\{ (p_3 a), (p_4 a), (w a) \}$. Therefore, only one of the two vectors $w^\mu$ or $\tilde{w}^\mu$ is needed to encompass all the possible contractions of the spin vector with polarization vectors (recall that only two were independent since the complex conjugates of $\epsilon_{3,h}, \epsilon_{4,-h}$ are related in this gauge). From \cref{eq:B.3} we can also see that the unphysical poles appearing in $wa$, $\tilde{w}a$ are related to kinematic poles for $\xi \to 1$. In \cite{Bautista:2022wjf} it was shown how this is related to a pole in real kinematics, for $\theta \to \pi$ in a specific reference frame where one of the massive particles (the black hole) is at rest.

These two vectors have an interesting property on the $s$-pole. Since $s=2(p_3 p_4) = \langle 3 4 \rangle [43] = - \langle 3 4 \rangle [34] $:

\begin{align} 
	&w^\mu \to p_3^\mu, \qquad \text{as} \qquad  [3 4 ] \to 0  \qquad \quad  ,	  &w^\mu \to -p_4^\mu, \qquad \text{as} \qquad \langle 34 \rangle \to 0 . \label{eq:B.4}
 \end{align}
 
These are straightforward to show. For the first one we have:

\begin{align}
	w^\mu = \frac{ 2(p_1 p_3)}{ 2(p_1 \epsilon_{3,-})} \epsilon_{3,-}^\mu = \frac{\langle 3|p_1| 3] \langle 3|\sigma^\mu | 4]}{2 \langle 3 |p_1 | 4]} = \frac{- \langle 3| p_1 \bar{\sigma}^\mu | 3 \rangle [34]+\langle 3 | p_1 | 4] [3 | \bar{\sigma}^\mu | 3 \rangle }{2\langle 3 | p_1 | 4]} , \label{eq:B.6}
\end{align}
by using Schouten identities. Then, in the limit $[34] \to 0$ we have:

\begin{align}
	w^\mu \to \frac{[3|\bar{\sigma}^\mu |3 \rangle}{2} = p_3^\mu .
\end{align}

For the second case we follow the same line
after we have first made the replacement $(p_1 p_3) \to - (p_1 p_4)$ which is true in the classical limit 
%- corrections to this replacement are of order $\mathcal{O}(\hbar)$ 
as we saw in \cref{eq:B.1}. Then, the rest of the steps are essentially the same.

A similar procedure applies for the $\tilde{w}^\mu$ vector as well. In this case, for example:

\begin{align}
	\tilde{w}^\mu = \frac{ \langle 3 | p_1 | 3] \langle 4 | \sigma^\mu | 3]}{2 \langle 4 |p_1 | 3]} = \frac{- \langle 3 4 \rangle [3| p_1 \sigma^\mu | 3]+\langle 3 | \sigma^\mu 3] [3|p_1 |4\rangle }{2 \langle 4 | p_1| 3]} \xrightarrow{\langle 3 4 \rangle \to 0 } p_3^\mu  .
\end{align}

Similarly, it can be shown that $\tilde{w}^\mu \to - p_4^\mu$ in the limit of $[34] \to 0$ up to $\mathcal{O}(\hbar)$ corrections (use again that $p_1 p_3 \approx p_1 p_4$ in the classical limit). 

There is one more non-trivial combination that we should consider contracting the spin vector with: $a^2$. However it can be shown that in the classical limit we can write:

\begin{align}
 \frac{(t-m^2)(u-m^2)}{4 m^2} a^2 \approx \xi \big ( w\cdot  a - p_3 \cdot a \big ) \big ( w \cdot a + p_4 \cdot a \big ) + \big (w a)^2 . \label{eq:B.8}
\end{align}
 
There is a caveat here: the relation \cref{eq:B.8} shows that the operator $a^2$ can be expressed in terms of $\{ (p_3 a), (p_4 a), (w a) \}$ (it is argued in \cite{Bautista:2022wjf} that it corresponds to the quadratic Casimir of $su(2)$). However, this is not true for $|a|=\sqrt{-a^2}$. On the contrary this operator can be shown that is actually linearly independent from $\{ (p_3 a), (p_4 a), (w a) \}$. Thus it can be included, in the combination $|a|'=\frac{(t-m^2)}{2m} |a| = \frac{(p_1 p_3)}{m} |a|$, so that the whole expression scales as $\hbar^0$ (as the other terms of this basis do). It was discussed in \cite{Bautista:2022wjf} that this operator's appearance signals non-conservative effects (i.e. black hole horizon absorption), which were also discussed from different approaches in \cite{Aoude:2023fdm, Jones:2023ugm, Chen:2023qzo, Cangemi:2023bpe}.

In total, our basis of spin dependent operators is: $\{ (p_3 a), (p_4 a), (w a), |a|' \}$ and with that at hand we want to write down the spin dependent part of the amplitude. We first discuss on the part that comes from unitarity. 

In \cite{Bautista:2022wjf}, the authors have minimally coupled gravity into consideration, that's why they only consider this specific amplitude. We will keep the discussion general and we will write our amplitude in a factorized manner:

\begin{align}
A_4^S = A_4^0 \bigg( F( (p_3 a), (p_4 a), (w a), |a|') + P_{\xi} ( (p_3 a), (p_4 a), (w a), |a|') \bigg ).  \label{eq:B.9}
\end{align}

Some explanations are needed to justify the form of \cref{eq:B.9}. Firstly, $A_4^0$ is the scalar part of the amplitude, which factorizes for minimal coupling as has been already observed in all the cases examined in the literature
This function carries all the little group weight as it should and because of that it will be proportional to $(p_1 \epsilon_{3, h_3})^n (p_1 \epsilon_{4, h_4})^m$ (equivalently replace $\epsilon \to \tilde{\epsilon}$ to include the other helicity configurations), where $n$ and $m$ are fixed by the little group transformations, as we have argued also before. 

The terms in the parenthesis are in general polynomials of our basis truncated at order $a^{2S}$, being effectively polynomials for finite spin quantum number. Here, $ F( (p_3 a), (p_4 a), (w a), |a|')$ denotes the function of the spin operator that will come only from the gluing on the residues and we are being as general as possible. This will be one of the exponentials in the case of (conservative) minimal coupling in gravity for either the helicity-preserving or helicity-flipping case for example. We assume that this function scales as $F( (p_3 a), (p_4 a), (w a), |a|') \sim \hbar^0$, which is reasonable since all higher $\hbar$ contributions would be subleading compared to the scalar part of the amplitude. 

The arbitrary polynomial, $P_\xi$, that we have inserted actually corresponds to the inclusion of contact terms. Before we comment on what restrictions this polynomial should obey, we are going to justify the subscript $\xi$: In general, if we are to add an arbitrary function to modify our amplitude, this can in principle also depend on the kinematics, parametrized by the Mandelstam invariants. However, we know that in classical Physics there is only one free parameter for the scattering of the objects, the scattering angle $\theta$. Therefore, we can choose only one combination of the Mandelstam invariants which has all the information concerning the kinematics of the scattering. We choose this to be the optical parameter $\xi$, since we also want a combination which is of order $\sim \hbar^0$ as well, so that it doesn't change the $\hbar$ scaling of the polynomial. Therefore, we conclude that the most general form of $P_\xi$ is a Laurent expansion in $\xi$:

\begin{align}
	P_\xi ( (p_3 a), (p_4 a), (w a), |a|') = \sum_m \xi^m p^{(m)} (  (p_3 a), (p_4 a), (w a), |a|') .
\end{align}

Now we must constrain the polynomial $p^{m}$. One constraint is that this polynomial should still only have contact terms, i.e. terms whose residue on the physical channels vanishes. We will impose that condition in the classical limit, in which $u-m^2 \approx - (t-m^2)$. Because of that we will no longer refer to poles in the $u$-channel, but only in $s$ and $t$ channels. This kind of poles can appear either from $A_4^0$ (in fact it will contain at least one of them) or from powers of $\xi$, with negative powers of $\xi$ giving poles in the $t$-channel and with positive powers of $\xi$ giving poles in $s$. 

We consider first the $t$-channel poles. One such pole can be cancelled by the inclusion of $(w a)$, since it scales like $\sim (t-m^2)$. For the case of a $\xi$ insertion, a factor $(w a)^2$ term is actually required to cancel both poles. 

On the other hand, poles in $s$ can be cancelled by the inclusion of the combination $( w \cdot a - p_3\cdot a) (w \cdot a+ p_4\cdot  a)$. The reason is \cref{eq:B.4}: Using this, we can see that if we multiply by this operator the residue at the $s$-pole is identically zero. The same thing applies for $w^\mu \to \tilde{w}^\mu$.

Now there remains one last thing, which is to cancel the unphysical poles inside $(w a)$, which are proportional to $\sim (p_1 \epsilon_{3,-})$ (or equivalently in $(\tilde{w} a)$, which are proportional to $\sim(p_1 \epsilon_{3,+})$). These poles can be rewritten in terms of spinors as $\sim \langle 3 | p_1 | 4] $ and $\sim \langle 4 | p_1 | 3] $ (which one is which is determined from the choice of helicity of the particles). 
These two poles are also connected through the identity:
\begin{align}
	\frac{\langle 3 | p_1 | 4] \langle 4 | p_1 | 3] }{m^2 s} = \xi -1 \label{eq:B.11}
.\end{align}
Equations \cref{eq:B.3} and \cref{eq:B.11} can be used to cancel the unphysical pole that appears in the amplitude when $\xi \to 1$. In the case of GR for example this pole also appears in the exponential as was discussed in \cite{Bautista:2022wjf}.

With this in mind we can build order by order the polynomial which will give us the classical contributions from the contact terms for any 4-point amplitude of 2 equal mass and 2 massless particles. 

\subsection{Contact terms for $\phi FF$ and $\phi F \tilde{F}$ interactions}

\label{appQEDCONT}

Now we wish to apply what learned to the case of the QED toy model with dilaton/axion interactions as presented in \cref{sec:dilaxion} to check consistency with our proposal in \cref{sec:GBCS_amplitudes} for the contact terms in at linear order in spin. Based on that discussion in the previous appendix we first write the amplitudes in \cref{eq:3.10} in terms of polarization vectors, by using the following relations:

\begin{align}
    -\frac{1}{2\sqrt{2}} \langle 3 | \sigma^\mu \bar{\sigma}^\nu | 3 \rangle = p_3^\mu \epsilon^\nu_- - p_3^\nu \epsilon^\mu_-  , \nnl
	-\frac{1}{2\sqrt{2}} [3| \bar{\sigma}^\mu \sigma^\nu | 3 ] = p_3^\mu \epsilon^\nu_+ - p_3^\nu \epsilon^\mu_+ , \label{eq:B.14}
\end{align}
and in this gauge:

\begin{align}
    &\epsilon^\mu_{3,-} = \frac{\langle 3 | \sigma^\mu | 4]}{\sqrt{2}[3 4]} ,\nnl
    &\epsilon^\mu_{3,+} = \frac{\langle 4 | \sigma^\mu | 3]}{\sqrt{2}\langle 3 4 \rangle } . \label{eq:B.15}
\end{align}
Then, the amplitudes read:

    \begin{align}
&\M_U[1_\Phi 2_{\bar{\Phi}} 3^-_\gamma 4_\phi ] = -  \frac{4 Q e  \hat \alpha}{ \Lambda  } \frac{( p_3 \cdot q) (p_1 \cdot \epsilon_-)}{s} e^{q \cdot a_1} , \nnl
&\M_U[1_\Phi 2_{\bar{\Phi}} 3^+_\gamma 4_\phi ] = -  \frac{4 Q e \hat \alpha^*}{ \Lambda  } \frac{( p_3 \cdot q) (p_1 \cdot \epsilon_+)}{s} e^{-q \cdot a_1} ,
    \end{align}
where we have dropped subscript on the polarization vector for simplicity and we made use of the fact that $p_4 \cdot \epsilon_{\pm} = 0$ in this gauge. Now, using that $p_3 \cdot q = - p_3 \cdot p_4 = - \frac{s}{2}$, we arrive in a final form for the amplitude:

    \begin{align}
&\M_U[1_\Phi 2_{\bar{\Phi}} 3^-_\gamma 4_\phi ] =  \frac{2 Q e \hat \alpha  (p_1 \cdot \epsilon_-)}{ \Lambda  }  e^{q \cdot a_1} , \nnl
&\M_U[1_\Phi 2_{\bar{\Phi}} 3^+_\gamma 4_\phi ] =  \frac{2 Q e \hat \alpha^* (p_1 \cdot \epsilon_+)}{ \Lambda  } e^{-q \cdot a_1} . \label{eq:B.17}
    \end{align}

Before moving on, a comment is in place: this last expression looks like it has no physical factorization poles. However, this is not true, but merely an artifact of the gauge we are working with, that let us drop the $ p_4 \cdot \epsilon_{\pm}$ terms. In fact, there are still poles in this answer hidden inside the polarization tensors as can be seen from \cref{eq:B.15}, which make each of the amplitudes in \cref{eq:B.17} scale as $\sim \frac{1}{\sqrt{s}}$. It is exactly these poles that should be cancelled when considering contact terms' deformations of these amplitudes.

Now let's first focus on the minus helicity case and examine what happens when we include contact terms. We deform that amplitude by including the most general polynomials on the basis $\{ p_3 \cdot a_1, p_4 \cdot a_1, w \cdot a_1, |a_1|' \}$, with $w^{\mu} = \frac{(t-m^2)}{2(p_1 \epsilon_-)} \epsilon^\mu_- = \frac{(p_1 p_3)}{(p_1 \epsilon_-)} \epsilon^\mu_-$:

\begin{align}
    \M[1_\Phi 2_{\bar{\Phi}} 3^-_\gamma 4_\phi ] =  \frac{2 Q e \hat \alpha  (p_1 \cdot \epsilon_-)}{ \Lambda  } \bigg \{ e^{q \cdot a_1} + P_\xi (p_3 \cdot a_1, p_4 \cdot a_1, w \cdot a_1, |a_1|') \bigg \} \label{eq:B.18}
\end{align}

Now, following the discussion in \cref{app:GeneralCONT} we can write down the polynomial at any $\xi$ order. Restricting to the $\xi^{-1}, \xi^0$ and $\xi$ pieces, the answer reads:

\begin{align}
	P_\xi = ... &+ \xi^{-1} (wa_1)^2 \bigg ( \sum_{i,j,k} A^{-1}_{ijk} (p_3 a_1)^i (p_4 a_1)^j (w a_1)^k \bigg )   
     \nonumber \\
	&+\xi^0 ((wa_1)-(p_3 a_1)) (   \bigg ( \sum_{i,j,k} A^0_{ijk} (p_3 a_1)^i (p_4 a_1)^j (w a_1)^k \bigg )       
    \nonumber \\
	&+ \xi  ((wa_1) - (p_3 a_1))^2 ((wa_1)+(p_4 a_1))  \bigg ( \sum_{i,j,k} A^1_{ijk} (p_3 a_1)^i (p_4 a_1)^j (w a_1)^k \bigg )    
 +...  ,
 \label{eq:B.19}
\end{align}
where $A^n_{ijk}$ are coefficients in front of all the possible independent four-vector contractions we can have. The indices ${i,j,k}$ are running from $0$ up to $2S-m$, where $m$ is the power of the spin vector outside the triple sum, such that the whole amplitude doesn't include terms with $a_1^{m}$ for $m>2S$. Notice that since there is no notion of crossing symmetry like in the gravity case \cite{Bautista:2022wjf}, there is no condition on the form of this polynomial. However, keep in mind that $\sim (w \cdot a_1)$ terms contain a spurious poles, which we impose that should be cancelled at each order in the spin, thus giving some extra conditions between some of the coefficients inside these polynomials. 

A small comment is in place: in a QED-like theory there are in fact no such objects like black holes. In turn, we do not consider any absorption effects, which is the reason why the polynomial does not include any terms containing $|a_1|$. This will not be the case in gravity as we will see in \cref{appGRCONT}.

Let's now focus at first order in the spin. We can see that there is only one independent term at this order that reads:

\begin{align}
    A^0_{000} \left( w \cdot a_1 -  p_3 \cdot a_1 \right).
\end{align}

Thus, rewriting the linear in spin part of \cref{eq:3.8} using \cref{eq:B.17} for the unitarity piece, we can write:

\begin{align}
    \M[1_\Phi 2_{\bar{\Phi}} 3^-_\gamma 4_\phi ] =  \frac{2 Q e \hat \alpha  (p_1 \cdot \epsilon_-)}{ \Lambda  } (q \cdot a_1) - \hat A_1 \frac{4 Q e }{ \Lambda  } (p_1 \cdot \epsilon_-)  ( w \cdot a_1 -  p_3 \cdot a_1 ), \label{eq:B.21}
\end{align}
where we now parametrize the Wilson coefficient in front as $\hat A_1=-\frac{1}{2} \hat{a} A^0_{000}$ and $[A_1]=0$. This contact term doesn't contribute when $[34] \to 0$, since $w^\mu \to p_3^\mu$ in this limit as shown in the \cref{app:GeneralCONT}. Not only that, but this is in fact the same contact term as in \cref{eq:3.12}. In order to see that, we rewrite the spinor contraction in \cref{eq:3.12} as:

\begin{align}
    \langle 3 | p_1 a_1 |  3  \rangle  = -  2 \sqrt{2} ( (p_1 \cdot p_3) ( \epsilon_- \cdot a_1) - (p_1 \cdot \epsilon_-)( p_3 \cdot a)  ) = - 2 \sqrt{2} (p_1 \epsilon_-) ( w \cdot a_1 - p_3 \cdot a_1 ). \label{eq:B.23}
\end{align}

So we can clearly see that the contact term we derived by inspection in \cref{eq:3.12} is the same one in \cref{eq:B.21} and the Wilson coefficients in these two expression are identical $\hat A_1 \equiv \hat C_1$. The same thing essentially happens for the other helicity configuration, which can be derived from \cref{eq:B.21} using crossing symmetry.

We could go on and explicitly calculate contact terms to higher orders in the spin vector using this construction, but we will not be interested in them in this work.

\subsection{Contact terms for Scalar-Gauss Bonnet and Cherns Simons Gravity} \label{appGRCONT}

Turning to the case of interest, we now examine the case of Scalar-Gauss Bonnet and Chern-Simons Gravity. These theories can be treated simultaneously, similar to the case of the QED toy model with dilaton and axion interactions. However, in this case we will also have to check that the contact terms preserve the shift symmetry of the theory; we will see that this is indeed the case, at least up to quadratic order.

We thus use again \cref{eq:B.14} to rewrite the amplitudes in \cref{eq:GBCS_M4pointPhiExp} and consider their contact deformations:

\begin{align}
	&\M[1_{\Phi_i} 2_{\bar\Phi_i} 3^-_h 4_\phi ]= \frac{4 \hat{\alpha}}{\mpl^2 \Lambda^2} s (p_1 \epsilon_-)^2 \bigg \{ e^{q \cdot a} + P_\xi (p_3a_1, p_4a_1, wa_1, |a_1|') \bigg \}.  \\
	&\M[1_{\Phi_i} 2_{\bar\Phi_i} 3^+_h 4_\phi ] =  \frac{4 \hat{\alpha}} {\mpl^2 \Lambda^2} s (p_1 \epsilon_+)^2 \bigg \{ e^{-q \cdot a} + P'_\xi (p_3a_1, p_4a_1, wa_1, |a_1|') \bigg \} .
\end{align}

In here we used that in the chosen gauge $q \cdot \epsilon_s =0$ and that 
$q \cdot p_3 = - \frac{s}{2}$. 
Notice one more time that, because of the appearance of the $s=q^2$ factor in front, the whole Amplitude scales as $\M%_U
[1_{\Phi_i} 2_{\bar\Phi_i} 3^{s}_h 4_\phi ] \to \hbar^2 \M%_U
[1_{\Phi_i} 2_{\bar\Phi_i} 3^{s}_h 4_\phi ]$. 

Note that, if we write explicitly the polarization vector $\epsilon^\mu_- \equiv \epsilon^\mu_{3,-} = \frac{\langle 3 | \sigma^\mu | 4]}{\sqrt{2} [34]}$ (respectively $\epsilon^\mu_+ \equiv \epsilon^\mu_{3,+} = \frac{\langle 4 | \sigma^\mu | 3]}{\sqrt{2} \langle 34 \rangle}$), we can clearly see the poles of the Amplitudes:

\begin{align}
	&\M[1_{\Phi_i} 2_{\bar\Phi_i} 3^-_h 4_\phi ] =  \frac{2 \hat{\alpha}}{\mpl^2 \Lambda^2} \frac{ \langle 4 3 \rangle \langle 3 | p_1 | 4 ]^2 }{[34]}  \bigg \{ e^{q \cdot a} + P_\xi (p_3a_1, p_4a_1, wa_1, |a_1|') \bigg \} \\
	&\M[1_{\Phi_i} 2_{\bar\Phi_i} 3^+_h 4_\phi ] =  \frac{2\hat{\alpha}}{\mpl^2 \Lambda^2} \frac{ [43]\langle 4 | p_1 | 3 ]^2 }{\langle 3 4 \rangle }  \bigg \{ e^{-q \cdot a} + P'_\xi (p_3a_1, p_4a_1, wa_1, |a_1|') \bigg \} .
\end{align}

Thus, in our construction of the contact terms we have only a $\sqrt{s}$ pole to cancel, just like in the previous case of \cref{appQEDCONT}. Focusing on $P_\xi$ momentarily we have:

\begin{align} & 
	P_\xi  (p_3a_1, p_4a_1, wa_1, |a_1|') =  ... + \xi^{-1} (wa_1)^2 \bigg ( \sum_{i,j,k} C^{-1}_{ijk} (p_3 a_1)^i (p_4 a_1)^j (w a_1)^k \bigg ) (1 + C^{-1} |a_1|') \nonumber \\
	&+  \xi^{-1} (wa_1) |a_1|' \bigg ( \sum_{i,j,k} D^{-1}_{ijk} (p_3 a_1)^i (p_4 a_1)^j (w a_1)^k \bigg ) \nonumber \\
	&+\xi^0 ((wa_1)-(p_3 a_1)) (   \bigg ( \sum_{i,j,k} C^0_{ijk} (p_3 a_1)^i (p_4 a_1)^j (w a_1)^k \bigg )(1 + C^0 |a_1|') \nonumber \\
	&+ \xi  ((wa_1) - (p_3 a_1))^2 ((wa_1) +(p_4 a_1))  \bigg ( \sum_{i,j,k} C^1_{ijk} (p_3 a_1)^i (p_4 a_1)^j (w a_1)^k \bigg )(1 + C^1 |a_1|')  +...  .
\end{align}

We see that we get exactly the same expansion as in \cref{eq:B.19} for the case of QED, as expected. Up to order $\mathcal{O}(a_1^2)$:

\begin{align}
	P_\xi \left( (p_3 a_1), (p_4 a_1), (w a_1), |a_1|' \right)&\big|_{\mathcal{O}(a_1^2)} = \xi^{-1} \big ( C^{-1}_{000}  (wa_1)^2 + D^{-1}_{000}  (wa_1) |a_1|' \big ) \nonumber \\
	&+ \xi^0 \bigg ( C^0_{000} ((wa_1)-(p_3 a_1)) + C^0_{100} ((wa_1)-(p_3 a_1)) (p_3 a_1) \nonumber \\
	& + C^0_{010} ((wa_1)-(p_3 a_1)) (p_4 a_1) + C^0_{001} ((wa_1)-(p_3 a_1)) (wa_1) \nonumber \\
	&+ C^0_{000} C^0 ((wa_1)-(p_3 a_1)) |a_1|' \bigg ).
\end{align}

At this spin order there is no restriction for the coefficients, since absorption effects can be present and any spurious poles cancel in the final form of the amplitude. The reason for the latter is simple: The only spurious pole that can appear comes from $w^\mu$ (or equivalently $\tilde w^\mu$ if we construct the other helicity configuration) and is proportional to $\sim \frac{1}{(p_1 \cdot \epsilon)}$. However, in the gravity case we have a factor of $(p_1 \cdot \epsilon)^2$ in front of the polynomial, so that makes the Amplitude spurious-pole free for any value of the contact term coefficients.

Let us now write down explicitly the contact terms, first for the case of the emission of a negative helicity graviton. Before that we  write down how the terms appearing inside the polynomial look like in terms of spinors:

\begin{align}
	&p_3 a = \frac{\langle 3|a_1|3]}{2} &,& &p_4 a = \frac{\langle 4|a_1|4]}{2} & &, & & wa = \frac{\langle 3|p_1|3] \langle 3|a_1|4]}{2 \langle 3|p_1|4] } . 
\end{align}

At linear order in spin there is only one structure available, which can be written in terms of spinors following \cref{eq:B.23}. The contact term is then:

\begin{align}
	\M[1_{\Phi_i} 2_{\bar\Phi_i} 3^-_h 4_\phi ] \Big |_{\mathcal{O}(a_1), C_1} &= - \frac{\hat{C}_1}{ \mpl^2 \Lambda^2} \langle 3 |p_1 q |3 \rangle \langle 3 |p_1 a_1 |3 \rangle    , \label{eq:11.60}
\end{align}
and $\hat{C}_1 = - \hat \alpha C^0_{000}$ is a Wilson coefficient, again of dimensions $[C_1]=0$. For the other helicity configuration, applying crossing symmetry, we have:

\begin{align}
	\M[1_{\Phi_i} 2_{\bar\Phi_i} 3^+_h 4_\phi ] \Big |_{\mathcal{O}(a_1), C_1}  =  { \hat C_1^*  \over \mpl^2 \Lambda^2} [ 3| p_1 q | 3]  [ 3| p_1 a_1 | 3]. \label{eq:11.61}
\end{align}

We see that the term generated at linear order obeys crossing symmetry and is in fact exactly the one that we had guessed in \cref{eq:GBCS_M4pointPhiContact1}

We will also check the shift symmetry of the quadratic order terms, by explicitly writing them in terms of spinors. We have:

 \begin{align}
	\M[1_{\Phi_i} 2_{\bar\Phi_i} 3^-_h 4_\phi ] \Big |_{\mathcal{O}(a_1^2), \hat C_{2,1}} &= \frac{ \hat C_{2,1}}{\mpl^2 \Lambda^2} \frac{\langle 4 3 \rangle \langle 3 | p_1 | 4]^2} { [34] }  \xi^{-1} (wa_1)^2  \nonumber \\
	& = - \frac{\hat C_{2,1} m^2}{4 \mpl^2 \Lambda^2}   \langle 3 | a_1 p_4|3 \rangle^2  ,
	\end{align}

\begin{align}	
	\M[1_{\Phi_i} 2_{\bar\Phi_i} 3^-_h 4_\phi ] \Big |_{\mathcal{O}(a_1^2), \hat C_{2,2}} &= \frac{\hat C_{2,2}}{\mpl^2 \Lambda^2} \frac{ \langle 43 \rangle \langle 3 | p_1 | 4]^2 } {[34]}   \xi^0 (w\ \cdot a_1 - p_3 \cdot a_1) (p_3 a_1 ) \nnl 
    &=   \frac{\hat C_{2,2}}{4 \mpl^2 \Lambda^2} \langle 3|p_1 a_1|3\rangle \langle 3 |  p_4 p_1 |3  \rangle (p_3 a_1) ,
	\end{align}

\begin{align}
	\M[1_{\Phi_i} 2_{\bar\Phi_i} 3^-_h 4_\phi ] \Big |_{\mathcal{O}(a_1^2), \hat C_{2,3}} &= \frac{\hat C_{2,3}}{\mpl^2 \Lambda^2} \frac{ \langle 43 \rangle \langle 3 | p_1 | 4]^2 } {[34]} \xi^0 (w\ \cdot a_1 - p_3 \cdot a_1) (p_4 a_1 ) \nnl
	& = - \frac{\hat C_{2,3}}{4 \mpl^2 \Lambda^2}  \langle 3|p_1 a_1|3\rangle \langle 3 | p_1 p_4 a_1 p_4  |3  \rangle , 
	\end{align}

\begin{align} 
	\M[1_{\Phi_i} 2_{\bar\Phi_i} 3^-_h 4_\phi ] \Big |_{\mathcal{O}(a_1^2), \hat C_{2,4}} &= \frac{\hat C_{2,4}}{\mpl^2 \Lambda^2} \frac{ \langle 43 \rangle \langle 3 | p_1 | 4]^2 } {[34]}  \xi^0 (w \cdot a_1 - p_3 \cdot  a_1) (w  a_1) \nnl
    &=  \frac{\hat C_{2,4}}{4 \mpl^2 \Lambda^2} (p_1 p_3) \langle 3 | p_4 a_1 | 3 \rangle \langle 3 | p_1 a_1 | 3 \rangle,
	\end{align}

\begin{align} 
	\M[1_{\Phi_i} 2_{\bar\Phi_i} 3^-_h 4_\phi ] \Big |_{\mathcal{O}(a_1^2), \hat C_{2,5}} &= \frac{\hat C_{2,5}}{\mpl^2 \Lambda^2} \frac{ \langle 43 \rangle \langle 3 | p_1 | 4]^2 } {[34]}  \xi^0 (w \cdot a_1 - p_3 \cdot  a_1) |a_1|' \nnl
    &= \frac{\hat C_{2,5}}{ 4 m \mpl^2 \Lambda^2} (p_1 p_3 )  \langle 3| p_1 q |3 \rangle \langle 3 | p_1 a_1 | 3 \rangle |a_1|,
	\end{align} ,

\begin{align} 
	\M[1_{\Phi_i} 2_{\bar\Phi_i} 3^-_h 4_\phi ] \Big |_{\mathcal{O}(a_1^2), \hat C_{2,6}} &= \frac{\hat C_{2,6}}{\mpl^2 \Lambda^2} \frac{ \langle 43 \rangle \langle 3 | p_1 | 4]^2 } {[34]}  \xi^{-1} (w a_1) |a_1|' \nnl
	&= \frac{\hat C_{2,6} m}{ 2 \sqrt{2}  \mpl^2 \Lambda^2} \langle 3| p_1 p_4 | 3 \rangle [ 4 | p_3 a_1 p_4 | 3 \rangle |a_1|,
	\end{align} ,
 
and the definition of the Wilson coefficients in this case is just:

\begin{align}
    \hat C_{2,1} &= 2 \hat \alpha C^{-1}_{000}, &  \hat C_{2,2} &=2 \hat \alpha C^{-1}_{100} , & \hat C_{2,3} &= 2 \hat \alpha C^{-1}_{010} , \nnl
     \hat C_{2,4} &= 2 \hat \alpha C^{-1}_{010}  ,  & \hat C_{2,5} &= 2 \hat \alpha C^{-1}_{000} C^0 ,  & \hat C_{2,6} &= 2 \hat \alpha D^{-1}_{000} .
\end{align}

It can be easily checked that all terms are shift symmetric. 

We can actually give an argument based on dimensional analysis why we believe this construction is meant to reproduce only terms which are shift symmetric: We had noticed before that the amplitude we constructed by residues had an extra $\hbar^2$ scaling compared to the case of the Compton amplitude in gravity. In fact, this actually encodes the double derivative structure of the theory, hidden inside the Gauss-Bonnet and the Chern Simons terms, which gives these interactions their shift symmetric nature. In other words, we can relate the extra $\hbar$ counting to derivatives acting on our massless fields - the scalar and the graviton in our case - drawing a map $\partial \leftrightarrow \hbar$. 

Moving on the construction the polynomial $P_\xi$, this is constructed in such a way such that it doesn't change the $\hbar$ scaling of the amplitude. Therefore, we can conjecture that it will therefore not affect the derivative structure of the theory leading to terms which could only come from an effective theory with two additional derivatives - like the theories we are considering - and it will end up contributing only shift symmetric terms inside the amplitude.

%////////////////////////////////////////
\section{Scalar Waveforms at all spin orders-Unitarity term}
\label{app:allorders}

In this appendix we are going to present results for the scalar waveforms at all spin orders in both scalar-Gauss Bonnet and Cherns Simons gravity theories by including only the contributions coming from the unitarity pieces. This essentially means setting all contact deformations' coefficients to zero. 

For that purpose we will employ our knowledge from \cref{app:INT} on how to compute the integrals appearing in the waveform calculations, extending the analysis to arbitrary spin orders. Therefore, we would need to consider a more general type of integrals. These are:

\begin{align}
    \mathcal{I}_{c,i}^{\mu_1,...,\mu_n} = \prod_{i=1,2} \int \hat{d}^4w_i \hat{\delta}(u_i \cdot w_i) \hat{\delta}^{(4)} (w_1+w_2 - k) e^{i b_i \cdot w_i} \frac{w_i^{\mu_1}...w_i^{\mu_n}}{w_i^2} \cosh\left(w_1 a_1 + w_2 a_2 \right) , 
\end{align}

\begin{align}
    \mathcal{I}_{s,i}^{\mu_1,...,\mu_n} = \prod_{i=1,2} \int \hat{d}^4w_i \hat{\delta}(u_i \cdot w_i) \hat{\delta}^{(4)} (w_1+w_2 - k) e^{i b_i \cdot w_i} \frac{w_i^{\mu_1}...w_i^{\mu_n}}{w_i^2} \sinh\left(w_1 a_1 + w_2 a_2 \right) ,
\end{align}

\begin{align}
    \mathcal{J}_{c,i}^{\mu_1,...,\mu_n} = \prod_{i=1,2} \int \hat{d}^4w_i \hat{\delta}(u_i \cdot w_i) \hat{\delta}^{(4)} (w_1+w_2 - k) e^{i b_i \cdot w_i} \frac{w_i^{\mu_1}...w_i^{\mu_n}}{w_1^2 w_2^2} \cosh\left(w_1 a_1 + w_2 a_2 \right) ,
\end{align}

\begin{align}
    \mathcal{J}_{s,i}^{\mu_1,...,\mu_n} = \prod_{i=1,2} \int \hat{d}^4w_i \hat{\delta}(u_i \cdot w_i) \hat{\delta}^{(4)} (w_1+w_2 - k) e^{i b_i \cdot w_i} \frac{w_i^{\mu_1}...w_i^{\mu_n}}{w_1^2 w_2^2} \sinh\left(w_1 a_1 + w_2 a_2 \right) ,
\end{align}

These integrals can be computed similarly to the integrals defined in \cref{SINGLE:POLE:INT} and \cref{DOUBLE:POLE:INT}. For convenience, one can define:

\begin{align}
	&A_k = (k a_1) + (\hat u_1 k)[-\hat u_1^\mu + y \hat u_2^\mu + z \tilde b^\mu] (a_{2\mu}-a_{1\mu}), \nnl
	&B_k = (\hat u_1 k) \sqrt{z^2+1} \tilde{\upsilon}^\mu (a_{2 \mu} - a_{1\mu}) . \nnl
    &\Gamma_k = (k a_2) + (\hat{u}_2 k)[ \hat{u}_2^\mu - y \hat{u}_1^\mu + z \tilde{b}^\mu] (a_{2\mu}-a_{1\mu}) , \nnl
	&\Delta_k = (\hat{u}_2 k) \sqrt{z^2+1} \tilde{\upsilon}^\mu (a_{2 \mu} - a_{1\mu}) .
\end{align} 

We note that these are related under $1 \leftrightarrow 2$ exchange:

\begin{align}
	A_k \Big |_{1 \leftrightarrow 2} = \Gamma_k & & , & & B_k \Big |_{1 \leftrightarrow 2} = - \Delta_k &  & , & &  \Gamma_k \Big |_{1 \leftrightarrow 2}= A_k & & , & & \Delta_k \Big |_{1 \leftrightarrow 2} = - B_k .
\end{align}

Following the same procedure as before we find:

\begin{align}
  \mathcal{I}_{c,2}^{\mu_1,...,\mu_n} = & - \frac{(\hat{u}_1k)^n  }{2 \sqrt{y^2-1}}\int \hat{d} z \frac{e^{i (b_1 k+ b (\hat{u}_1 k) z)}}{\sqrt{z^2+1}} \nonumber \\ 
&\times \text{Re} \Big [ [ - \hat{u}_1^{\mu_1} + y \hat{u}_2^{\mu_u1} + z \tilde{b}^{\mu_1} + i \sqrt{z^2+1} \tilde{\upsilon}^{\mu_1}]... [\mu_1 \to \mu_n] \cosh \left( A_k + iB_k \right) \Big ] , \label{IC2:INT}
\end{align}

\begin{align}
  \mathcal{I}_{s,2}^{\mu_1,...,\mu_n} = & - \frac{(\hat{u}_1k)^n  }{2 \sqrt{y^2-1}}\int \hat{d} z \frac{e^{i (b_1 k+ b (\hat{u}_1 k) z)}}{\sqrt{z^2+1}} \nonumber \\ 
&\times \text{Re} \Big [ [ - \hat{u}_1^{\mu_1} + y \hat{u}_2^{\mu_u1} + z \tilde{b}^{\mu_1} + i \sqrt{z^2+1} \tilde{\upsilon}^{\mu_1}]... [\mu_1 \to \mu_n] \sinh \left( A_k + iB_k \right) \Big ] . \label{IS2:INT}
\end{align}

Again, in the case of $w_1^{\mu_k}$ insertions, one should replace the $k$-th square bracket in the first line  by 
$ \big [ k^{\mu_k}/(\hat u_1 k) + \hat u_1^{\mu_k} -    \gamma    \hat u_2^{\mu_k}    - z \tilde b^{\mu_k}  -  i \sqrt{ z^2   + 1 }  \tilde v^{\mu_k}  \big ]$.

\begin{align}
	 \mathcal{J}_{c,2}^{\mu_1,...,\mu_n} =&  \frac{1}{4\sqrt{y^2-1}} \int  \frac{\hat{d}z }{\sqrt{z^2+1}}  \Bigg \{ e^{i((b_1 k) + z (\hat{u}_1 k) b)} (\hat{u}_1k)^{n-1} \nonumber \\
	& \qquad \times  \text{Re} \bigg \{  \frac{[ - \hat{u}_1^{\mu_1} + y \hat{u}_2^{\mu_1} + z \tilde{b}^{\mu_1} + i \sqrt{z^2+1} \tilde{\upsilon}^{\mu_1}]... [\mu_1 \to \mu_n] }{ [y (\hat{u}_2 k) - (\hat{u}_1 k) + (\tilde{b} k ) z + i (\tilde{\upsilon} k) \sqrt{z^2+1}] } \cosh \left( A_k +i B_k \right) \bigg \} \nonumber \\
    & + e^{i((b_2 k) + z (\hat{u}_2 k) b)} (\hat{u}_2 k)^{n-1} \nonumber \\
    &\qquad \times \text{Re} \bigg \{   \frac{[ k^{\mu_1}/{(\hat{u}_2k)} + \hat{u}_2^{\mu_1} - y \hat{u}_1^{\mu_1}  + z \tilde{b}^{\mu_1} + i \sqrt{z^2+1} \tilde{\upsilon}^{\mu_1}] ... [\mu_1 \to \mu_n]  }{(\hat{u}_2 k) [y (\hat{u}_1 k) - (\hat{u}_2 k) - (\tilde{b} k ) z - i (\tilde{\upsilon} k) \sqrt{z^2+1}] } \cosh \left( \Gamma_k + i \Delta_k \right) \bigg \} \Bigg \} . \label{JC2:INT}
\end{align}

\begin{align}
	 \mathcal{J}_{s,2}^{\mu_1,...,\mu_n} =&  \frac{1}{4\sqrt{y^2-1}} \int  \frac{\hat{d}z }{\sqrt{z^2+1}}  \Bigg \{ e^{i((b_1 k) + z (\hat{u}_1 k) b)} (\hat{u}_1k)^{n-1} \nonumber \\
	& \qquad \times  \text{Re} \bigg \{  \frac{[ - \hat{u}_1^{\mu_1} + y \hat{u}_2^{\mu_1} + z \tilde{b}^{\mu_1} + i \sqrt{z^2+1} \tilde{\upsilon}^{\mu_1}]... [\mu_1 \to \mu_n] }{ [y (\hat{u}_2 k) - (\hat{u}_1 k) + (\tilde{b} k ) z + i (\tilde{\upsilon} k) \sqrt{z^2+1}] } \sinh \left( A_k +i B_k \right) \bigg \} \nonumber \\
    & + e^{i((b_2 k) + z (\hat{u}_2 k) b)} (\hat{u}_2 k)^{n-1} \nonumber \\
    &\qquad \times \text{Re} \bigg \{   \frac{[ k^{\mu_1}/{(\hat{u}_2k)} + \hat{u}_2^{\mu_1} - y \hat{u}_1^{\mu_1}  + z \tilde{b}^{\mu_1} + i \sqrt{z^2+1} \tilde{\upsilon}^{\mu_1}] ... [\mu_1 \to \mu_n]  }{(\hat{u}_2 k) [y (\hat{u}_1 k) - (\hat{u}_2 k) - (\tilde{b} k ) z - i (\tilde{\upsilon} k) \sqrt{z^2+1}] } \sinh \left( \Gamma_k + i \Delta_k \right) \bigg \} \Bigg \} . \label{JS2:INT}
\end{align}

And one more time, to consider $w_1^{\mu_k}$ insertions we replace the $k$-th square bracket in the first line by
 $ \big [ k^{\mu_k}/(\hat u_1 k) + \hat u_1^{\mu_k} -    \gamma    \hat u_2^{\mu_k}    - z \tilde b^{\mu_k}  -  i \sqrt{ z^2   + 1 }  \tilde v^{\mu_k}  \big ]$, 
 and in the second line by  
 $ \big [  \gamma  \hat u_1^{\mu_k}  -     \hat u_2^{\mu_k}    -    z \tilde b^{\mu_k}  +  i \sqrt{ z^2  + 1 } \tilde v^{\mu_k} \big ] $.

In fact, we comment that only integrals with up to one $w_i^\mu$ insertions appear in the amplitude for these theories. In the process of computing the waveforms we also identified some recurring functions of our final integration variable $z$ which we list here:

\begin{align}
	a_1(z) = \frac{1}{\sqrt{z^2+1}} \bigg [ \frac{  (\hat{u}_2 n)^2 (\gamma^2-1)^2 \big [ \gamma(\hat{u}_2 n) - (\hat{u}_1 n) + (\tilde{b} n) z \big ] }{ (\gamma (\hat{u}_2 n) - (\hat{u}_1 n) + (\tilde{b} n) z)^2 + (\tilde{\upsilon} n)^2 (z^2+1)} &- (\gamma^2-1) [2 \gamma (\hat{u}_2n) - ( \hat{u}_1 n)]  \nnl
	&+ (\gamma^2-\frac{1}{2})[-(\hat{u}_1 n) + \gamma ( \hat{u} _2 n) + z (\tilde{b} n)]  \bigg ] , \label{eq:C.1}
\end{align}

\begin{align}
	a_2(z) = \frac{  (\hat{u}_2 n)^2 (\gamma^2-1)^2 (\tilde{\upsilon} n)  }{ (\gamma (\hat{u}_2 n) - (\hat{u}_1 n) + (\tilde{b} n) z)^2 + (\tilde{\upsilon} n)^2 (z^2+1)} -  (\gamma^2-\frac{1}{2}) (\tilde{\upsilon} n) ,\label{eq:C.2}
\end{align}

\begin{align}
	a_3(z) = \frac{1}{\sqrt{z^2+1}} \bigg [\gamma \sqrt{\gamma^2-1} (\tilde{\upsilon} n ) z -  \frac{ (\hat{u}_2 n) (\tilde{\upsilon} n) (\gamma^2-1)^{3/2} \Big ( z [ - (\hat{u}_1 n) + \gamma (\hat{u}_ 2 n) ]-   (\tilde{b} n) \Big )  }{ (\gamma (\hat{u}_2 n) - (\hat{u}_1 n) + (\tilde{b} n) z)^2 + (\tilde{\upsilon} n)^2 (z^2+1)} \bigg ] \label{eq:C.3}
\end{align}

\begin{align}
	a_4(z) = \gamma \sqrt{\gamma^2-1} (\tilde{b} n)  - \frac{ (\hat{u}_2 n) (\gamma^2-1)^{3/2} \Big ( z [ (\tilde{\upsilon} n)^2 + (\tilde{b} n)^2 ] + (\tilde{b} n) [- (\hat{u}_1 n) + \gamma (\hat{u}_2 n)] \Big )  }{ (\gamma (\hat{u}_2 n) - (\hat{u}_1 n) + (\tilde{b} n) z)^2 + (\tilde{\upsilon} n)^2 (z^2+1)} \label{eq:C.4}
\end{align}

We also redefine the rescaled $\tilde{A}_n$, $\tilde{B}_n$ functions as:

\begin{align}
    &\tilde{A}_n (z) = \frac{1}{b((\hat{u}_1 n)} A_n (z) = \frac{1}{b((\hat{u}_1 n)} \bigg \{(n a_1) + (\hat u_1 n)[-\hat u_1^\mu + \gamma \hat u_2^\mu + z \tilde b^\mu] (a_{2\mu}-a_{1\mu}) \bigg \}, \nnl
    & \tilde{B}_n (z) = \frac{1}{b((\hat{u}_1 n)} B_n (z) = \frac{1}{b((\hat{u}_1 n)} \bigg \{ (\hat u_1 n) \sqrt{z^2+1} \tilde{\upsilon}^\mu (a_{2 \mu} - a_{1\mu}) \bigg \}, \nnl
\end{align}
and similarly one can define $\tilde{\Gamma}_n(z)$ , $\tilde{\Delta}_n(z)$, but they will not appear in the final result as it was shown before that they are related to $A_n(z)$, $B_n(z)$ by $1 \leftrightarrow 2$ exchange.

We first focus on the result for the dilaton. For that purpose we take the amplitude in \cref{eq:GBCS_M5U} and set $\tilde{\alpha}=0$. The waveform for dilaton emission can be written as a double sum over derivatives of the aforementioned functions:

\begin{align}
W^U_{\phi} =- & \frac{\alpha m_1 m_2}{4 \pi^2 \mpl^3 \Lambda^2 \sqrt{\gamma^2-1}  (\hat{u}_1 n)^2 b^3}  \sum_{\ell=0}^{\infty} \sum_{m=0}^{\infty} \Bigg \{ \frac{(-1)^m i^{2(\ell+ m)}}{(2 \ell)! (2m)!} \frac{\partial^{2+2(\ell+ m)}}{\partial z^{2 +2(\ell+ m)}} \left[ a_1(z) \tilde{A}_n^{2\ell}(z) \tilde{B}_n^{2m}(z) \right]  \nnl
	&\qquad \qquad  +  \frac{(-1)^m i^{2(\ell +m)+2}}{(2\ell+1)! (2m+1)!} \frac{\partial^{4+2(\ell + m)}}{\partial z^{4 +2(\ell +m)}} \left[ a_2(z) \tilde{A}_n^{2\ell+1}(z) \tilde{B}_n^{2m+1}(z) \right] ] \nnl
	&\qquad \qquad  + i  \frac{(-1)^m i^{2(\ell + m)+1}}{(2\ell+1)!(2m)!} \frac{\partial^{3+2(\ell+m)}}{\partial z^{3 +2(\ell+ m)}} \left[ a_3(z) \tilde{A}_n^{2\ell+1}(z) \tilde{B}_n^{2m}(z) \right] \nnl
	&\qquad \qquad   + i   \frac{(-1)^m i^{2(\ell +m)+1}}{(2\ell)! (2m+1)!} \frac{\partial^{3+2(\ell +m)}}{\partial z^{3 +2(\ell +m)}} \left[ a_4(z) \tilde{A}_n^{2\ell}(z) \tilde{B}_n^{2m+1}(z) \right]  \Bigg \} \bigg |_{z=T_1 } \nnl
	& \qquad  \qquad \qquad \qquad \qquad \qquad \qquad \qquad \qquad \qquad \qquad \qquad \qquad \qquad \qquad+  (1 \leftrightarrow 2) ,
\end{align}
and the superscript $U$ is there to remind us that this is the waveform after completely neglecting contact terms' contributions.

We can write this in a more compact form by resumming this form of the waveform. Denoting $\frac{\partial}{\partial z}f(z) = f'(z)$ - and so on for higher derivatives - and exploiting the fact that $\tilde{A}''_n(z) = 0$ we have:

\begin{align}
	W^U_{\phi} =& - \frac{\alpha m_1 m_2}{4 \pi^2 \mpl^3 \Lambda^2 \sqrt{\gamma^2-1} (\hat{u}_1 n)^2 b^3}  \nnl
	& \qquad \times \Bigg \{  \text{Re} \Big [ \cosh ( \tilde B_n(T_1) \partial_z + i \tilde A_n (T_1) \partial_z)  a''_1(z)   +  a_1(T_1)    \cosh (  \partial_z  \tilde  B''_1(z) + i \tilde A_n (T_1) \overleftarrow{\partial_z} )  \Big ] \nnl
	& \qquad  \qquad - \text{Im} \Big  [ \cosh ( \tilde B_n(T_1) \partial_z + i \tilde A_n (T_1) \partial_z)  a''_2(z)   +  a_2(T_1)    \cosh (  \partial_z  \tilde  B''_1(z) + i \tilde A_n (T_1) \overleftarrow{\partial_z} ) \Big ] \nnl
	&  \qquad  \qquad - \text{Im} \Big [ \sinh(\tilde  B_n(T_1) \partial_z + i \tilde A_n (T_1) \partial_z)  a''_3(z)  +  a_3(T_1) \sinh(\tilde  B''_1(z) \partial_z + i \tilde A_n (T_1) \overleftarrow{\partial_z})  \Big ] \nnl
	& \qquad  \qquad -\text{Re} \Big [ \sinh(\tilde  B_n(T_1) \partial_z + i \tilde A_n (T_1) \partial_z)  a''_4(z)  +  a_4(T_1) \sinh(\tilde  B''_1(z) \partial_z + i \tilde A_n (T_1) \overleftarrow{\partial_z})  \Big ]  \Bigg \} \bigg |_{z=T_1}  \nnl
    &\qquad \qquad \qquad \qquad \qquad \qquad \qquad \qquad \qquad \qquad \qquad \qquad \qquad \qquad \qquad \qquad \qquad \qquad + (1 \leftrightarrow 2).
\end{align}

In this last form it is also manifest that the leading order  waveform is a real function.

The waveform for the axion emission in Chern Simons gravity is in fact closely related to the one for the dilaton we just calculated. This can be easily seen even at the level of the amplitude in \cref{eq:GBCS_M5U}. We can, actually, get the amplitude in Chern-Simons gravity from the one in dilaton-Gauss Bonnet by interchanging $\alpha \to \tilde{\alpha}$ and $\cosh(w_i a_i) \to i \sinh (w_i a_i) $ in the first line of \cref{eq:GBCS_M5U} and $ \sinh(w_i a_i) \to i\cosh(w_i a_i)$. This change just amounts to a shift of the functions defined in \cref{eq:C.1,eq:C.2,eq:C.3,eq:C.4}, $a_1(z) \to - a_3(z)$, $a_2 (z) \to - a_4(z)$ and $a_3(z) \to a_1(z)$, $a_4 (z) \to a_2(z)$. We can then write that the waveform in Chern Simons gravity at all spin orders is:

\begin{align}
W^U_{ \tilde \phi} = & \frac{ \tilde \alpha m_1 m_2}{4 \pi^2 \mpl^3 \Lambda^2 \sqrt{\gamma^2-1}  (\hat{u}_1 n)^2 b^3}  \sum_{\ell=0}^{\infty} \sum_{m=0}^{\infty} \Bigg \{ \frac{(-1)^m i^{2(\ell+ m)}}{(2 \ell)! (2m)!} \frac{\partial^{2+2(\ell+ m)}}{\partial z^{2 +2(\ell+ m)}} \left[ a_3(z) \tilde{A}_n^{2\ell}(z) \tilde{B}_n^{2m}(z) \right]  \nnl
	&\qquad \qquad  +  \frac{(-1)^m i^{2(\ell +m)+2}}{(2\ell+1)! (2m+1)!} \frac{\partial^{4+2(\ell + m)}}{\partial z^{4 +2(\ell +m)}} \left[ a_4(z) \tilde{A}_n^{2\ell+1}(z) \tilde{B}_n^{2m+1}(z) \right] ] \nnl
	&\qquad \qquad  - i  \frac{(-1)^m i^{2(\ell + m)+1}}{(2\ell+1)!(2m)!} \frac{\partial^{3+2(\ell+m)}}{\partial z^{3 +2(\ell+ m)}} \left[ a_1(z) \tilde{A}_n^{2\ell+1}(z) \tilde{B}_n^{2m}(z) \right] \nnl
	&\qquad \qquad   - i   \frac{(-1)^m i^{2(\ell +m)+1}}{(2\ell)! (2m+1)!} \frac{\partial^{3+2(\ell +m)}}{\partial z^{3 +2(\ell +m)}} \left[ a_2(z) \tilde{A}_n^{2\ell}(z) \tilde{B}_n^{2m+1}(z) \right]  \Bigg \} \bigg |_{z=T_1 } \nnl
	& \qquad  \qquad \qquad \qquad \qquad \qquad \qquad \qquad \qquad \qquad \qquad \qquad \qquad \qquad \qquad+  (1 \leftrightarrow 2) ,
\end{align}
or 

\begin{align}
	W^U_{\tilde \phi} =&  \frac{\tilde \alpha m_1 m_2}{4 \pi^2 \mpl^3 \Lambda^2 \sqrt{\gamma^2-1} (\hat{u}_1 n)^2 b^3}  \nnl
	& \qquad \times \Bigg \{  \text{Re} \Big [ \cosh ( \tilde B_n(T_1) \partial_z + i \tilde A_n (T_1) \partial_z)  a''_3(z)   +  a_3(T_1)    \cosh (  \partial_z  \tilde  B''_1(z) + i \tilde A_n (T_1) \overleftarrow{\partial_z} )  \Big ] \nnl
	& \qquad  \qquad - \text{Im} \Big  [ \cosh ( \tilde B_n(T_1) \partial_z + i \tilde A_n (T_1) \partial_z)  a''_4(z)   +  a_4(T_1)    \cosh (  \partial_z  \tilde  B''_1(z) + i \tilde A_n (T_1) \overleftarrow{\partial_z} ) \Big ] \nnl
	&  \qquad  \qquad + \text{Im} \Big [ \sinh(\tilde  B_n(T_1) \partial_z + i \tilde A_n (T_1) \partial_z)  a''_1(z)  +  a_1(T_1) \sinh(\tilde  B''_1(z) \partial_z + i \tilde A_n (T_1) \overleftarrow{\partial_z})  \Big ] \nnl
	& \qquad  \qquad + \text{Re} \Big [ \sinh(\tilde  B_n(T_1) \partial_z + i \tilde A_n (T_1) \partial_z)  a''_2(z)  +  a_2(T_1) \sinh(\tilde  B''_1(z) \partial_z + i \tilde A_n (T_1) \overleftarrow{\partial_z})  \Big ]  \Bigg \} \bigg |_{z=T_1}  \nnl
    &\qquad \qquad \qquad \qquad \qquad \qquad \qquad \qquad \qquad \qquad \qquad \qquad \qquad \qquad \qquad \qquad \qquad \qquad + (1 \leftrightarrow 2).
\end{align}

\bibliographystyle{JHEP}
\bibliography{classRad}

\end{document}